\documentclass{emulateapj}

\usepackage{color}
\usepackage{xspace}

\begin{document}

\title{Application of an {\sl XMM-Newton} EPIC Monte Carlo Technique to Analysis and 
Interpretation of Data for the Abell 1689, RX J0658-55, and Centaurus Clusters of Galaxies}

\author{ K.~Andersson$^{1,2}$, J.R.~Peterson$^{3,4}$, G.~Madejski$^{2,4}$}

\affil{$^1$ Stockholm University, S-106 91, Stockholm, Sweden}

\email{kanderss@physto.se}

\affil{$^2$ Stanford Linear Accelerator Center, Menlo Park, CA 94025}

\affil{$^3$ Purdue University, West Lafayette, IN 47907}

\affil{$^4$ KIPAC, Stanford University, Stanford, CA 94309}


\begin{abstract}  

We propose a new Monte Carlo method to study extended X-ray sources 
with the European Photon Imaging Camera (EPIC) aboard {\sl XMM-Newton}. The smoothed 
particle inference (SPI) technique, described in a companion paper, is 
applied here to the EPIC data for the clusters of galaxies Abell 1689, 
Centaurus, and RX J0658-55 (the ``bullet cluster'').  
We aim to show the advantages of this method 
of simultaneous spectral and spatial modeling over traditional X-ray spectral 
analysis. In Abell 1689 we confirm our earlier findings about structure 
in the temperature distribution and produce a high-resolution temperature map. 
We also find a hint of velocity structure within the gas, consistent with 
previous findings. 
In the bullet cluster, 
RX J0658-55, we produce the highest resolution temperature map ever to be 
published of this cluster, allowing us to trace what looks like the trail of the motion 
of the bullet in the cluster. We even detect a south-to-north temperature 
gradient within the bullet itself.  In the Centaurus cluster 
we detect, by dividing up the luminosity of the cluster 
in bands of gas temperatures, a striking feature to the north-east of the 
cluster core. We hypothesize that this feature is caused by a subcluster 
left over from a substantial merger that slightly displaced the core.
We conclude that our method is very powerful in determining the spatial 
distributions of plasma temperatures and very useful for systematic 
studies in cluster structure. 

\end{abstract}



\section{Introduction}

Early work on X-ray emission from clusters of galaxies, based on
X-ray data obtained with instruments of modest angular and spectral 
resolution, implied that the profiles of X-ray emission are smooth 
and that the spectra can be adequately described as nearly isothermal hot 
plasma, generally indicating relaxed structure.  This picture has 
changed markedly with precise imaging data from the {\sl Chandra} and {\sl XMM-Newton} 
instruments:  X-ray images of clusters reveal complex intensity 
distributions, where in general, the surface brightness lacks circular 
symmetry.  The cluster emission cannot be described by a single-
temperature plasma. Neither is the distribution of temperatures 
spherically symmetric, and it cannot simply be described as radially dependent 
\citep[e.g.\ ][]{markevitch00}.
In addition, simple but well-motivated models 
such as ``cooling flows'' fail to adequately describe the observations 
\citep[e.g.\ ][]{peterson01}, even for clusters that are 
otherwise ``relaxed.''  This is likely due to a complex history and 
physical processes associated with the cluster formation, which is 
yet to be fully understood.  Such 
complexity may include effects of recent merger activity; 
large-scale ``bubbles,'' presumably due to the interaction 
of the outflows produced by the central active galaxy; 
sharp abundance gradients associated with recently triggered star formation;  
or other, still unknown processes \citep{sanders}.  
Clearly, an analysis technique not relying 
on symmetry of the flux or temperature distribution is needed.  

The methods considered for this task include fitting
isothermal spectra to fixed or adaptively binned 
grids of photons across the detector plane
\citep[e.g.\ ][]{markevitch00,sanders}.  Another approach is 
imaging deprojection or ``onion-peeling'' methods 
\citep[e.g.\ ][]{fabian_depro}, which in turn have been extended 
to spectroscopic deprojection \citep{arabadjis,arnaud,kaastra,andersson}.  
An alternative method developed recently by some of us and termed 
``smoothed particle inference'' or SPI (\citet{peterson06};  hereafter PMA07) 
relies on a description of a cluster as a 
large set of ``primitives,'' which in our case are smoothed particles, 
or ``blobs.''  Each of those is described by its overall luminosity 
and a spatial position, but also by a Gaussian width, a single 
temperature, and a set of elemental abundances.  A large set 
(well upward of a hundred) of such primitives is then propagated 
through the instrument response using Monte Carlo (MC) techniques, and 
their parameters are adjusted via the use of Markov chain 
methods to map out the likelihood of the distribution as compared 
to the observation.  The resulting distribution is then a 
good ``nonparametric'' description of the cluster.  

In this paper, we describe the implementation of this method to 
observations obtained with the {\sl XMM-Newton} imaging detectors 
known collectively as
the European Photon Imaging Camera (EPIC).  We apply the method to data 
from three clusters of galaxies, namely Abell 1689, 
RX J0658-55, and the Centaurus cluster.  All three clusters exhibit 
a substantial degree of complexity, and furthermore, all three have 
good-quality, deep {\sl XMM-Newton} observations, which in turn can be 
compared against previous X-ray analyses as well as data from other 
instruments, such as the {\sl Chandra} observatory.  In particular, 
Abell 1689 is likely a merger or a superposition of two components 
aligned close to the line of sight \citep[e.g.\ ][]{andersson};  
RX J0658-55 reveals an ongoing merger 
close to the plane of the sky \citep{markevitch02};  
and the Centaurus cluster shows abundance gradients as well as 
filaments and bubbles, presumably caused by the energy deposited in 
the cluster by the central radio source \citep{fabian05}.   
The choice of objects should provide a good illustration 
(or test) of the method for three quite different cases.  
To perform this analysis, we constructed a Monte Carlo for the {\sl XMM-Newton} 
EPIC detectors that is analogous to the MC that some of us developed for the 
Reflecting Grating Spectrometer \citep[RGS;][]{peterson3}.  However, 
here we use the new Markov chain-based method, SPI, for parameter iteration, 
as described above.

Below, in Section 2, we describe the previous and current observations 
of the three clusters and summarize the data reduction procedures. 
In Section 3, we summarize the SPI method and describe the 
specifics of the {\sl XMM-Newton} EPIC response function used here. We also 
describe the application of the method to the data.  
In Section 4 we discuss the results of the analysis, including 
the spatially resolved maps of the spectral parameters. 
In Section 5 we summarize the paper and discuss the advantages and disadvantages 
of the method and, finally, we discuss the future applications of the method 
in Section 6.  


\section{Choice of targets and data reduction}
Here we describe the choice of targets selected to 
demonstrate the capabilities and versatility of our method. 
The three chosen targets span a broad range of complexity of the 
flux, temperature, and redshift. Note that the exact 
same analysis chain is applied to all objects. 

\subsection{Abell 1689} 
This cluster is one of the objects most studied with gravitational 
lensing techniques, as the optical images clearly reveal arcs and arclets, 
allowing strong-lensing analysis \citep{tyson95}.  Deep studies
with the ESO MPG Wide Field Imager provide additional constraints towards 
the determination of the cluster gravitational potential, via the use of weak 
lensing \citep{clowe01,king02}. More recently, the mass profile has been 
detailed further with combined strong and weak lensing, using {\sl Hubble Space Telescope (HST)} ACS and 
Subaru images \citep{broadhurst05}.
The optical data (including studies of galaxy velocities) indicate 
that the cluster contains sub-structures \citep{miralda95,lokas06}.  
The early X-ray data for this cluster 
indicated an X-ray intensity profile that implied a single, 
relaxed system, but the mass determination from the X-ray analysis 
indicated a lower mass (by about a factor of 2) than the mass value  
determined from lensing \citep[e.g.\ ][]{miralda95}.  

The analysis of the {\sl XMM-Newton} observation used in this paper, 
but using more traditional techniques than the SPI method,  
was presented by \citet{andersson}, and we refer the reader to that
paper for more extensive discussion of Abell 1689's properties as well as 
previous X-ray observations.  In summary, the X-ray emission appears 
to be symmetric, and the average temperature inferred for the region of 3$\arcmin$ 
in radius is $9.3 \pm 0.2$ keV, assuming the value of the Galactic column 
of $1.8 \times 10^{20}$ cm$^{-2}$ derived by \citet{dickey90}.  However, 
the spatial analysis indicates that the temperature of the emitting plasma is 
clearly not uniform, ranging from $\sim 7$ to $\sim 10$ keV, with a hint of 
a temperature gradient in the southwest - northeast direction.  In addition, 
the redshift of the emitting gas as inferred from the 
X-ray spectrum varies across the image, with a high-redshift structure to 
the east, with $z = 0.185 \pm 0.006$, separated from the rest of 
the cluster at $z = 0.17$.  This strongly suggests that the cluster 
consists of two components in projection, which either have started to merge, 
or are falling towards each other.  One of the premises of our study is 
to determine if the two sub-components are indeed related, or if they 
should be treated as two separate clusters that happen to be located 
close to the same line of sight in the sky and are observed in 
projection against each other.

\subsection{RX J0658-5557} 

RX J0658-5557 (or 1ES 0657-55.8), at $z=0.296$ was first discovered as an 
extended X-ray source by the {\sl Einstein} IPC \citep{tuc95} and was 
later found from {\sl Advanced Satellite for Cosmology and Astrophysics (ASCA)} 
data \citep{tuc98} to have a temperature 
of about 17 keV, although subsequent simultaneous 
analysis of {\sl ASCA} and {\sl ROSAT} PSPC data by \citet{liang00} 
suggested a lower value of $14.5^{+2.0}_{-1.5}$ keV.  Still, even the 
revised value of the temperature makes it one of the hottest known 
clusters.  The disturbed profile of the X-ray emission seen in the 
{\sl ROSAT} observation suggests an ongoing merger \citep{tuc98}.  
Furthermore, it is associated with a powerful radio halo \citep{lia02}, 
probably radiating via the synchrotron process, which suggests  
the presence of a population of ultra-relativistic particles.  
RX J0658-5557 was first observed by {\sl Chandra} in 2000 October with 24.3 ks of 
usable data \citep{markevitch02}, and now has a total of over 500 ks 
of {\sl Chandra} exposure clearly revealing the bow shock in front of 
the ``bullet'' emerging from the merger. The collision is clearly 
supersonic, with a Mach number of 3 deduced from the angle of the Mach cone. 

The {\sl XMM-Newton} data for the cluster have been analyzed previously in 
\citet{zhang04} and \citet{finoguenov05}, who find an average temperature of 
$13.6 \pm 0.7$ keV, using data in the 2-12 keV range and fixing 
$n_H$ to the Galactic value of $6.5 \times 10^{20}$ cm$^{-2}$ 
\citep{dickey90}.  In addition, the {\sl XMM-Newton} data have been  
used in joint spectral fits with the {\sl Rossi X-Ray Timing Explorer (RXTE)} 
data (in the context of 
the search for hard X-ray emission) by \citet{petro06},
where the hard X-ray flux would be by Compton scattering of the cosmic 
microwave background by the same relativistic 
particles that are responsible for the radio emission.  
In their analysis, \citet{petro06} use 
MOS and {\tt pn} data in the range 1.0 - 10.0 keV, 
and, adopting the absorbing column to be that measured for our Galaxy 
of $4.6 \times 10^{20}$ cm$^{-2}$ by \citet{lia02}, they infer 
an average temperature for the region within a 4$\arcmin$ radius 
of $12.0 \pm 0.5$ keV, in marginal agreement with that determined 
by \citet{finoguenov05}, with the difference probably being due to the use 
of different bandpasses, calibration files, and the assumed absorbing 
column density.  

Similarly to Abell 1689, the spatial structure of the gravitational 
potential in this cluster has been extensively studied 
using both strong and weak gravitational lensing.  
Weak-lensing analysis by \citet{clowe} and, more recently, 
joint weak- and strong- lensing analysis by \citet{clo06} and \citet{bra06}, 
reveals a striking offset between the peaks in the gravitating material and 
X-ray-luminous matter. This suggests a scenario in which the subclusters 
have collided head-on and the gas is being slowed 
down by ram pressure, while the 
dark matter is able to pass through more or less without resistance.  As such, 
this cluster offers one of the most compelling arguments for 
dark matter, which interacts only via gravitation, being distinct 
from the baryonic material, which is responsible for the X-ray luminosity.  

\subsection{The Centaurus cluster}

The Centaurus cluster, also known as Abell 3526, is one of the most 
nearby clusters, and because of this, it enables good spatially resolved studies 
of cluster structure.  It is a relatively cool cluster, at an average 
temperature of $kT = 3.68 \pm 0.06$ keV \citep{fukazawa98}, with a modest 
luminosity.  One of its distinguishing characteristics is the 
detection of spatially resolved gradients of elemental abundances, 
even with data of modest angular resolution (see, e.g., Fukazawa 
et al. 1994, 1998).  This, as well as a wide distribution of 
temperatures of the emitting plasma, is clear in the {\sl Chandra} data 
\citep{fabian05}, where the X-ray image shows a ``swirl'' in the central 
structure, with a filament extending towards the north-east. 
This feature could have been shaped by a strong magnetic field or 
a bulk flow within the intracluster medium. 
In \citet{sanders06}, the {\sl XMM-Newton} data are analyzed in combination with the 
{\sl Chandra} data in order to derive maps of abundances for separate elements. 
They find that the element ratios are consistent with the solar ratios, 
with a metallicity of up to twice the solar value. 
A recent observation using the XIS detector (the X-ray Imaging Spectrometer) 
aboard the {\sl Suzaku} 
satellite finds no evidence for bulk motion of the cluster gas and 
put an upper limit of $|\Delta v| < 1400~$km s$^{-1}$ on any line-of-sight 
velocity difference \citep{ota}.

\subsection{Observation details and data reduction}

The data were reduced using standard pipeline processing as of
{\sl XMM-Newton} \verb=SAS= version 6.5 producing photon event lists.
For the screening of soft proton
flares, we create light curves in the 10 - 12 keV band for MOS and in the 12 - 14 keV
band for {\tt pn} in 100 s bins. We discarded the data when the total flux reached
$3\sigma$ above the quiescent level \citep[cf.\ ][]{pratt02}.
After this cut, we perform a similar second cut on the soft flux
light curves in the 0.3-10 keV band for MOS and the 0.3-12 keV band for {\tt pn},
binned by 10s.
We use all event patterns (singles, doubles, triples and quadruples) for MOS
and singles and doubles only for {\tt pn}. We also require \verb=XMMEA_EM= for MOS and
\verb=XMMEA_EP= for {\tt pn}, and also \verb-FLAG-=0.
The resulting effective exposure times and gross count rates in the $0.3 - 10~$keV band 
for the three observations are shown in Table \ref{obstab}.

\begin{deluxetable}{llrr}
\tabletypesize{\scriptsize}
\tablecaption{Observation parameters \label{obstab}}
\tablewidth{0pt}
\tablehead{
\colhead{Name / ObsID} &
\colhead{Detector} &
\colhead{Exposure (ks)} & 
\colhead{Count Rate (s$^{-1}$)} \\
}
\startdata
{\it A1689} & \quad & \quad & \quad \\
\hline 
0093030101 & MOS & 37 & 4.4 \\
\quad & pn & 29 & 15 \\
\quad & \quad & \quad & \quad \\ 
{\it RX J0658-5557} & \quad & \quad & \quad \\
\hline
0112980201 & MOS & 25 & 3.3 \\
\quad & pn & 22 & 9.9 \\
\quad & \quad & \quad & \quad \\ 
{\it Centaurus} & \quad & \quad & \quad \\
\hline
0046340101 & MOS & 46 & 18 \\
\quad & pn & 42 & 60 \\
\enddata
\end{deluxetable}

In the case of A1689, we use the same data set as reported in 
\citet{andersson} but reprocessed with the latest software and 
calibration data.  
For the ``bullet'' cluster, we extracted and analyzed data collected 
during the {\sl XMM-Newton} pointing on 2000 October 20 - 21;  this is the 
same {\sl XMM-Newton} observation as was reported in \citet{zhang04}, 
\citet{finoguenov05} and \citet{petro06} (see above).   
Finally, the Centaurus observation was collected on 2002 January 3 
and is the same as the one used by \citet{sanders06}.

Using the SAS command \verb=eexpmap=, we create exposure maps with 
$1\arcsec \times 1\arcsec$ bins in detector coordinates for all detectors. 
These account for bad pixels and columns in the data and 
correct for varying exposure over the CCDs. 


\section{Application of SPI to {\sl XMM-Newton} data}

\subsection{Smoothed Particle Inference}
\label{spisec}

The method of Smoothed Particle Inference (SPI), which was constructed  
to model diffuse X-ray-emitting astrophysical sources, is 
described in PMA07. Here we only give a 
brief summary of the basic features of the method.

In order to describe currently observed diffuse X-ray sources, a 
model with thousands of parameters is required.  
Our choice of model is a set of spatially Gaussian X-ray 
emitters of an assigned spectral type. 
Each Gaussian ``blob'' is described by a spatial position, 
a Gaussian width, and a set of spectral parameters. 

In the Monte Carlo simulation, the flux of the astrophysical source at 
each energy and spatial position is converted to a prediction 
of the number of photons detected at a given detector position 
and energy.  We calculate the probability of detection, $D$, given 
the instrument response $R$ and the source model flux $F$:

\begin{equation}
D(x,y,p)=\int dE~d\theta~d\phi~R(x,y,p | \theta,\phi,E) 
~\frac{d^2F}{dE~d\theta~d\phi}.
\end{equation}

Here $(x,y)$ is the position on the detector, $p$ is the observed 
pulseheight, $(\theta,\phi)$ are sky coordinates and $E$ is the 
photon energy. 
This integral is calculated by simulating photons sequentially while 
taking into account mirror and detector characteristics 
(described in the next section, 3.2).

The goodness of fit of the model is calculated from the likelihood 
function of the model parameters. The model data consist of 
a finite number of simulated photons, and we use a two-sample likelihood 
statistic to assess the goodness of fit. 
We explore the parameter space of the model with the Markov 
chain Monte Carlo (MCMC) method. The high dimensionality of 
the parameter space requires a method that is capable of exploring this 
space without being trapped in local minima. 
The Markov chain step is Gaussian, with a width that varies 
depending on parameter history. 
We also find it necessary to make some parameters coupled. These are 
the same for all particles (global) and vary simultaneously. 


\subsection{The XMM-Newton EPIC response function}

The ESA {\sl XMM-Newton} satellite consists of three co-aligned X-ray telescopes 
and an optical/UV telescope \citep{jansen01}. 
The telescopes focus X-rays onto two reflection grating spectrometers (RGS) 
and onto three CCD arrays for imaging spectroscopy: MOS1, MOS2 and {\tt pn}. 
The imaging detectors are collectively referred to as the European Photon 
Imaging Camera \citep[EPIC, ][]{turner01}.

The full details of EPIC are covered in \citet{ehle06}, and its 
latest calibration is described in \citet{kirsch06}.
This section briefly describes the EPIC response function as calculated 
by us using the {\sl XMM-Newton} Current Calibration Files (CCF) 
library\footnote{All calibration data were obtained from 
ftp://xmm.vilspa.esa.es/pub/ccf/constituents/} and explains how it is 
used in the Monte Carlo. A detailed description of the 
response calculation and interpolation is described in \citet{peterson3} 
for the {\sl XMM-Newton} RGS Monte Carlo. F
or EPIC the detector response is different 
in structure, but the Monte Carlo process is the same.  

In summary, the response probability function of the EPIC cameras can 
be written as
\begin{eqnarray}
R(x,y,p|\theta,\phi,E) = A(E) v(\theta,\phi,E) 
\mathrm{psf}(x,y|\theta,\phi,E)  v_R(E) \times \nonumber \\ 
f(E)  q(E)  r(p|x,y,E)  i(x,y)
\label{responsefunc}
\end{eqnarray}
where $A(E)$ is the mirror effective area, $v(\theta,\phi,E)$ is 
the mirror vignetting, 
psf$(x,y|\theta,\phi,E)$ is the energy-dependent point-spread function (PSF), 
$v_R(E)$ is the Reflection Grating Array (RGA) vignetting, 
$f(E)$ is the filter transmission, $q(E)$ is the quantum efficiency, 
$r(p|x,y,E)$ is the pulse-height response, and $i(x,y)$ is 
an exposure map correcting 
for bad pixels and differences in exposure time between different CCDs.

This function gives the probability of detecting a photon of energy $E$ 
originating at sky coordinates $(\theta,\phi)$ as an event of pulse height 
$p$ at detector coordinates $(x,y)$.  The details of these parts of the detector response are given below. \\

{\bf Effective area and vignetting:} 
The effective mirror collection area of the XMM mirrors is a function of 
energy and decreases with off-axis angle. This decrease is known as vignetting 
and is caused by shadowing from neighboring mirror shells.
We obtain the on-axis effective area $A(E)$ and the 
vignetting $v(\theta,\phi,E)$ for 
each telescope from the XMM calibration 
files.\footnote{XRT1\_XAREAEF\_0008.CCF XRT2\_XAREAEF\_0009.CCF 
and XRT3\_XAREAEF\_0010.CCF.} 
The response files are interpolated linearly and 
rebinned in order to optimize the performance of the Monte Carlo simulation. 
The bin size for 
$A(E)$ is set to 50 eV, ranging from 0 to 10.45 keV. 
The $v(\theta,\phi,E)$ bins are 0.75 keV wide in energy and $0.01^{\circ}$ 
wide in off-axis angle. 

{\bf Point-spread function:}
We use a circular symmetric approximation for the PSF as it 
is described in the calibration files. 
For on-axis extended sources of $< 10 \arcmin$ in radial extent, this is 
a sufficient approximation for our purposes.
Here we approximate the PSF 
by use of the encircled energy for different 
photon energies available from the 
CCF.\footnote{XRT1\_XENCIREN\_0003.CCF, XRT2\_XENCIREN\_0003.CCF and 
XRT3\_XENCIREN\_0003.CCF.} 
We use encircled energy samplings for photon energies 
from 0 to 9 keV with a 1.5 keV spacing. For each energy band the encircled 
energy is sampled at $1.37\arcsec$ intervals out to $9\arcmin$ from the center of 
the PSF. 

{\bf RGA vignetting:}
In the MOS cameras a fraction of the photons are intercepted by the 
RGA with some energy dependence. This loss of photons for MOS1 and 
MOS2 is tabulated in the calibration 
files.\footnote{RGS1\_QUANTUMEF\_0013.CCF and RGS2\_QUANTUMEF\_0014.CCF in 
FITS extension RGA\_OBSCURATE.} In our Monte Carlo model, we use a sampling 
of this energy dependence of 0.5 keV.  

{\bf Filter transmission:}
The transmission of the optical blocking filters aboard {\sl XMM-Newton} is modeled for the 
thin, medium and thick filter configurations as a function of energy. We rebin 
the transmission as found in the existing calibration 
files\footnote{EMOS1\_FILTERTRANSX\_0012.CCF, EMOS2\_FILTERTRANSX\_0012.CCF 
and EPN\_FILTERTRANSX\_0014.CCF.}
into bins of 50 eV. 

{\bf Quantum efficiency:}
The ability of the detector CCDs to detect photons as a 
function of energy, or quantum efficiency (QE), 
is tabulated in the CCF response 
files.\footnote{EMOS1\_QUANTUMEF\_0016.CCF, EMOS2\_QUANTUMEF\_0016.CCF and 
EPN\_QUANTUMEF\_0016.CCF.} We rebin the quantum efficiency for 
the Full Frame mode for MOS and both the Full Frame and Extended Full Frame 
modes for {\tt pn}. We bin the QE in the range from 0 to 12 keV, with 15 eV spacing. 

{\bf Pulse height redistribution function:}
EPIC {\tt pn} response matrices are available from the {\sl XMM-Newton} SAS 
Web site.\footnote{See ftp://xmm.vilspa.esa.es/pub/ccf/constituents/extras/responses/}
In the {\tt pn} detector, the CCD response of the 12 CCDs varies with the distance 
from the line separating the two CCD rows. The {\tt pn} response matrices are available 
for every 20 pixel rows from rows 0-20 (Y0, at the edge) to rows 181-200 
(Y9, at the center). In the Monte Carlo simulation, 
we calculate the distance from the detector 
center line and use the correct {\tt pn} response matrix accordingly. 
We only use matrices for the Full Frame and Extended Full Frame 
observing modes, 
and only for event patterns 0-4 (singles and doubles) for {\tt pn}.

EPIC MOS response matrices are dependent on the observing epoch and should be 
chosen according to the satellite revolution when the observation was taken.
We use the SAS command \verb=rmfgen= to generate MOS response matrices for 
14 different epochs from revolution 101 to revolution 1021. In the Monte Carlo we choose 
whichever epoch is closest to the observation. 
For MOS we use imaging mode matrices with all event patterns 
(0-12; singles, doubles, triples and quadruples). 
Recently, it has been discovered that the MOS response is also dependent 
on distance from the detector axis (see 
the {\sl XMM-Newton} EPIC Response and Background File Page, 
update 2005-12-15\footnote{See http://xmm.vilspa.esa.es/external/xmm\_sw\_cal/calib/ epic\_files.shtml} ). 
In the XMM CCFs, the response is modeled in three different regions: a ``patch,'' ``patch wings,'' and 
outside ``patch.'' Therefore, we also generate response matrices for all three regions for 
each epoch. In the MC simulation all three are read in for the epoch in question, and the 
correct one is chosen on the basis of the location of the detected photon. 

The {\tt pn} response matrices are rebinned to an $800 
\times 800$ matrix with a constant 15 eV binsize from 0.05 to 12.05 keV 
for both energy and pulseheight. 
MOS matrices have a 15 eV binsize and run from 0 to 12 keV.
These matrices are integrated over pulseheight to form cumulative 
distributions. 

\subsubsection{Response Calculation}
In the Monte Carlo, photons are generated via probability density 
functions normalized to unity. We calculate the cumulative distribution 
by integrating the probability functions and then draw a number 
from 0 to 1 at random to choose a particular photon property. 
First, photons are chosen from a model function with a given set 
of input model parameters. The output variables are the photon 
energy and sky coordinates $(E, \theta, \phi)$.
Second, we predict the detector coordinates and pulseheight 
$(p, x, y)$ by drawing photons using a response function $R$. In order 
to maintain the proper effective area and exposure normalization, 
photons are sometimes discarded according to the proper response 
functions (i.e. mirror effective area, filter transmission, vignetting, 
quantum efficiency, and exposure map).
The response functions in equation (\ref{responsefunc}) are in general not 
analytic and have to be stored in memory on grids. In order 
to limit the amount of used internal memory. we save the functions 
on coarse grids and interpolate linearly 
to get intermediate values.

\subsection{Application to the Data}

\subsubsection{Data Binning}
\label{binsec}
In the framework of SPI, a three-dimensional adaptive 
binning technique is used to bin the photon event lists. 
The details of this are described in PMA07.  
The bins are rectangular in shape and are created so that each bin with 
20 photons or more is split in two starting out from the entire 
dataspace. 
The choice of which dimension ($x$,$y$ or pulse height) 
should be split is random, with 
a possibility to split one dimension more often, on average, than 
another.
Here we choose to split the pulseheight dimension on average 10 times more 
often than the $x$ and $y$ dimensions.
This results in three-dimensional bins that are fine in pulseheight 
and coarse in $x$ and $y$, suitable for spatially resolved spectroscopy. 
The sizes of the bins are defined with respect to the full size 
of the data space in that dimension 
($20\arcmin$ for $x$ and $y$ and 9.7 keV for pulseheight). 

For RX J0658-55 this results in 14361 bins with 13.3 photons 
bin$^{-1}$ on average, and for the A1689 data we get 31096 bins 
which equals 13.5 photons bin$^{-1}$ on average.
For the Centaurus data, the binning resulted in 197092 bins 
with 13.7 photons bin$^{-1}$ on average.  

For analysis, we use MOS data in the 0.3 - 10.0 keV range and {\tt pn} 
data in the 1.1 - 10.0 keV range. The {\tt pn} low-energy data are cut off 
due to {\tt pn}-MOS disagreement at low energies \citep[cf.\ ][]{andersson}. 
In both analyses we restrict ourselves to a $20\arcmin \times 20\arcmin$ spatial 
region centered on the respective cluster centers. 
After the model photons have been generated, they are binned on the 
same grid as the data photons before the two-sample likelihood 
is calculated.

\subsubsection{Spatially Resolved Spectral Model}
To model the cluster emission, we use a multi-component model 
consisting of spatially Gaussian smoothed particles, or ``blobs,'' of 
cluster emission. 
Each of these is described by a spatial position, a Gaussian width, 
a single temperature, a set of elemental abundances, and an 
overall flux. Since each particle is described by a Gaussian, there will 
always be overlapping components. This means that the model everywhere 
describes a multi temperature plasma.

In this case we set the spectral emission model to be described
by the MEKAL 
\citep{mekal1,mekal2,mekal3,mekal4} thermal plasma model with solar abundances 
absorbed by a WABS \citep{wabs} absorption model.  
The prior ranges for all parameters are set to be flat for a 
fixed range, except for the spatial Gaussian sigma which has 
a logarithmic prior distribution. 
The midpoint of the spectral parameter ranges is determined from simple 
spectral analysis using the full cluster emission. The width 
of the range is chosen from the values that are expected for that 
parameter in the cluster. It is, in general, an advantage 
to choose a parameter range that is wider than that expected. 
We choose 
to let both the equivalent hydrogen column and the redshift of the 
cluster plasma be variable in the analyses, as well as temperature 
and metallicity with respect to solar. 
The ranges used for the different clusters are shown in Table \ref{paramtab}.

\begin{deluxetable}{llll}
\tabletypesize{\scriptsize}
\tablecaption{Allowed parameter ranges \label{paramtab}}
\tablewidth{0pt}
\tablehead{
\colhead{Parameter} &
\colhead{Min value} &
\colhead{Max value} & 
\colhead{Global?} \\
}
\startdata

{\it A1689} & \quad & \quad & \quad \\
\hline
$n_H$ ($10^{20} $cm$^2$) & 0 & 3.5 & Y \\
$T$ (keV) & 5 & 11 & N \\
$Z/Z_{\odot}$ & 0.15 & 0.35 & N \\
$z$ & 0.15 & 0.21 & N \\
R.A. \& decl.\footnote{W.r.t. the nominal pointing of XMM.} (arcmin) & $-10\arcmin$ & $+10\arcmin$ & N \\
ln $\sigma$ (arcsec) & $0.5\arcsec$ & $5.5\arcsec$ & N \\
\hline

{\it RX J0658-5557} & \quad & \quad & \quad \\
\hline
$n_H$ ($10^{20} $cm$^2$) & 2.7 & 3.7 & Y \\
$T$ (keV) & 1 & 19 & N \\
$Z/Z_{\odot}$ & 0.12 & 0.32 & N \\
$z$ & 0.28 & 0.30 & N \\
R.A. \& decl.$^a$ (arcmin) & $-10\arcmin$ & $+10\arcmin$ & N \\
ln $\sigma$ (arcsec) & $0.25\arcsec$ & $4.25\arcsec$ & N \\
\hline

{\it Centaurus} & \quad & \quad & \quad \\
\hline
$n_H$ ($10^{20} $cm$^2$) & 0 & 2.2 & Y \\
$T$ (keV) & 0.5 & 9.5 & N \\
$Z/Z_{\odot}$ & 0.1 & 1.5 & N \\
$z$ & 0 & 0.015 & N \\
R.A. \& decl.$^a$ (arcmin) & $-10\arcmin$ & $+10\arcmin$ & N \\
$\sigma$ (arcmin) & $0.25\arcsec$ & $4.25\arcsec$ & N \\
\hline

\enddata
\end{deluxetable}

In order to describe both the spatial and spectral properties of 
the clusters adequately, we choose to use 700 particles for the A1689 
analysis and 600 for RX J0658-55 and Centaurus. 
A justification for this number of particles for data sets of similar 
complexity can be found in PMA07.  
However, we also do an analysis using only 100 
components in order to 
check the consistency of the broad spatially varying spectral 
properties of the clusters. 

\subsubsection{Background Model}
\label{backsec}
The {\sl XMM-Newton} instrumental background consists of three different main parts: 
particle background (soft protons), cosmic-ray-induced internal line 
emission, and electronic noise. 
In our background model, \verb=epicback=, the particle background 
is approximated by a power-law spectrum with a variable spectral index and 
a separate normalization for each detector. 
This part of the model is not propagated through the mirror model, 
but is exposed directly onto the CCDs. In fact, this background 
component is scattered somewhat by the mirrors and does have a radial 
dependence \citep{read05}. It decreases to about 80 \% of its central 
value $10\arcmin$ from the pointing axis. This effect will be included 
in future papers. However, here, since we are dealing with bright clusters, 
we assume that a spatially flat modeling of this background is sufficient. 

We determine the best-fit parameters of our background model using 
several EPIC observations with the filter wheel in the closed 
position\footnote{Observation IDs 0073740101, 0134521601, 0134522401, 0134720401,
0136540501, 0136750301, 0150390101, 0150390301, 0154150101, \\ 
0160362501, 0160362601, 0160362801, 0160362901, and 0165160501}
and ``blank fields'' with removed point 
sources,\footnote{Observation IDs 0147511601 and 0037982001} 
as well as background files compiled by the {\sl XMM-Newton} Science Operations Centre 
\citep[SOC;][]{lumb}. 
We find the best-fit value of the  
power-law index to be approximately $-0.22$ (varying from -0.20 to -0.24 in the 
different observations), and we choose to fix it at this 
value in the remaining analysis. 

The internal line emission constitutes of fluorescent 
lines excited by high energy particles in various 
materials of the detector. These lines include the Al K, Si K,  
Au M, Cr K, Mn K, Fe K, Ni K, Cu K$\alpha$, Cu K$\beta$, 
Zn K, Au L$\alpha$ and Au L$\beta$ complexes. The lines are approximated 
with delta functions at the respective energies, and this is a good 
approximation, considering the limited energy resolution of the CCDs. 
The emission is assumed to be uniform across the detectors, except 
for the Cu K emission, which is highly nonuniform, with a hole, 
devoid of emission, in the center of the detector plane.
We approximate this hole with the intersection of a circle with 
a $390\arcsec$ radius and a rectangle $630\arcsec$ wide. 
The relative normalization of these lines is determined individually 
for the MOS and {\tt pn} detectors, using the filter wheel-closed observations.

The electronic noise background includes bright pixels and columns, 
readout noise, etc., and is modeled as an exponential, 
$F \propto e^{-(E/E_i)}$, with a variable value of $E_i$, 
where $E$ is the energy assigned to the noise 
event. We find an appropriate value 
for $E_i$  to be $\sim 150$ eV when using the MOS cutoff 
at 0.3 keV and the {\tt pn} cutoff at 1.1 keV. This value is a fixed parameter 
in the analysis. This noise background is assumed to be spatially 
uniform.  

The fraction of photons going to each of these model components 
(particle, lines, and noise) is variable in the analysis and 
all have a flat prior from 0 to 1.  Also, the fraction of photons 
for each model component going to each detector (MOS1, MOS2, and {\tt pn}) 
is variable.
The total normalization of the background model with 
respect to the cluster model, as well as the hard and soft 
X-ray background (XRB) 
components, is then set as a variable parameter 
also with a flat prior from 0 to 1.

We model the soft Galactic X-ray background using a uniform 
emission component consisting of a MEKAL 
spectral model with WABS absorption. 
The plasma temperature is fixed at 0.16 keV, 
with a metal abundance of $0.3 Z_\sun$ at $z=0$, and the 
absorption is fixed at $n_H = 1.5 \times 10^{20}~$cm$^{-2}$. 
Similarly, the hard XRB, presumably due to 
superposition of unresolved AGNs, is modeled using a 
power law with $\Gamma=1.47$ and with absorption fixed at 
$n_H = 1.2 \times 10^{21}~$cm$^{-2}$. 
In general, it is customary to use zero absorption 
for the soft XRB and Galactic absorption for the hard component 
in XRB analysis \citep[e.g.\ ][]{hickox}. We choose here to 
use the above values simply due to the fact that they give a 
better fit to the data in our analysis of the blank fields 
mentioned above, as well as source-free regions outside the 
clusters analyzed here. 

\begin{figure*}[!htb]
  \begin{center}
    \begin{tabular}{cc}
      \includegraphics[width=2.5in,angle=-90]{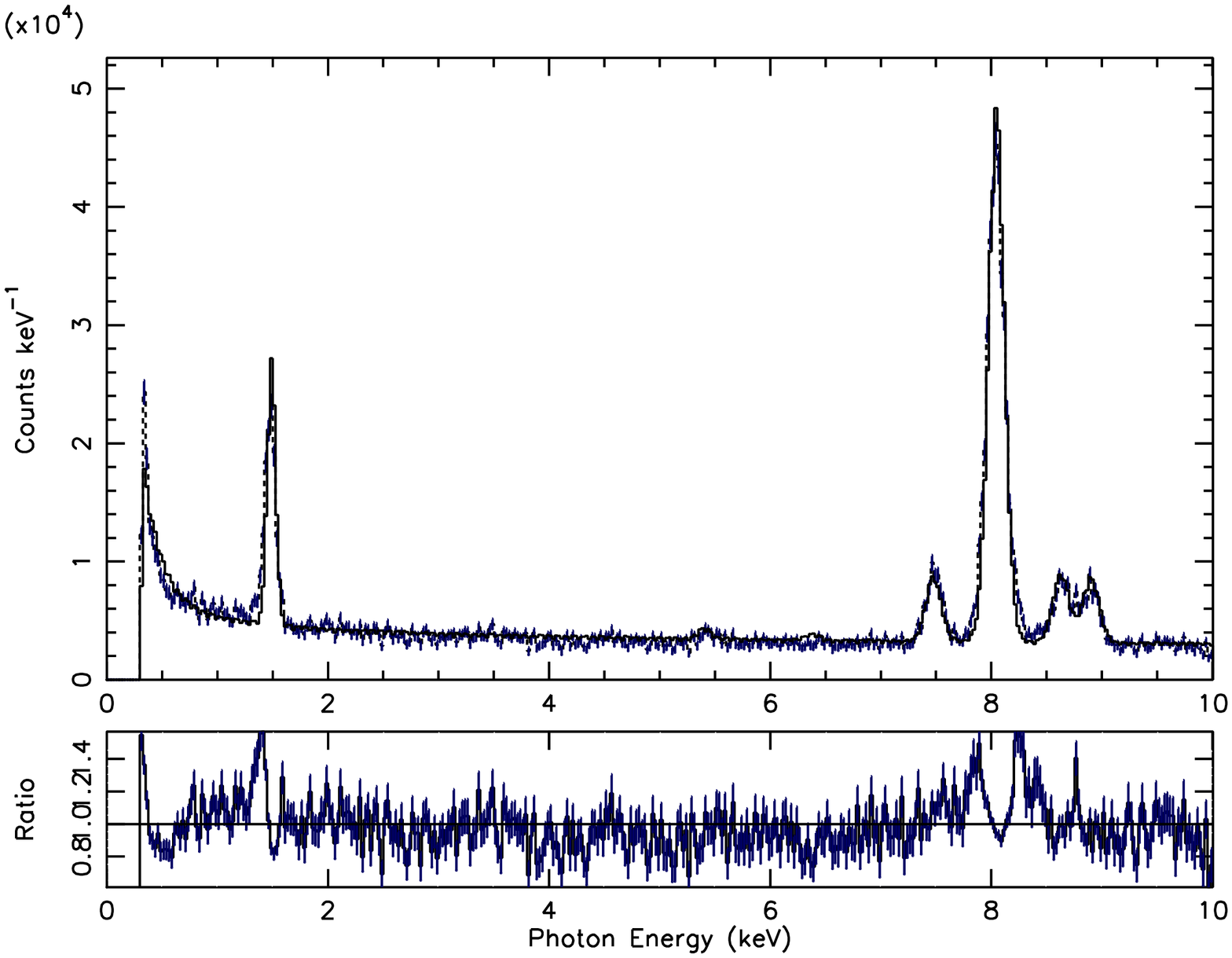} &
      \includegraphics[width=2.5in,angle=-90]{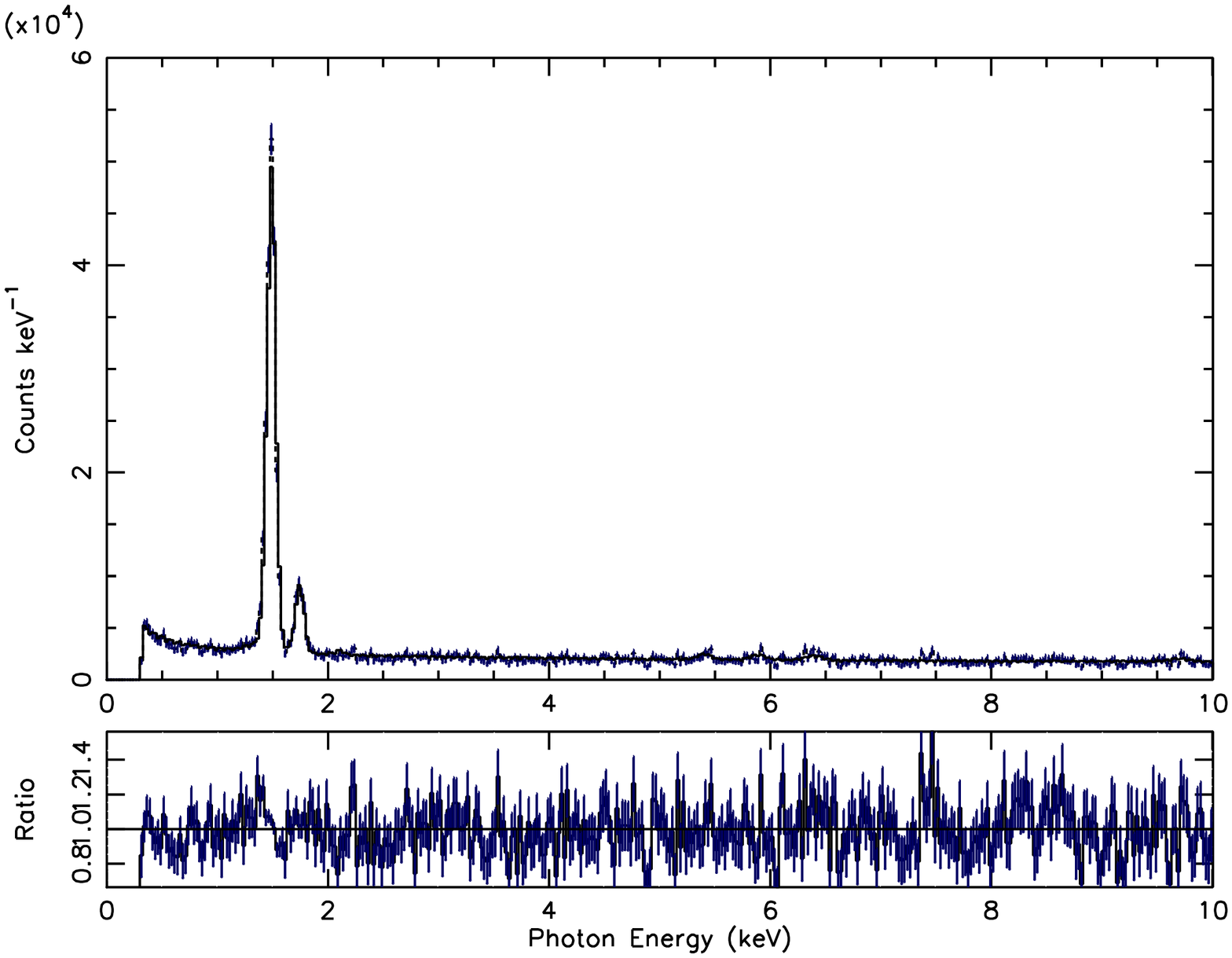} \\
    \end{tabular}
  \end{center}
  \caption{Typical background model including electronic noise, 
internal line emission, and particle background from a filter wheel-closed 
observation in {\tt pn} ({\sl left}) and for both MOS detectors ({\sl right}). The flux is 
shown in units of counts per 25 eV, with the data shown as dashed lines 
and the model as solid lines. The ratio of the two is shown in the lower panels. 
The effective exposure time for the {\tt pn} exposure is 28 ks, 
whereas for each MOS exposure it is 23 ks.
\label{epicback} }
\end{figure*}

An example of the background model spectrum for two observations 
with the filter wheel in the closed position is shown in 
Figure \ref{epicback}.
All parameters of the background models are global.

\subsubsection{Markov Chain Model Sample}
In the Monte Carlo, photons are simulated according to the probability 
functions given by the model parameters, as described in 
Section \ref{spisec}. The simulated photons are propagated through the 
detector model and binned on the grid determined by the three-dimensional 
photon density of the data, as described in Section \ref{binsec}.  
The two-sample likelihood 
Poisson statistic is calculated to provide a goodness of fit. 
In our analysis we use a model-to-data oversimulation factor 
of 10 for all data sets to reduce the model noise. 
In principle, it would 
be ideal to utilize a factor as large as possible, but we are limited by 
finite CPU speed and internal memory.  In PMA07 we compare 
different values of this factor and show that the results improve 
significantly when using a value of 10 or greater. Parameters are iterated 
by Markov chain sampling, as described in the previous Sections. 

\begin{figure*}[!htb]
  \begin{center}
    \begin{tabular}{ccc}
      \includegraphics[width=1.5in,angle=-90]{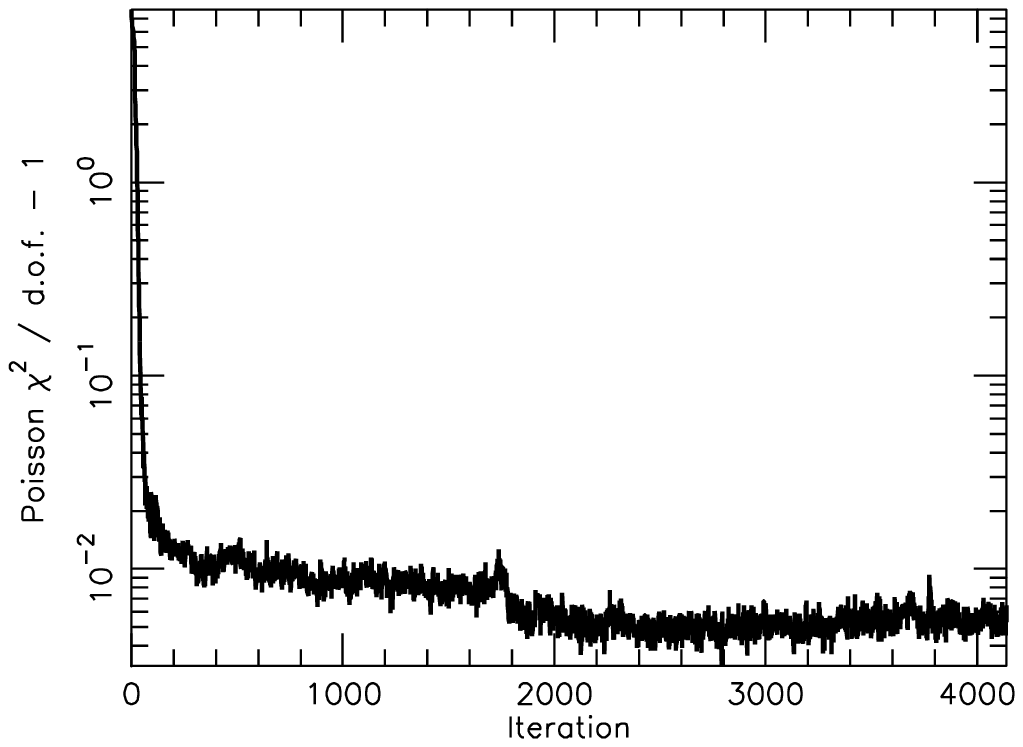} & 
      \includegraphics[width=1.5in,angle=-90]{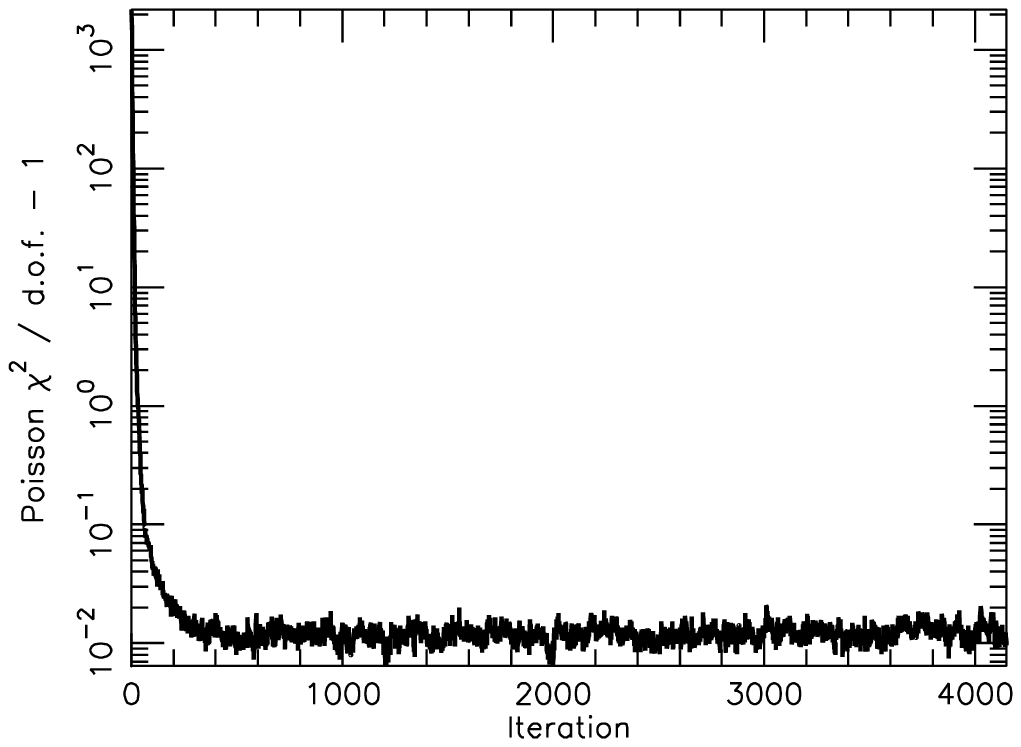} &
      \includegraphics[width=1.5in,angle=-90]{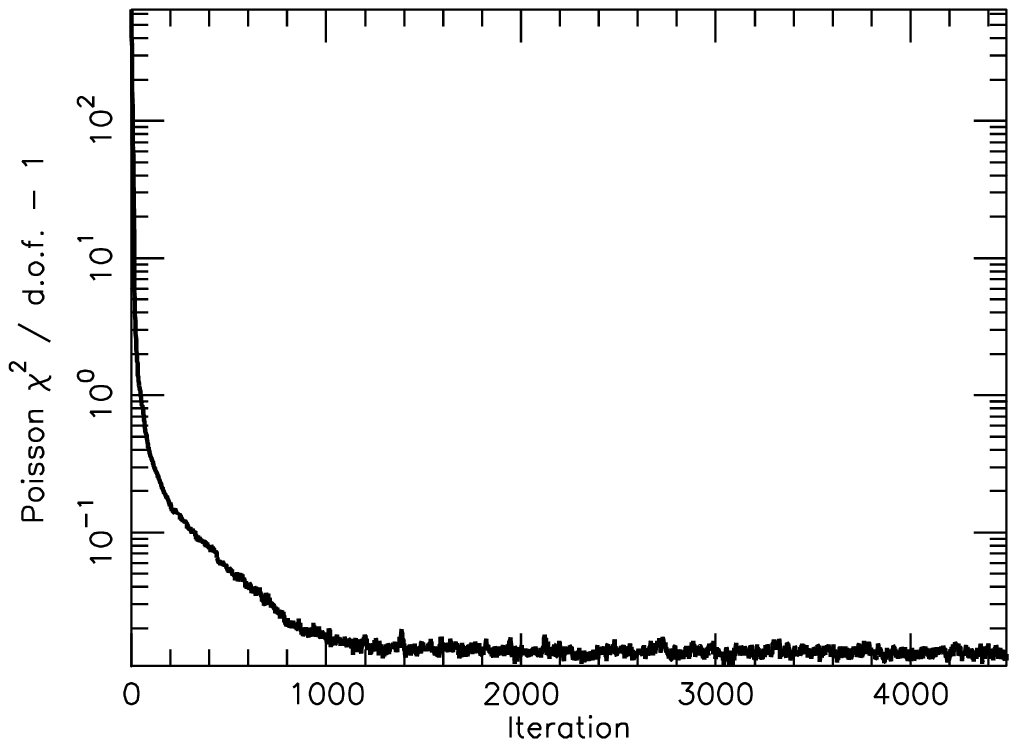} \\
      \includegraphics[width=1.5in,angle=-90]{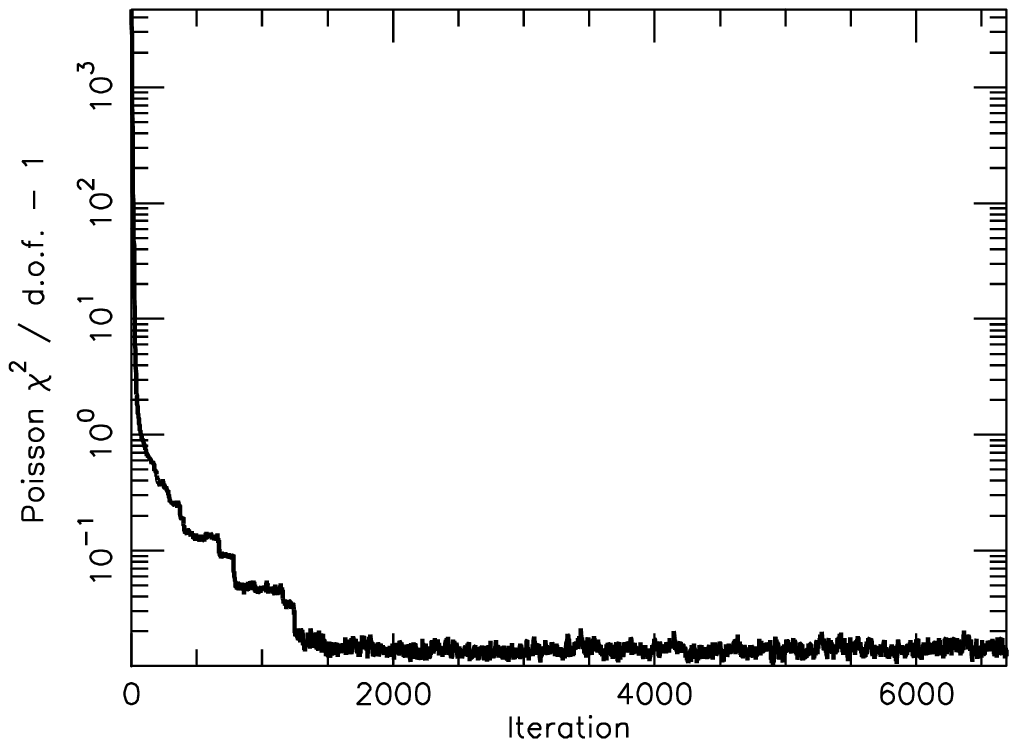} &
      \includegraphics[width=1.5in,angle=-90]{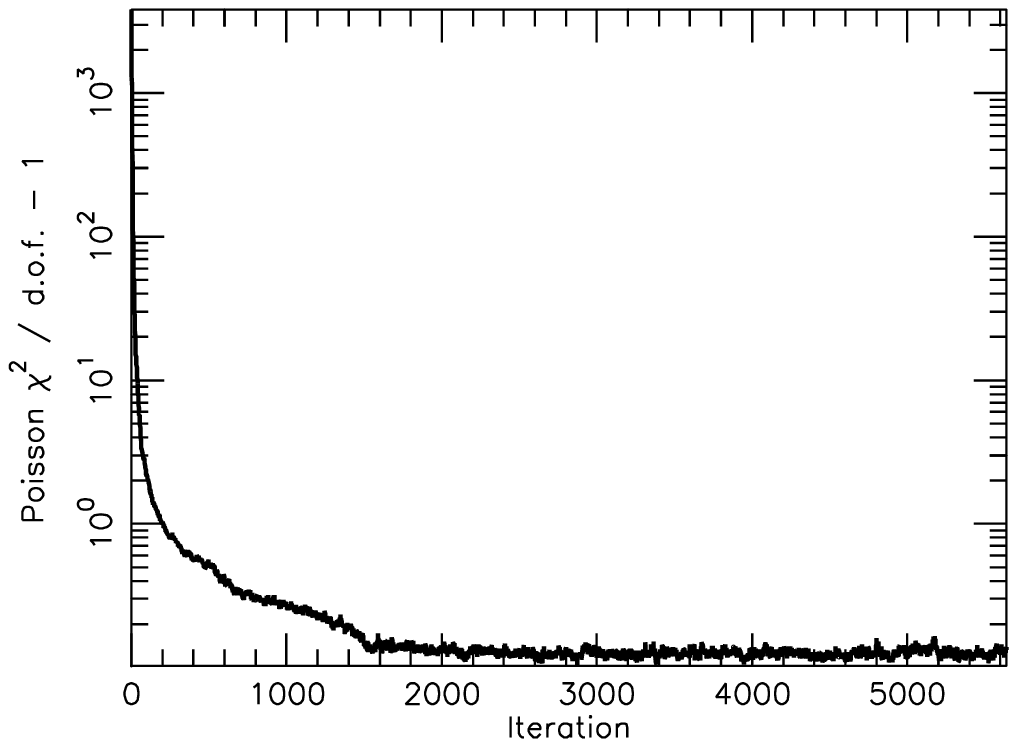} & 
      \includegraphics[width=1.5in,angle=-90]{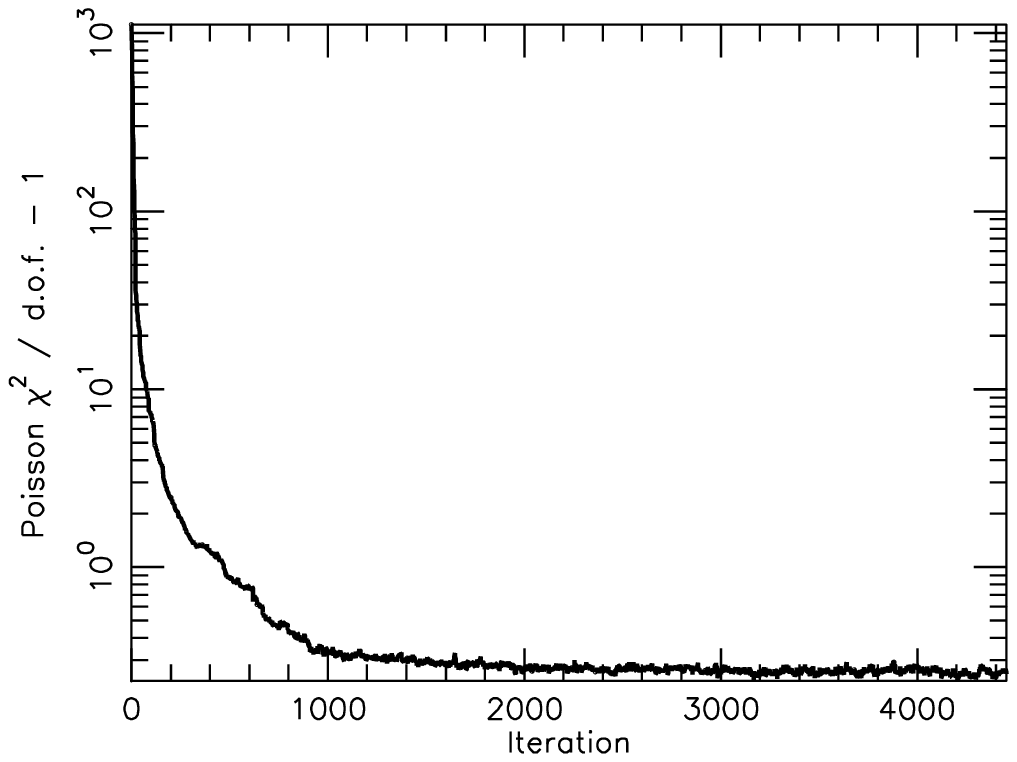} \\ 
    \end{tabular}
  \end{center}
  \caption{Poisson $\chi^2$ per degree of freedom for ({\sl top, from left to right}), 
the Abell 1689, RX J0658-5557, and Centaurus standard runs (700,600 and 600 blobs respectively),
and ({\sl bottom, from left to right}) the Abell 1689, RX J0658-5557, and Centaurus 100 blob runs 
\label{statrej}. }
\end{figure*}

Figure \ref{statrej} shows the evolution of the statistic 
(Poisson $\chi^2$/dof.)$ - 1$.  
Illustrated are ({\sl from left to right}) values for 
the 700 blob run for A1689, the 600 blob run for RX J0658-55 and the 
600 blob run for Centaurus. 
The statistic (Poisson $\chi^2$/dof) approaches a value very close to 1 
and stabilizes after $\sim 2000$ iterations. The value of the statistic 
at iteration 2000 for the three data sets (1.005, 1.013 and 1.016) is taken 
as an indicator that we have achieved an accurate fit.

In the bottom panels of the same figure we show the evolution of the (Poisson $\chi^2$/dof)-1  
value for the control runs with 100 smoothed particles. 
In general, the runs with more blobs reach a good fit faster and 
approach a lower value of the fit statistic. 
On the basis of these plots we consider the chain to be stable after 2000 iterations, 
when the statistic becomes stationary and close to 1. 
We only use samples from iteration 2000 on to deduce cluster properties in 
the following sections.  

\section{Final model}
To visualize the output model sample we use the methods described 
in PMA07. We stop the Markov chains after 4000 iterations and assume 
convergence after 2000 iterations, as described above.  We are left 
with 2000 models, each consistent with the data. Each iteration takes 
approximately 30 minutes on a single Intel Pentium 4 2.0 GHz CPU which 
results in about 3 months of computing time per cluster.  The models are 
filtered and marginalized in order to deduce the cluster properties 
that are discussed below.

\subsection{Abell 1689}
\begin{figure*}[!htb]
  \begin{center}
    \begin{tabular}{cc}
      \includegraphics[width=2.5in,angle=-90]{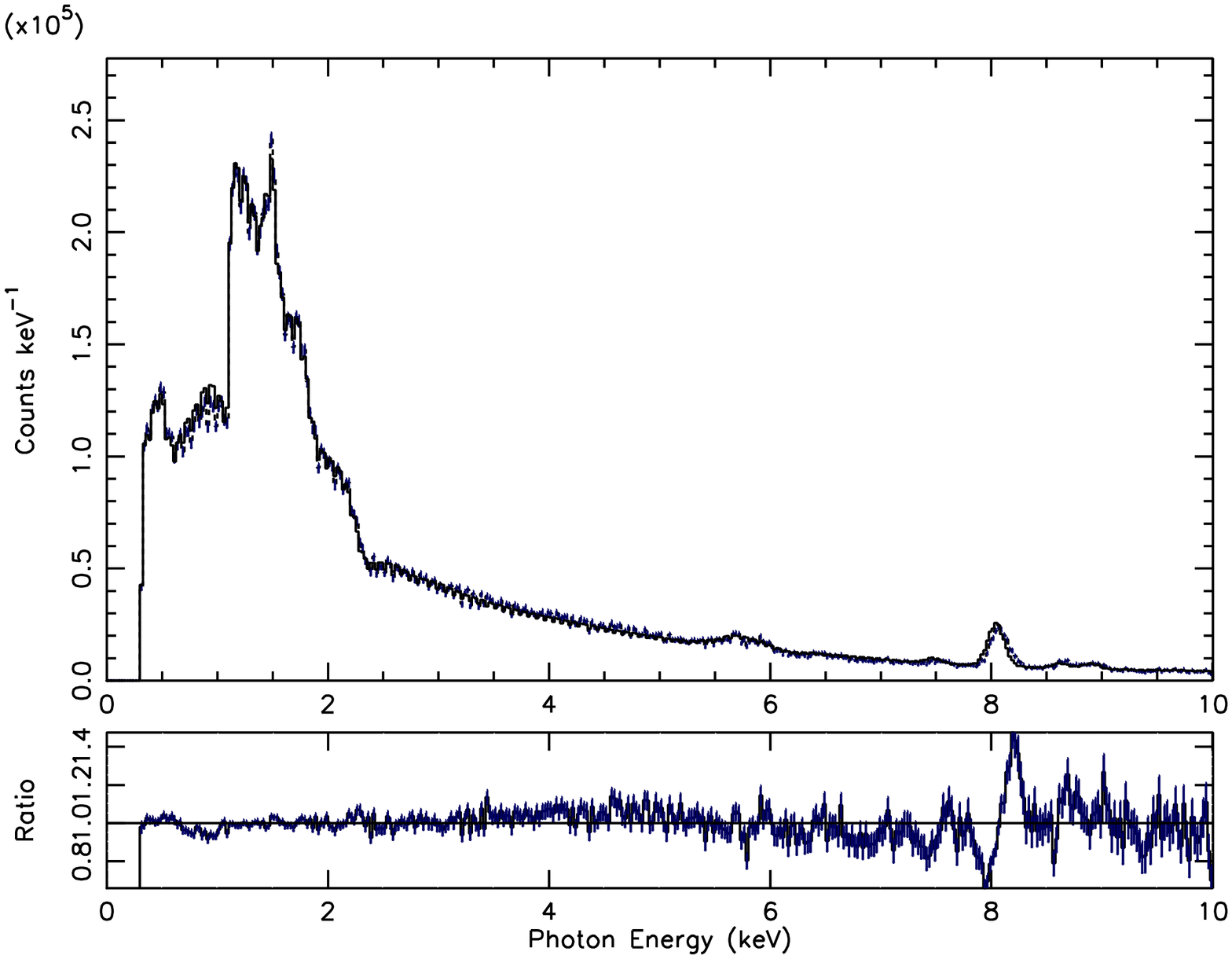} & 
      \includegraphics[width=2.5in,angle=-90]{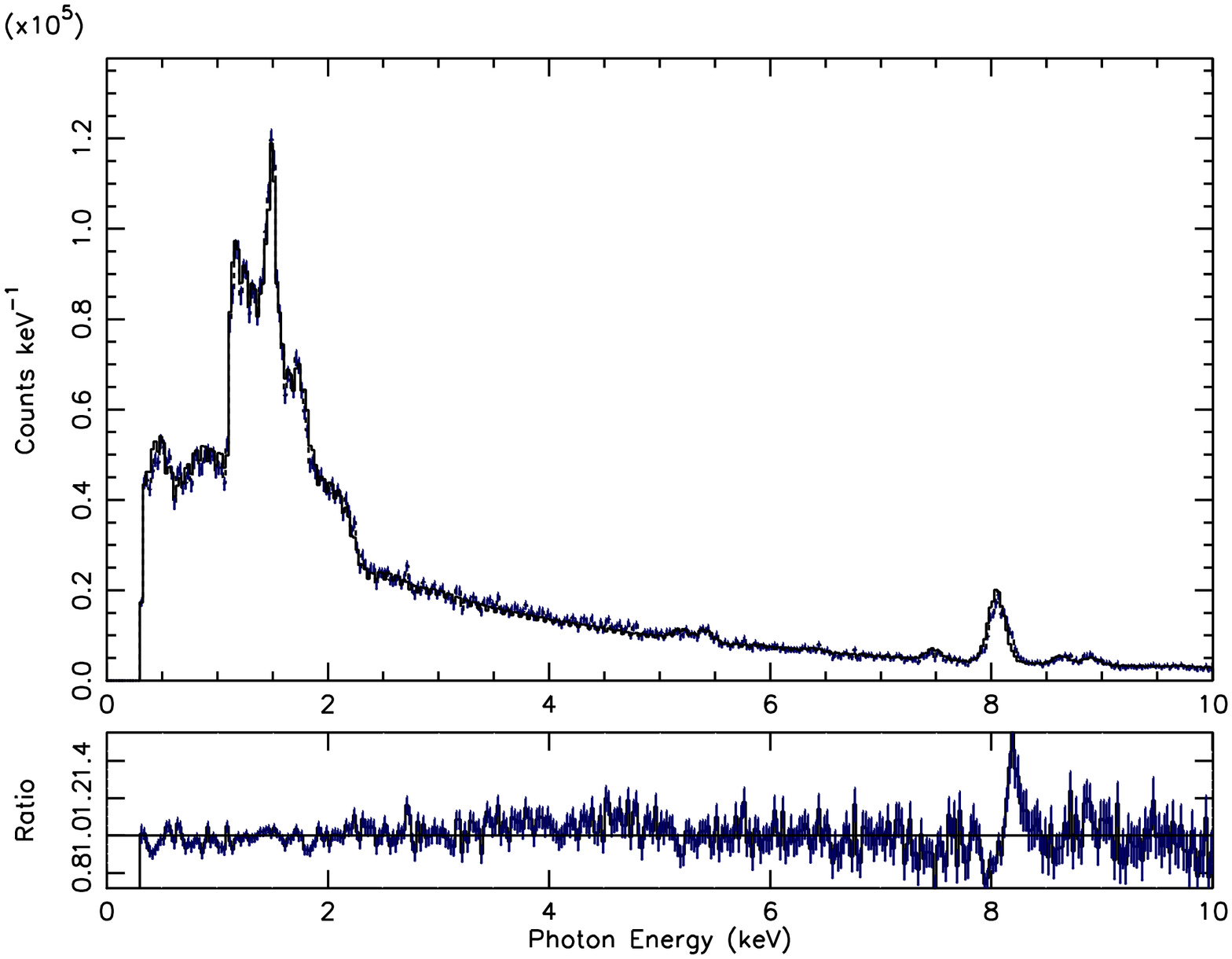} \\
      \includegraphics[width=2.5in,angle=-90]{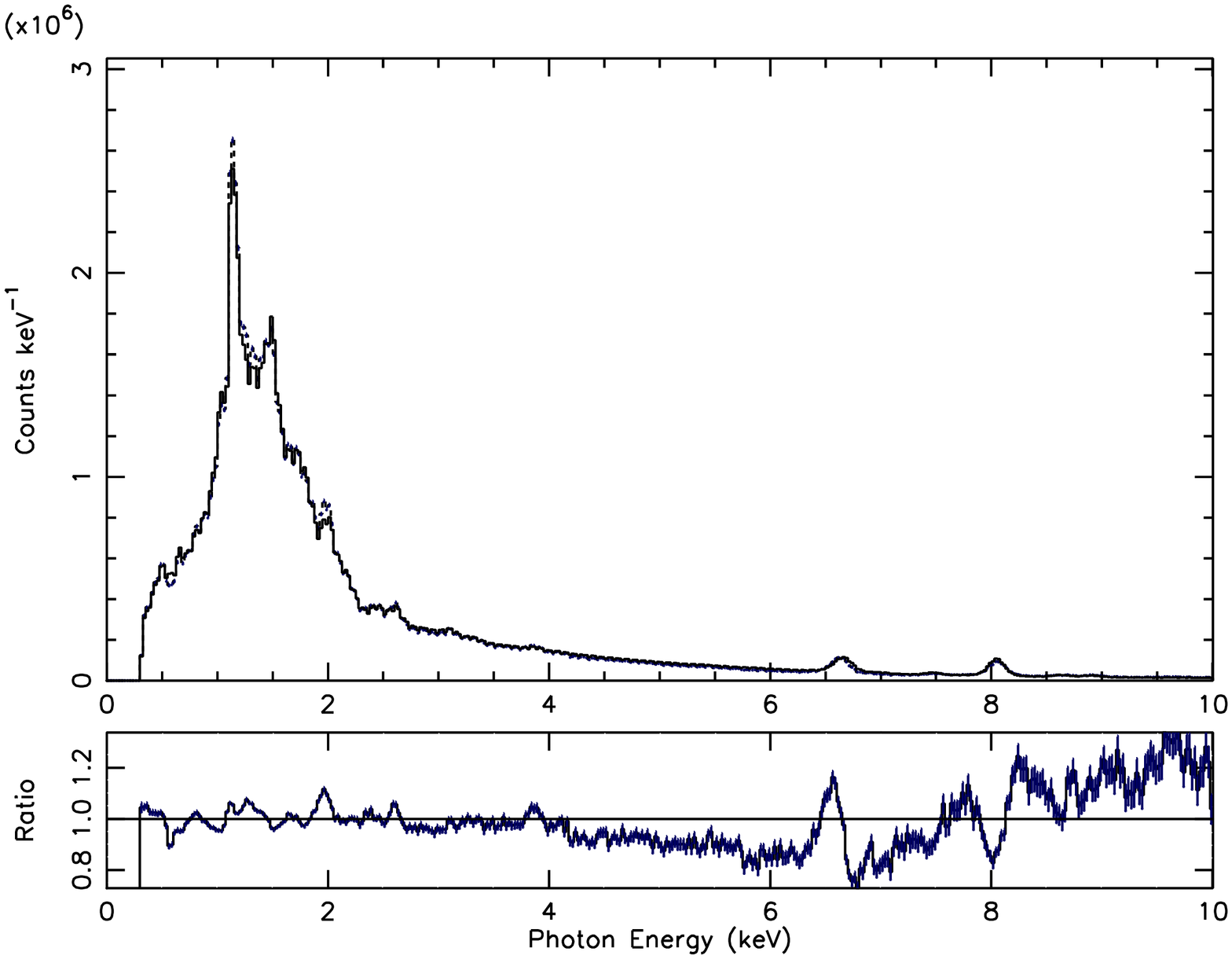} \\
    \end{tabular}
  \end{center}

  \caption{Spectral comparison of the data ({\sl dashed line}) to the model ({\sl solid line}) 
for the Abell 1689 700 blob analysis ({\sl top left}), the RX J0658-55 600 blob 
analysis ({\sl top right}), and the Centaurus 600 blob analysis ({\sl bottom}) for the full 
$20\arcmin \times 20\arcmin$ field used in the analysis.  
Flux is in units of photon counts per 25 eV bin, with the ratio of data to 
model shown in the lower panels. The model spectra are produced by averaging 
the spectrum from every 100th iteration from 2000 to 4000 and are 
renormalized to match the data counts. The spectra include all backgrounds and 
all three EPIC detectors. 
\label{modelvdata} }
\end{figure*}

To confirm that we have an acceptable overall spectral fit, we plot the 
model spectrum as inferred from the model sample versus the data 
and the ratio of the two in Figure \ref{modelvdata}. 
This plot shows an overall accurate spectral fit with the exception of the 
instrumental Cu K line, which appears to be detected at a slightly 
higher energy than expected. This mismatch is most likely due 
to gain variations in the {\sl XMM-Newton} CCDs.
The accuracy of the fit is seen by looking at the statistic in 
Figure \ref{statrej}.

\begin{figure*}[!htb]
\begin{center}
  \begin{tabular}{cc}
    \includegraphics[width=2.2in]{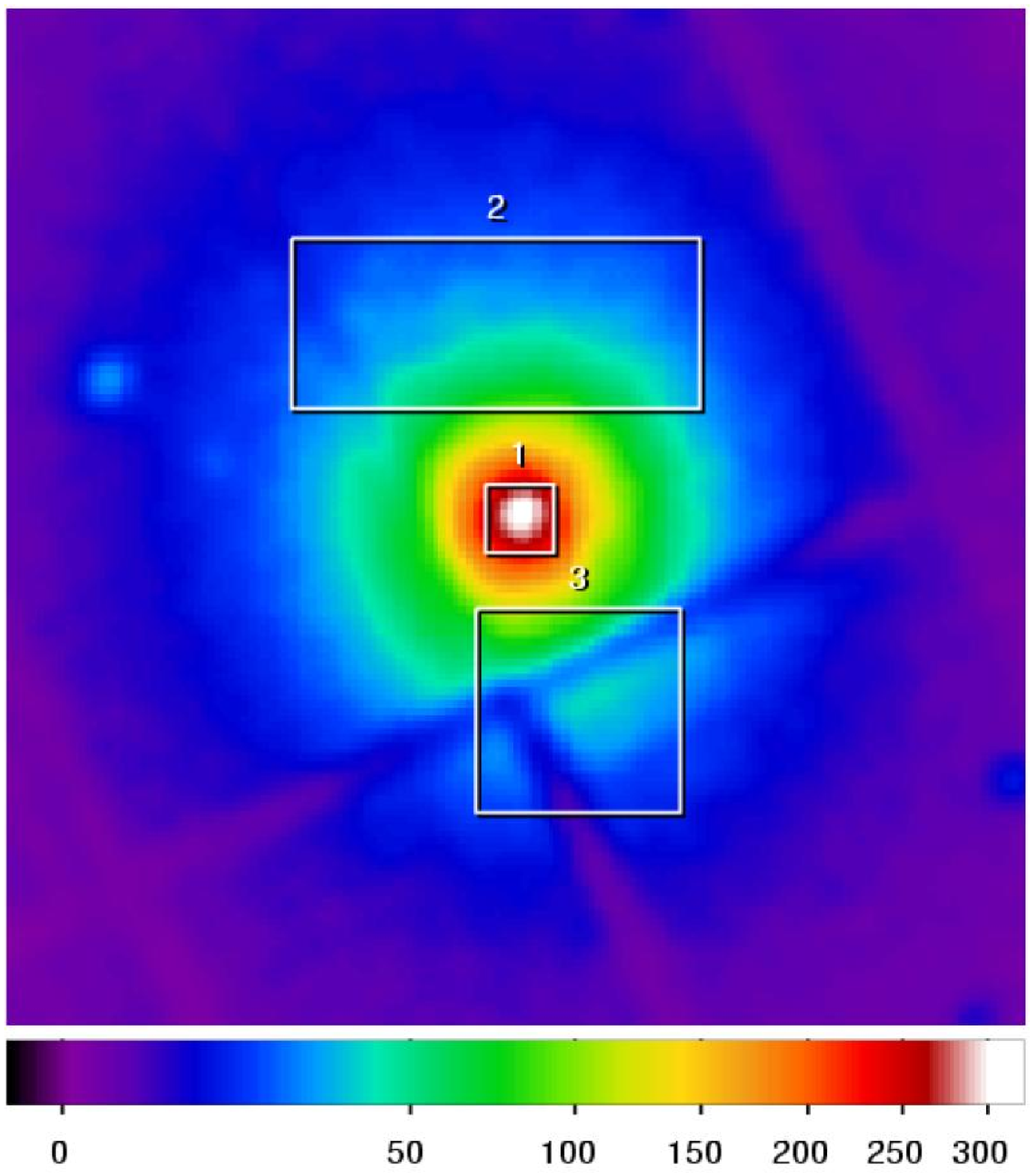} &
    \includegraphics[width=2.2in]{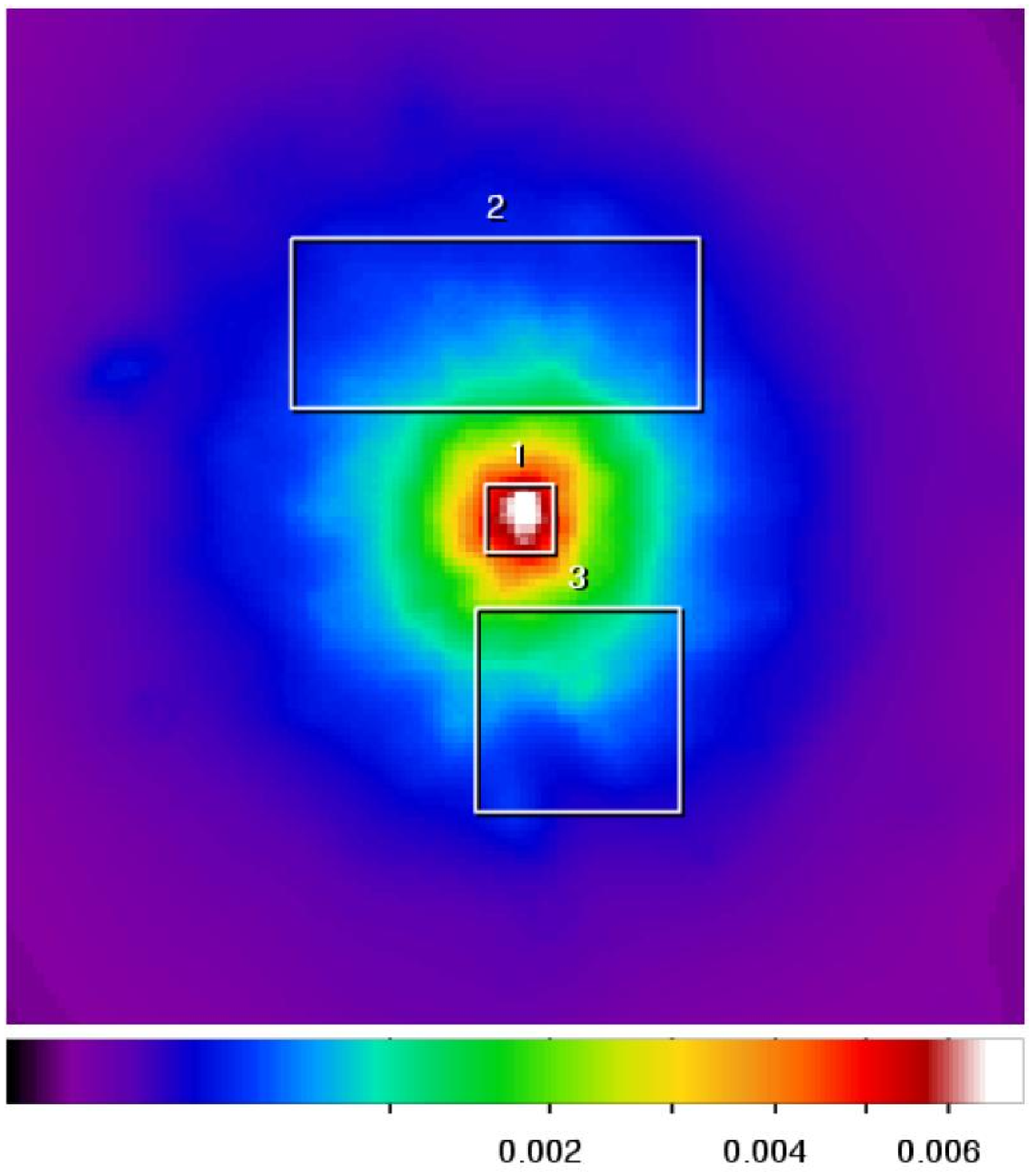} \\
  \end{tabular}
\end{center}
  \caption{Results of the analysis for Abell 1689, showing the central $5\arcmin \times 5\arcmin$ region.  
{\sl Left:} Raw data smoothed by a $4\arcsec$ kernel Gaussian 
(counts per $2\arcsec \times 2\arcsec$ pixel). 
{\sl Right:} Luminosity reconstruction using the median of all samples at 
each spatial point ($L_{bol}/10^{44}~$erg$~$s$^{-1}$ per $2\arcsec \times 2\arcsec$). \label{A1689lumtemp}}
\end{figure*}

To infer the luminosity distribution of the cluster, we generate the luminosity 
map for each model in the sample and take the median in each spatial pixel in 
order to make the median luminosity map shown in Figure \ref{A1689lumtemp} 
({\sl right}).  The bolometric luminosity, $L_{bol}$, is displayed in units of $10^{44}~$erg$~$s$^{-1}$ 
per $2\arcsec \times 2\arcsec$. 
The map of raw counts (counts per $2\arcsec \times 2\arcsec$) from the screened 
data file is shown on the left for comparison. 
There is good agreement between the reconstruction and the raw data;   
specifically, the reconstructed profile is sharper than the data 
due to PSF deconvolution and the fact that obvious chip gaps in the 
data are compensated by taking the exposure into account. 

We have selected three interesting regions (numbered in 
Figure \ref{A1689lumtemp}), motivated by our analysis in 
\citet{andersson} to study in more detail the distribution of 
plasma temperatures in those regions.  First, for a cluster that 
has a quite regular surface brightness distribution, similar to 
clusters with well-established ``cooling cores,'' the temperature of the 
core for Abell 1689 is quite high: $\sim 7.5 $ keV. 
Second, we have identified a region of 
hotter plasma, $\sim 9 $ keV, to the north of the 
cluster core, indicating possible shock heating of the gas in 
that region. Third, we have chosen to study in detail a region south of 
the core exhibiting a temperature almost as low as that of the core itself. 

\begin{figure*}[!htb]
  \begin{center}
    \begin{tabular}{cc}
      \includegraphics[width=2.2in]{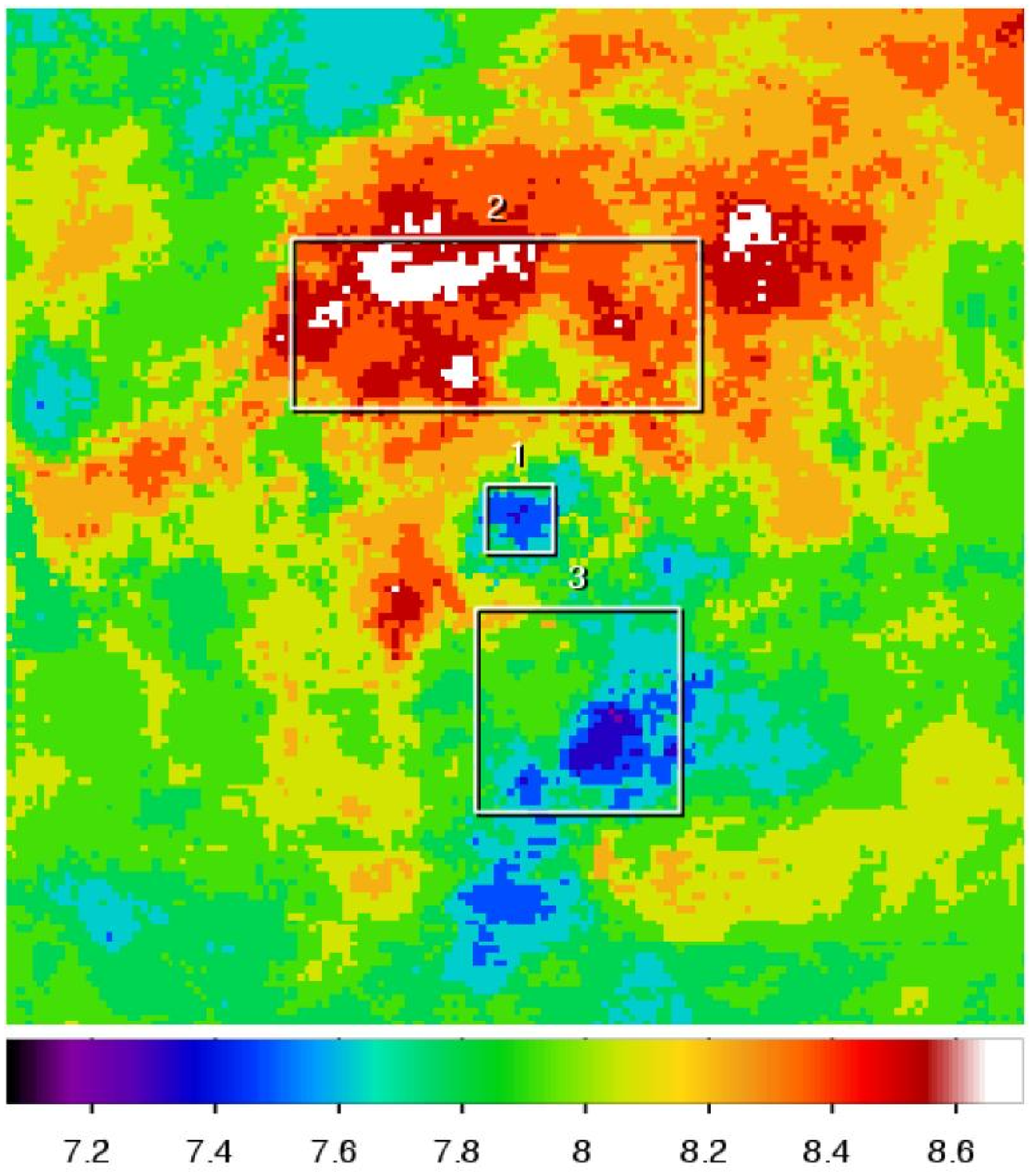} &
      \includegraphics[width=2.2in]{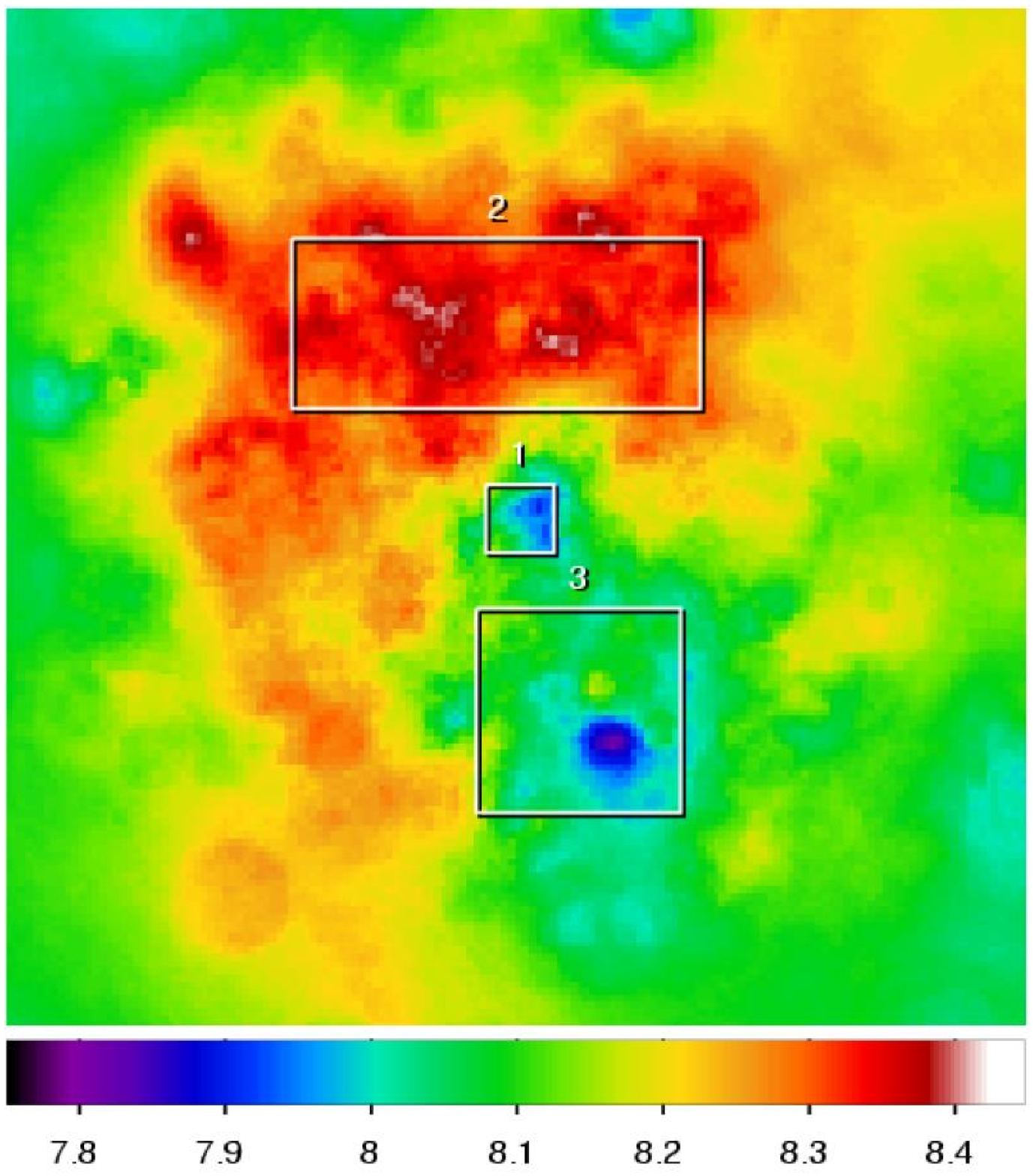} \\
      \includegraphics[width=2.2in,angle=0]{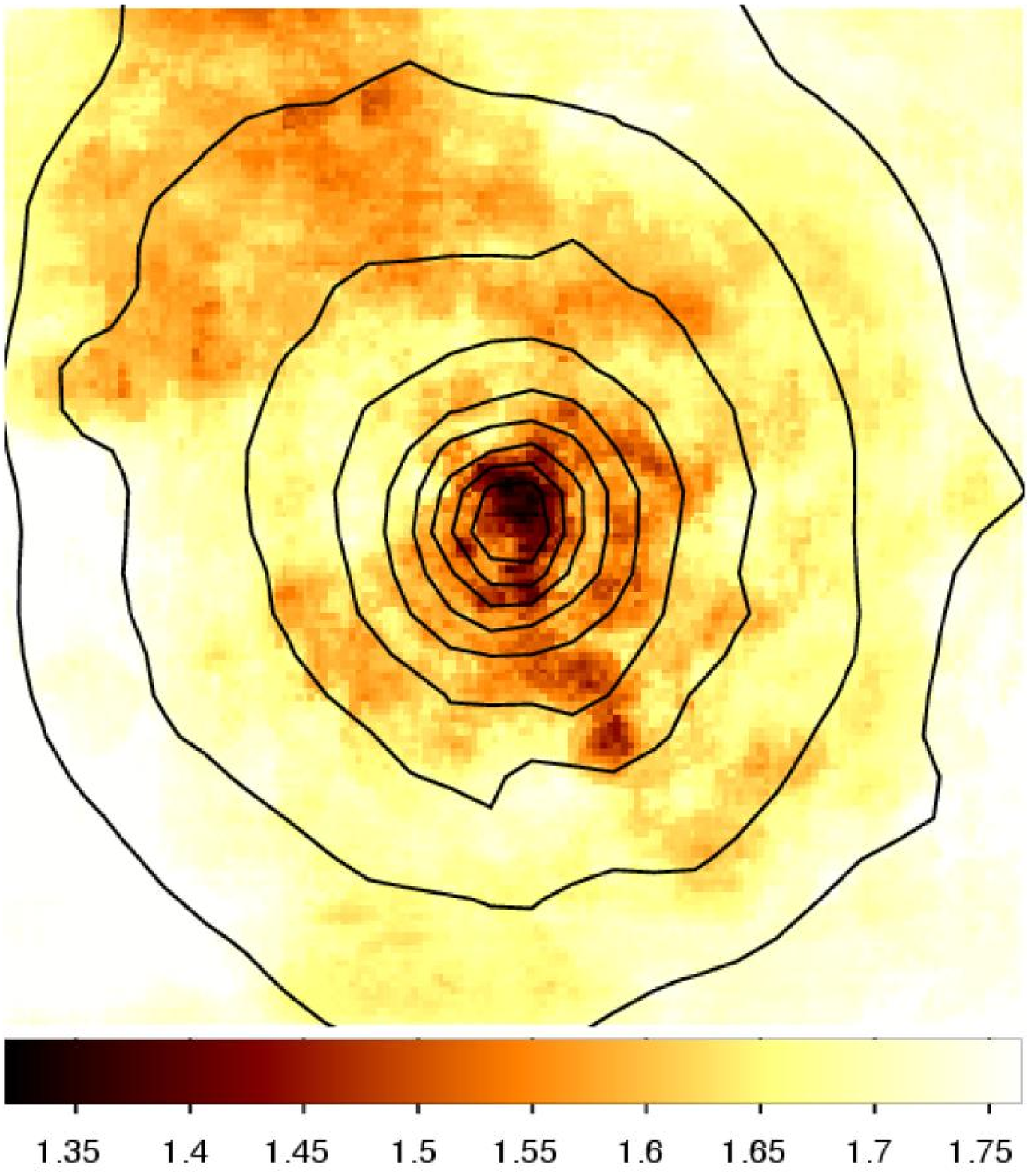} & 
      \includegraphics[width=2.2in,angle=0]{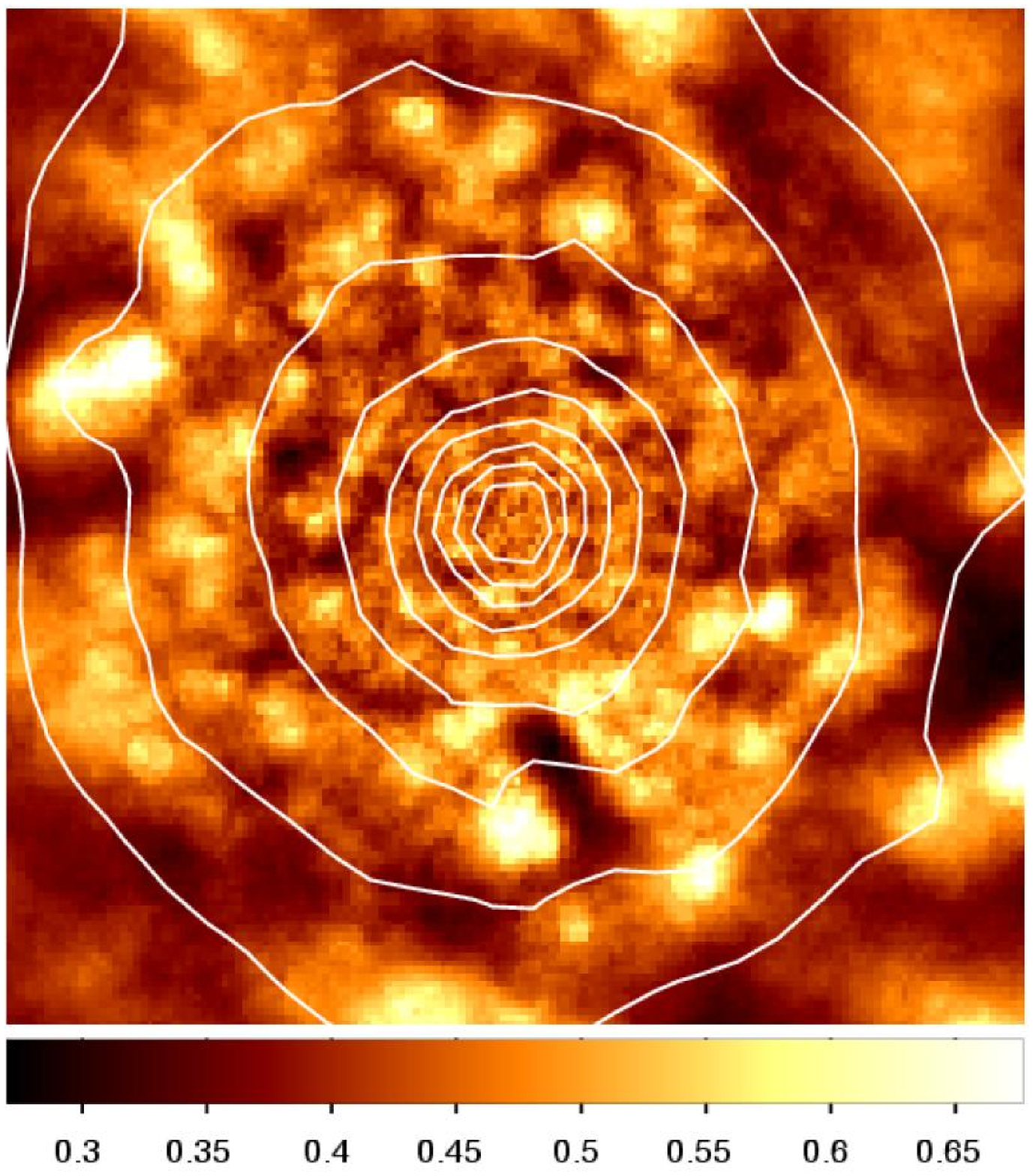} \\ 
    \end{tabular}
  \end{center}
  \caption{Results of the analysis for Abell 1689, showing the central $5\arcmin \times 5\arcmin$ region.
{\sl Top left,}
temperature map (in units of keV) constructed
using the mode of the distribution for all samples at each spatial point;
{\sl top right,}
temperature map (in units of keV) constructed using the median of the
distribution for all samples at each spatial point;
{\sl bottom left,}
temperature uncertainty map (in units of keV) showing the 1 $\sigma$ variation
of the temperature mode in the model sample;
{\sl bottom right,} 
temperature uncertainty map (in units of keV) showing the 1 $\sigma$ variation 
of the emission-weighted temperature in the model sample. 
\label{A1689Terr}}
\end{figure*}

Next, we form a median temperature map by first weighting the temperature of 
each model particle by its luminosity, creating a temperature map for each 
model in the 2000 model MCMC sample. 
This sample of maps is then averaged by taking the 
median of the distribution of temperatures for each spatial pixel, 
as is done in the creation of luminosity maps. 
Instead of taking the luminosity-weighted 
average, in another attempt to visualize the distribution 
of temperatures in the cluster, we bin the luminosity in bins of temperature 
in each spatial pixel, creating a three dimensional differential 
luminosity data cube. We do this for all sample 
models, and for each of them, we calculate the mode of the resulting distribution 
of each spatial pixel. Taking the median over all model samples of this mode 
accurately describes the dominant temperature at any given spatial position. 
We note that this will not give the same value of the temperature as one would get 
from traditional spectral analysis. 
The temperature measurement is biased due to the degeneracies 
present in spectral fitting when allowing for 700 separate temperature phases. 
However, we believe that it is the method of highest contrast when separating 
distinct temperature components of a galaxy cluster. 
The median distribution mode and the median emission-weighted temperature 
are shown in Figure \ref{A1689Terr} ({\sl top left and top right}).
To display an estimate of the error on the emission-weighted temperature 
we calculate the 1 $\sigma$ variation on these quantities over the model sample. 
These maps are shown in the same figure ({\sl bottom left and bottom right}). 
The uncertainty of the emission-weighted temperature is below 0.5 keV throughout 
most of the shown region. For all of the calculations above we have used the 
model run with 700 particles.

\begin{figure*}[!htb]
  \begin{center}
    \begin{tabular}{ccc}
      \includegraphics[width=1.5in,angle=-90]{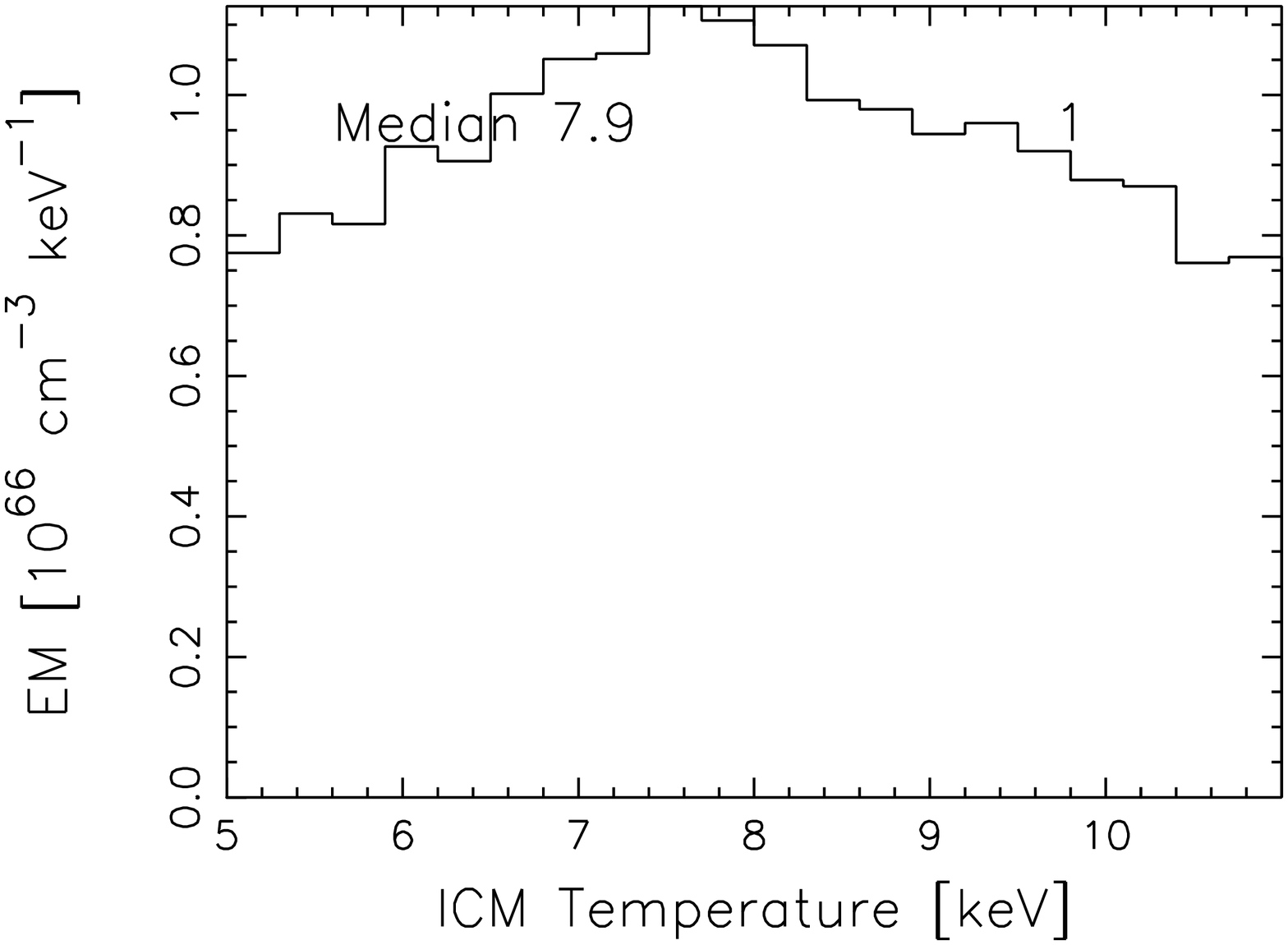} & 
      \includegraphics[width=1.5in,angle=-90]{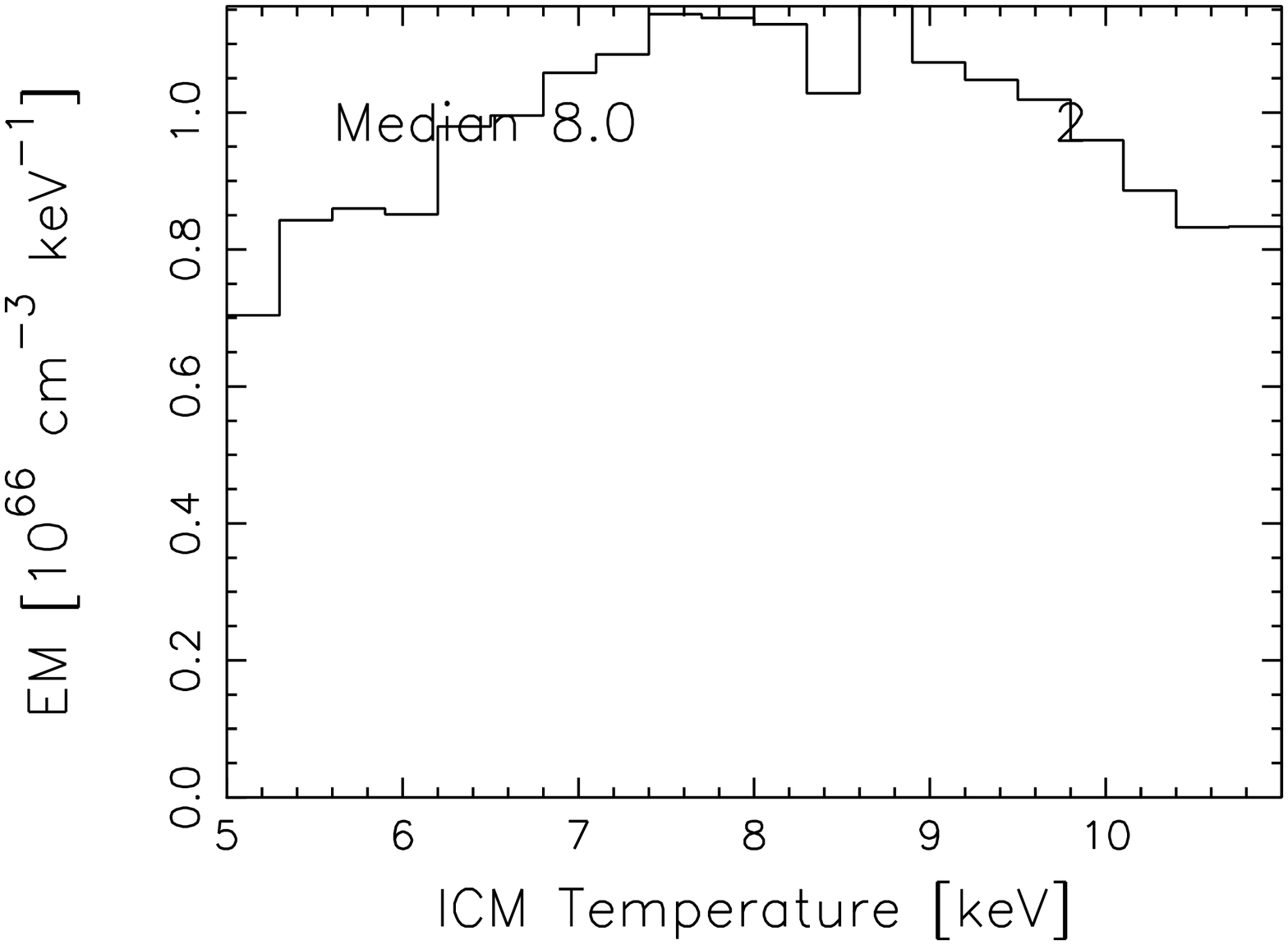} &
      \includegraphics[width=1.5in,angle=-90]{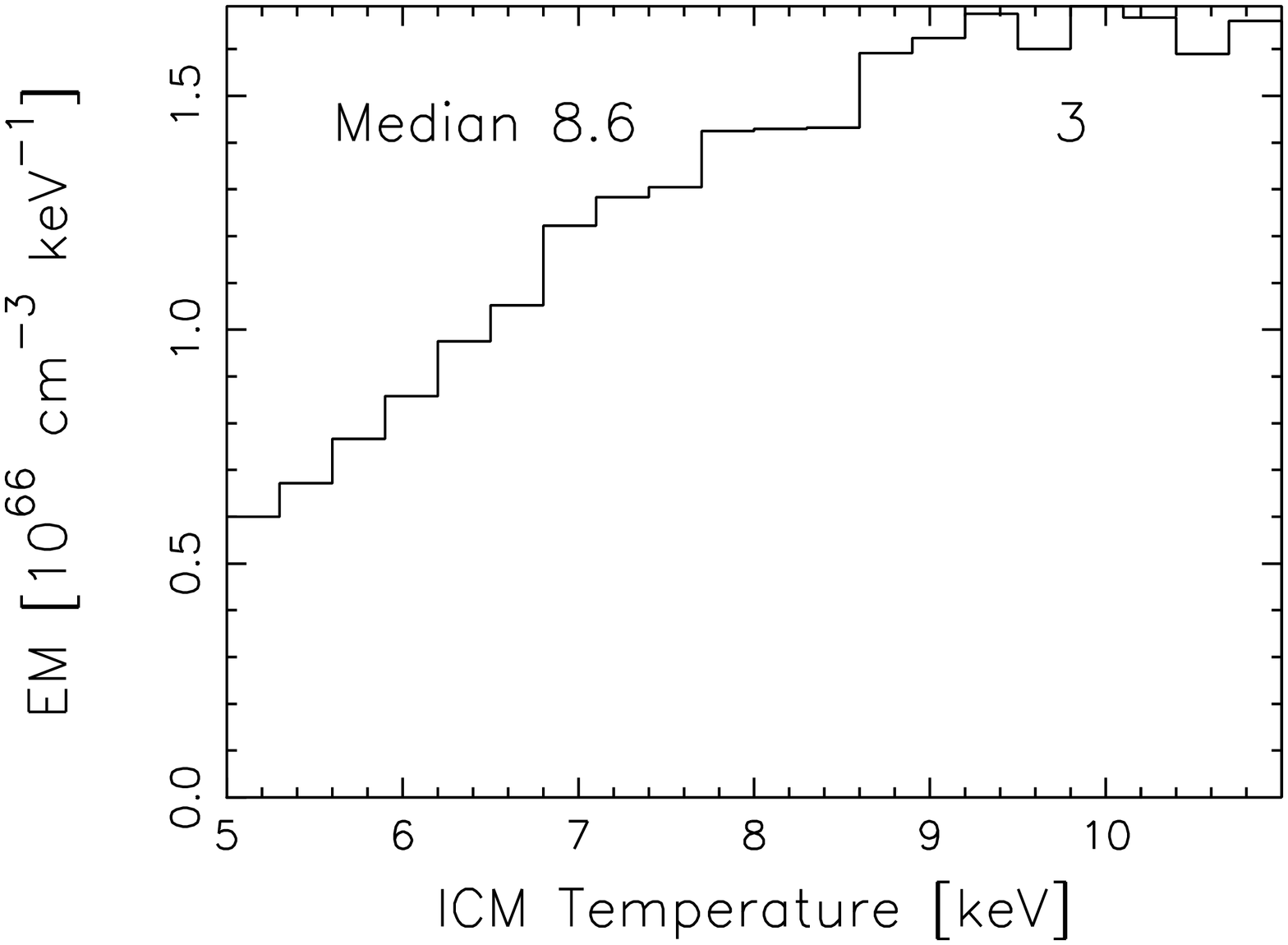} \\
    \end{tabular}
  \end{center}
  \caption{Iteration averaged distribution of temperatures for A1689 regions 1 (core; {\sl left}), 2 (south; {\sl middle}) and 3 (north, {\sl right}). \label{A1689Tdistr}}
\end{figure*}

In order to study the regions selected above in more detail, we have 
extracted the differential luminosity distribution in these regions 
and binned them into 20 bins over the 5-11 keV allowed range.  
We have chosen to use the model run with 100 particles for this in 
order not to over-complicate the problem.  While in this case 
the luminosity and temperature maps do not show the same level 
of detail, the distributions of spectral parameters become narrower. 

The differential luminosity, plotted as the emission measure, $\int{n_H n_e} $dV, 
per keV, along with the median 
of the distribution, is shown for these regions in Figure \ref{A1689Tdistr}. 
From these data alone we cannot detect any deviation from isothermality in 
these regions, mostly due to the inherent degeneracies associated with the 
attempt to fit high-temperature emission with a multi-phase model. Since most 
of the emission is from the bremsstrahlung continuum there are many 
combinations of gas phases (temperatures) 
that provide equally good fits to the data.

We have confirmed the existence of a region of hot plasma north of the cluster 
core as we reported in \citet{andersson}.  This hotter plasma appears 
to extend in an arc approximately halfway around the northern part. 
The technique verifies both the presence of the colder emission to the 
south and the lower than ambient (by $\sim 1$ keV) 
temperature of the cluster core.  
Our analysis reveals an apparent trail to the south, possibly a remnant 
tracing the path of the cluster core, and the shock-heated region to 
the north.
However, this could also be due to projection effects from filaments 
extending in the direction of the line of sight.

\begin{figure*}[!htb]
  \begin{center}
    \begin{tabular}{cc}
      \includegraphics[width=2.5in]{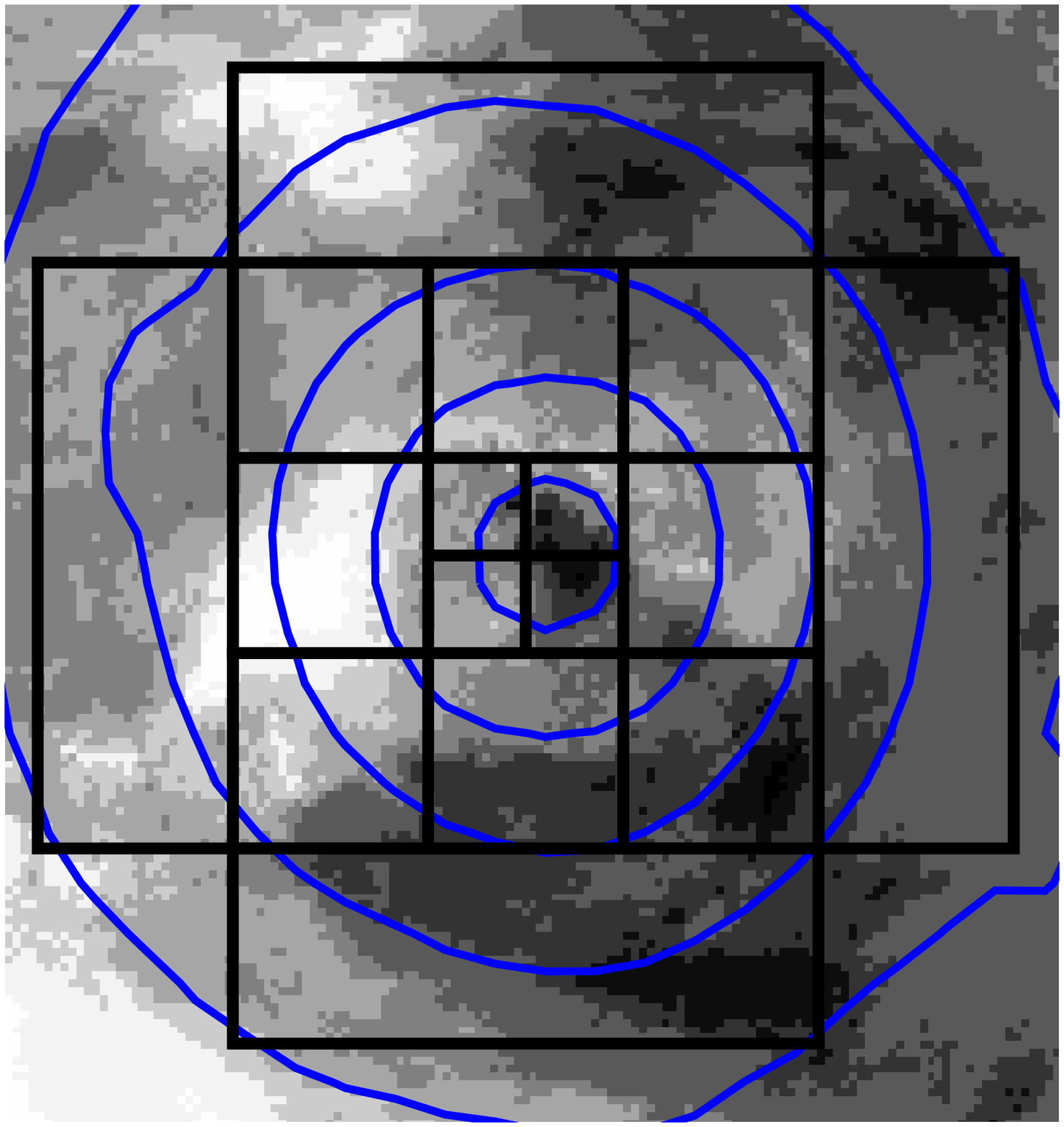} & 
      \includegraphics[width=2.5in]{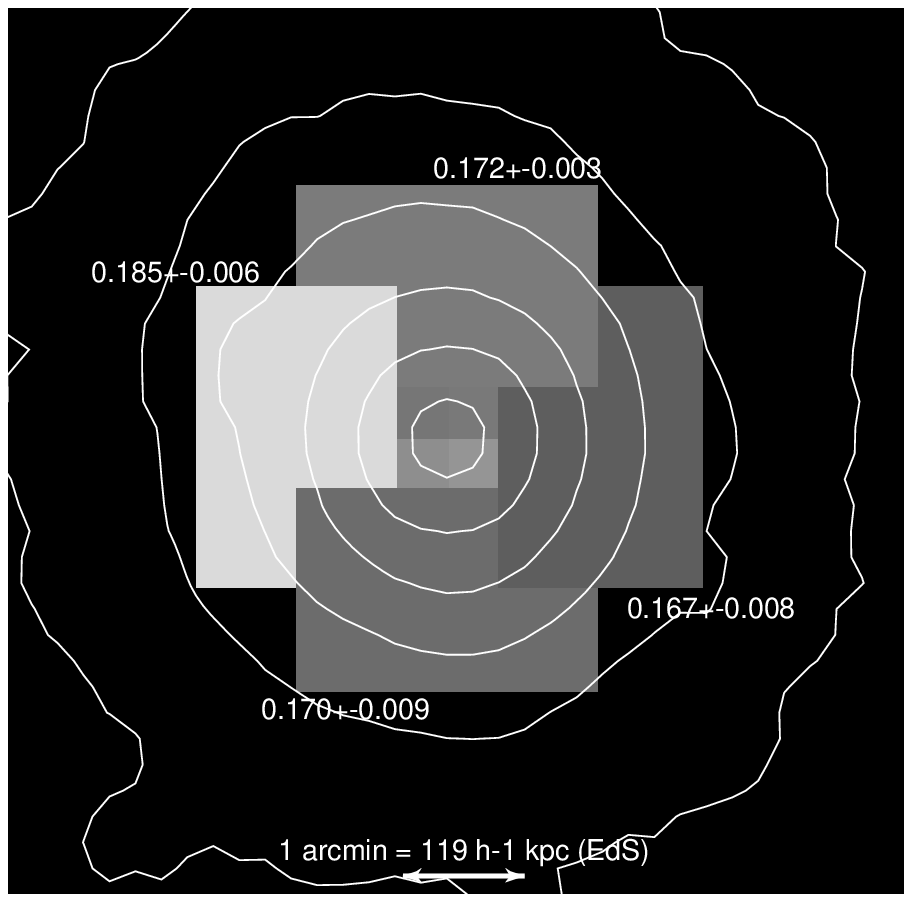} \\
    \end{tabular}
  \end{center}
  \caption{{\sl Left:} Redshift map of Abell 1689 
obtained by taking the median of the dominant redshift mode over 
the model sample. The gray scale from black to white corresponds to $z=0.175-0.185$. 
Superposed contours from the raw event list data and grid regions 
from \citet{andersson} are shown for comparison. {\sl Right:} Results from the 
analysis in \citet{andersson} \label{A1689redsh}}
\end{figure*}

In our earlier findings we infer intra-cluster gas motions from a shift in 
the position of the Fe K line complex corresponding to 
$\Delta z \approx 0.01$. Here we show a map of cluster redshift 
(Figure \ref{A1689redsh}, {\sl left}) that is similar to the temperature maps shown above. 
The map was created by calculating the 
dominant redshift mode at each spatial pixel and taking the median over 
the samples. The superposed contours are iso-photon count from the raw data, and 
the grid is shown for comparison with the regions used in \citet{andersson}. 
The results from \citet{andersson} are shown in Figure \ref{A1689redsh} ({\sl right}). 
The map clearly confirms the $\Delta z \approx 0.01$ east-west gradient found 
previously by us, corresponding to a velocity of $\sim 3000~$km s$^{-1}$. 
We also assess the significance of this 
by analyzing the redshift-mode variation in the model sample.  
The $1 \sigma$ model variance of the redshift mode in the regions 
of interest ranges from 0.016 to 0.019, and therefore our result is 
only $\sim 1 \sigma$ significant. 
Putting an upper threshold on the redshift discrepancy, we find that $\Delta z < 0.045$.
However, we note that it is not clear that the above is the 
most appropriate estimate of the true uncertainty, since inherent properties 
of the method could increase the variance; possibly, 
another means of error estimation could give a more accurate result. 
Also, this particular model was not designed specifically to deduce the 
redshift-structure of the cluster gas.  Devising the run only to measure 
this effect will likely prove more successful. 

\subsection{RX J0658-55}
\begin{figure*}[!htb]
  \begin{center}
    \begin{tabular}{cc}
      \includegraphics[width=2.2in]{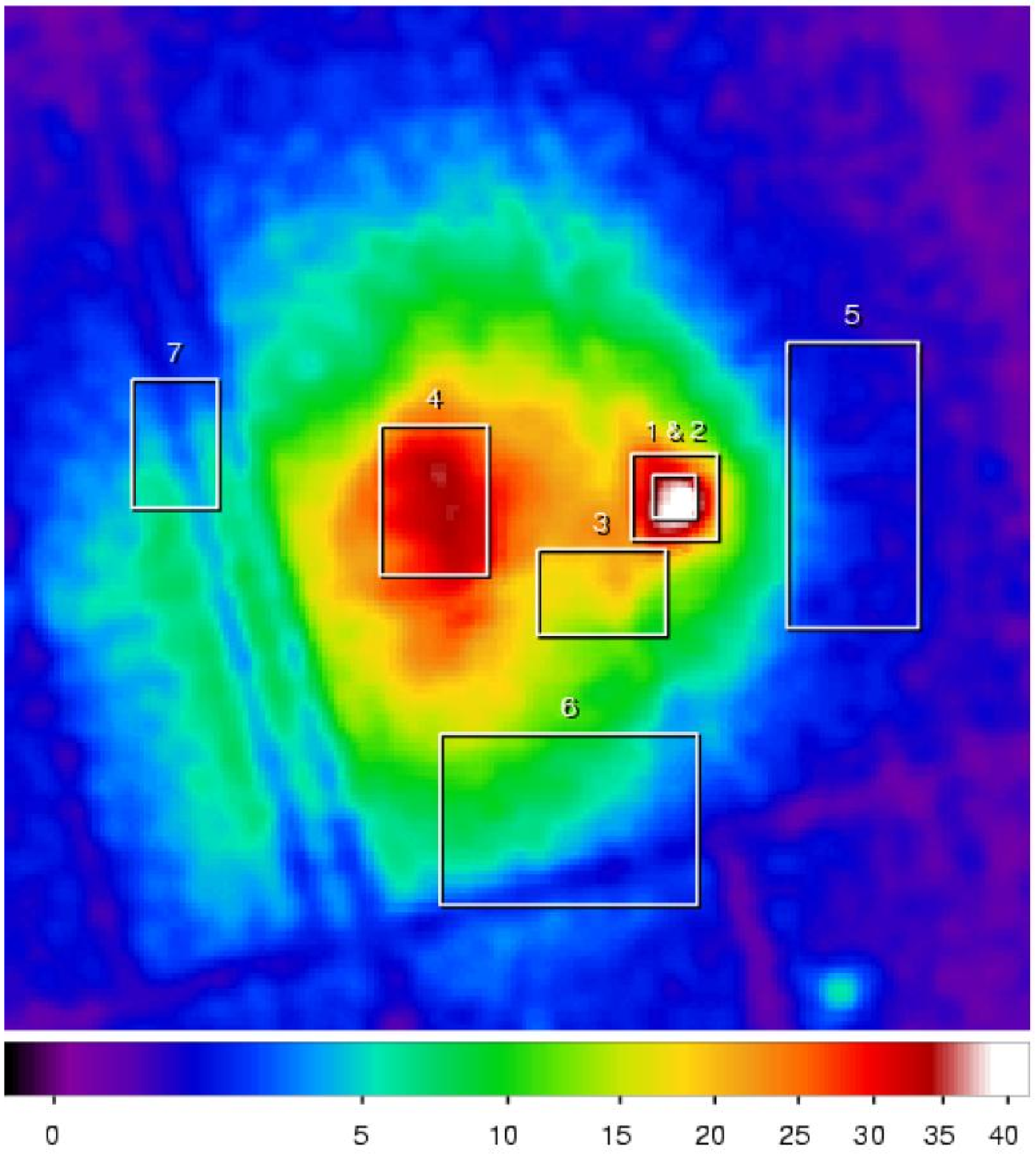} & 
      \includegraphics[width=2.2in]{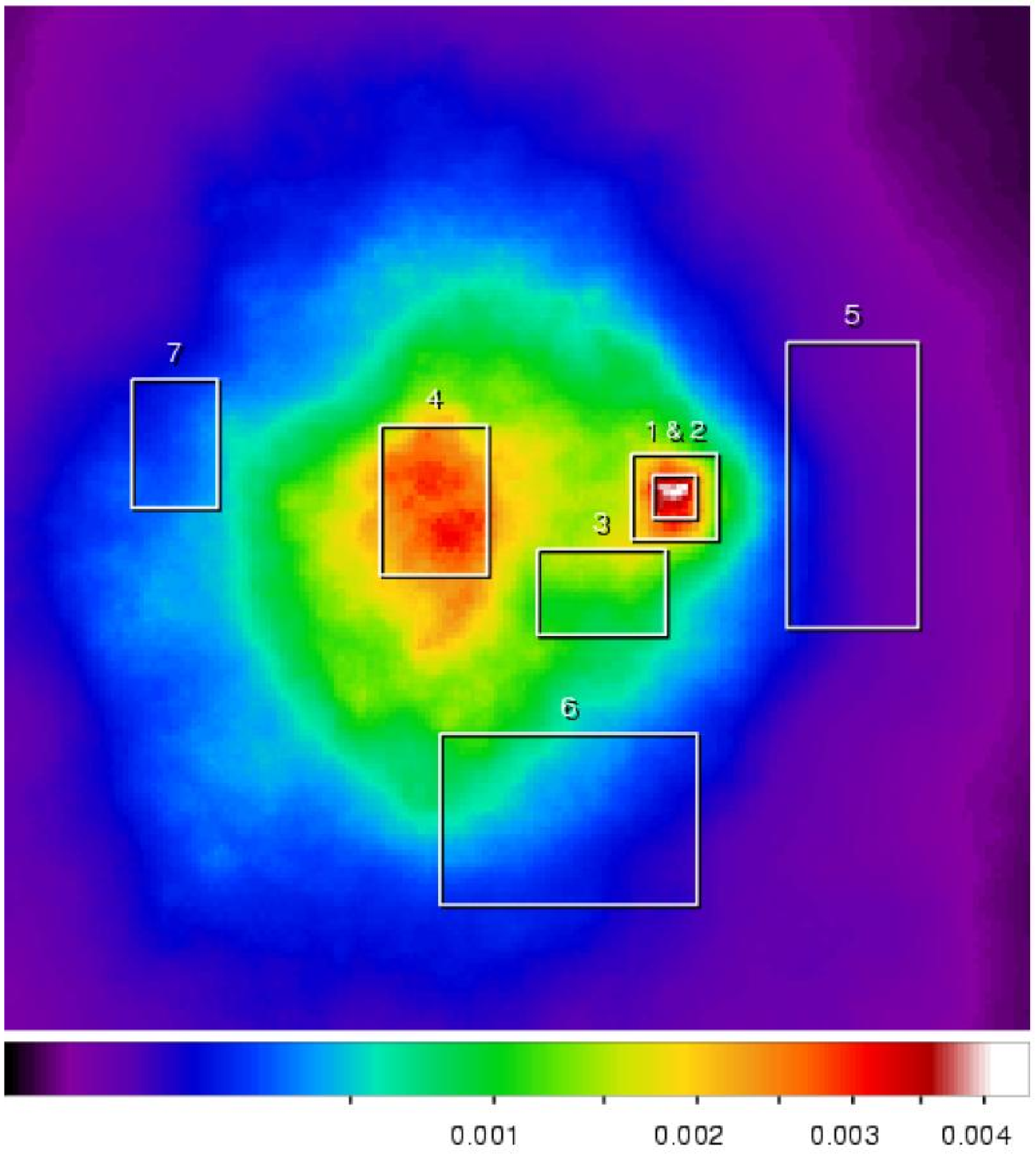} \\
    \end{tabular}
  \end{center}
  \caption{Results of the analysis for RX J0658-55 showing central $5\arcmin \times 5\arcmin$ region.
{\sl Left:} Raw data smoothed by a $4\arcsec$ kernel Gaussian
(counts per $2\arcsec \times 2\arcsec$ pixel).
{\sl Right:} Luminosity reconstruction using the median of all samples at
each spatial point ($L_{bol}/10^{44}~$erg$~$s$^{-1}$ per $2\arcsec \times 2\arcsec$). \label{bulletLumTemp}}
\end{figure*}

Similarly to the case of Abell 1689, we plot the model spectrum 
as inferred from a subsample of the model sample versus the data 
and the ratio of the two in Figure \ref{modelvdata}. 
This plot shows an acceptable fit, again with the exception of the 
Cu K line complex.  
The evolution of the Poisson $\chi^2$ 
as a function of Markov chain iteration is shown in Figure \ref{statrej}.
As described previously, 
we generate a median luminosity map from the model 
and compare it with a counts map smoothed by a $4\arcsec$ Gaussian 
(Figure \ref{bulletLumTemp} {\sl right and left}). It can be 
seen that the ``bullet'' and the main cluster features become somewhat 
sharper by the PSF deconvolution 
effect from the forward fitting that is inherent in the method. 

\begin{figure*}[!htb]
  \begin{center}
    \begin{tabular}{cc}
      \includegraphics[width=2.2in]{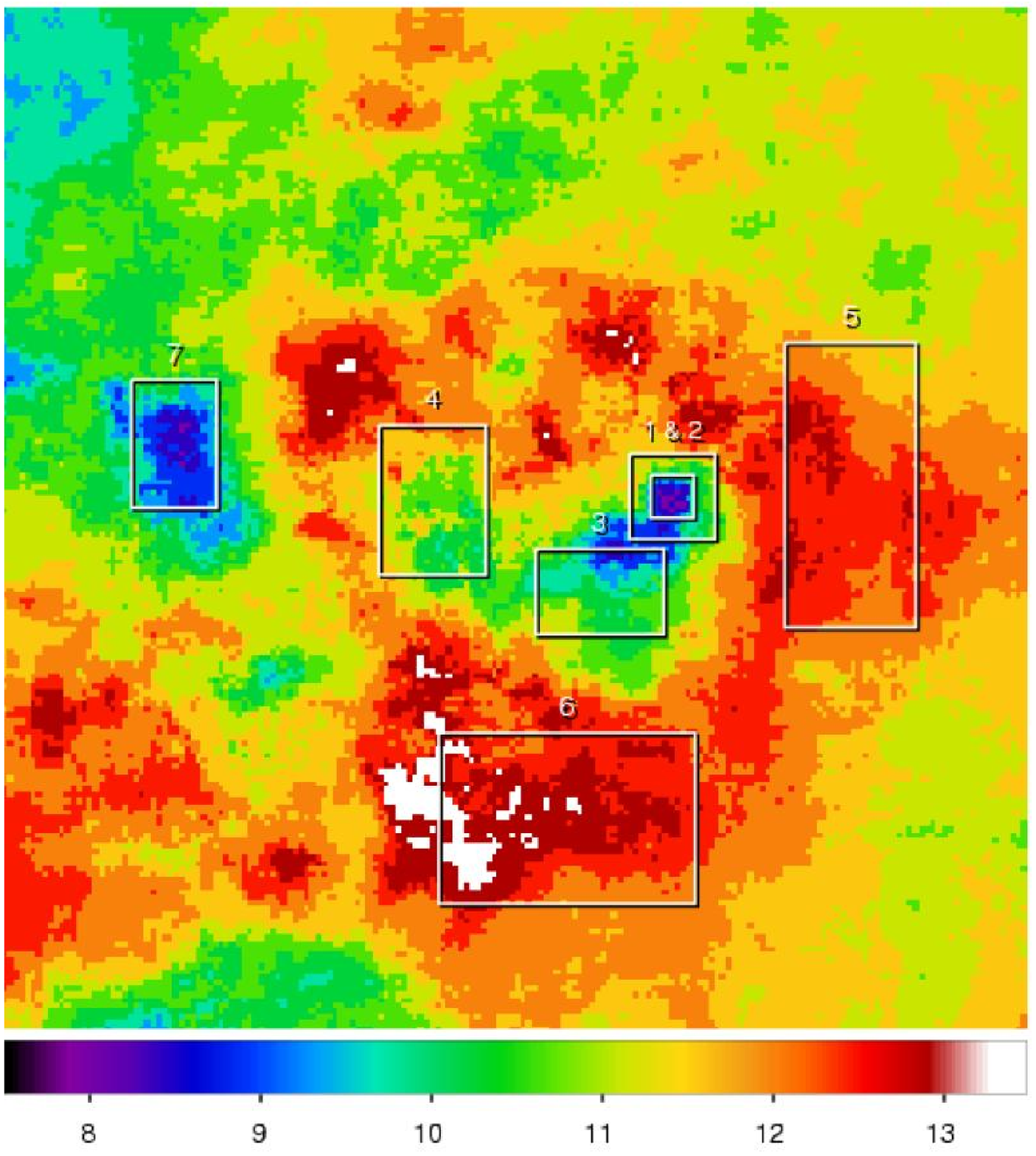} &
      \includegraphics[width=2.2in]{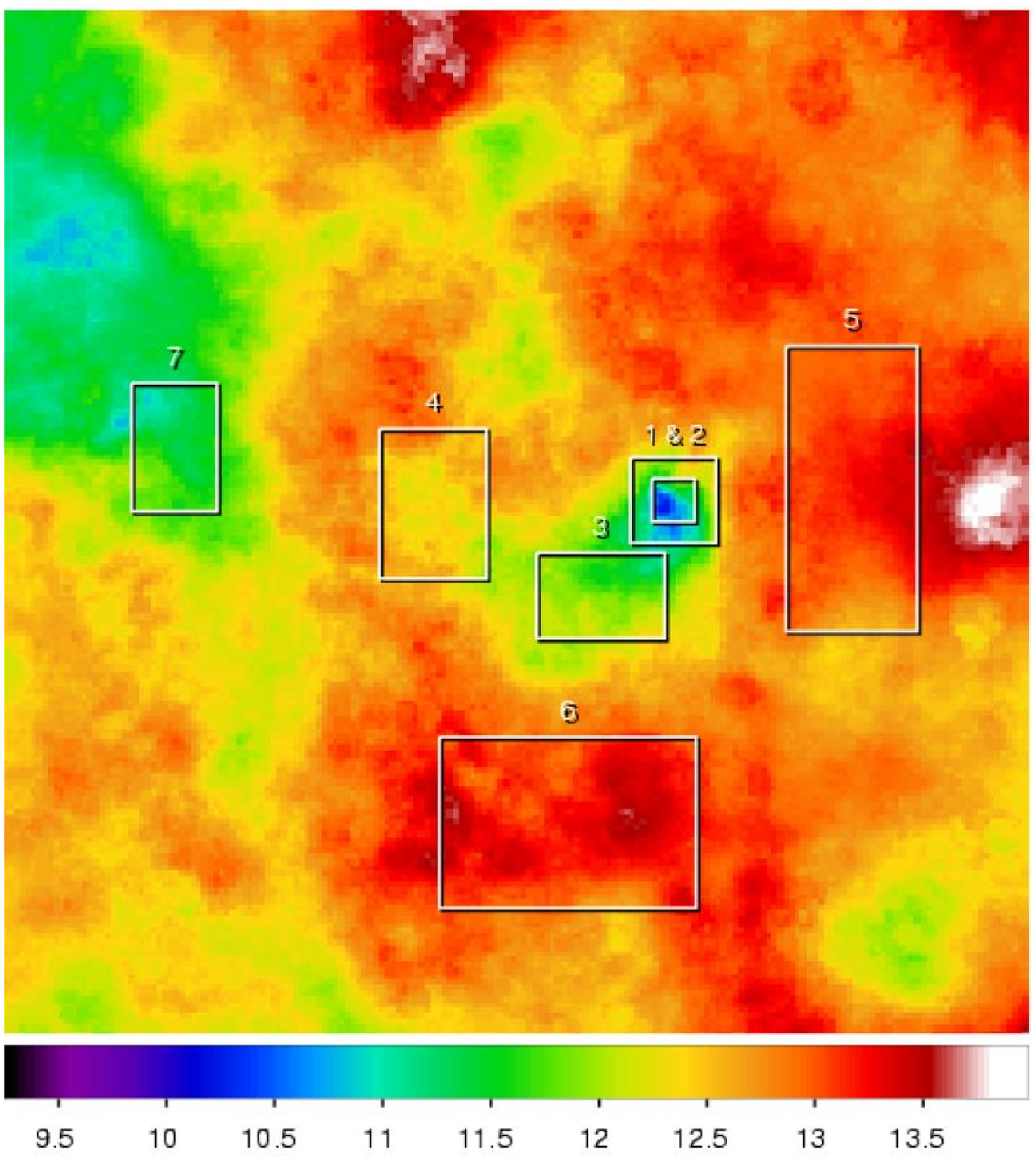} \\
      \includegraphics[width=2.2in,angle=0]{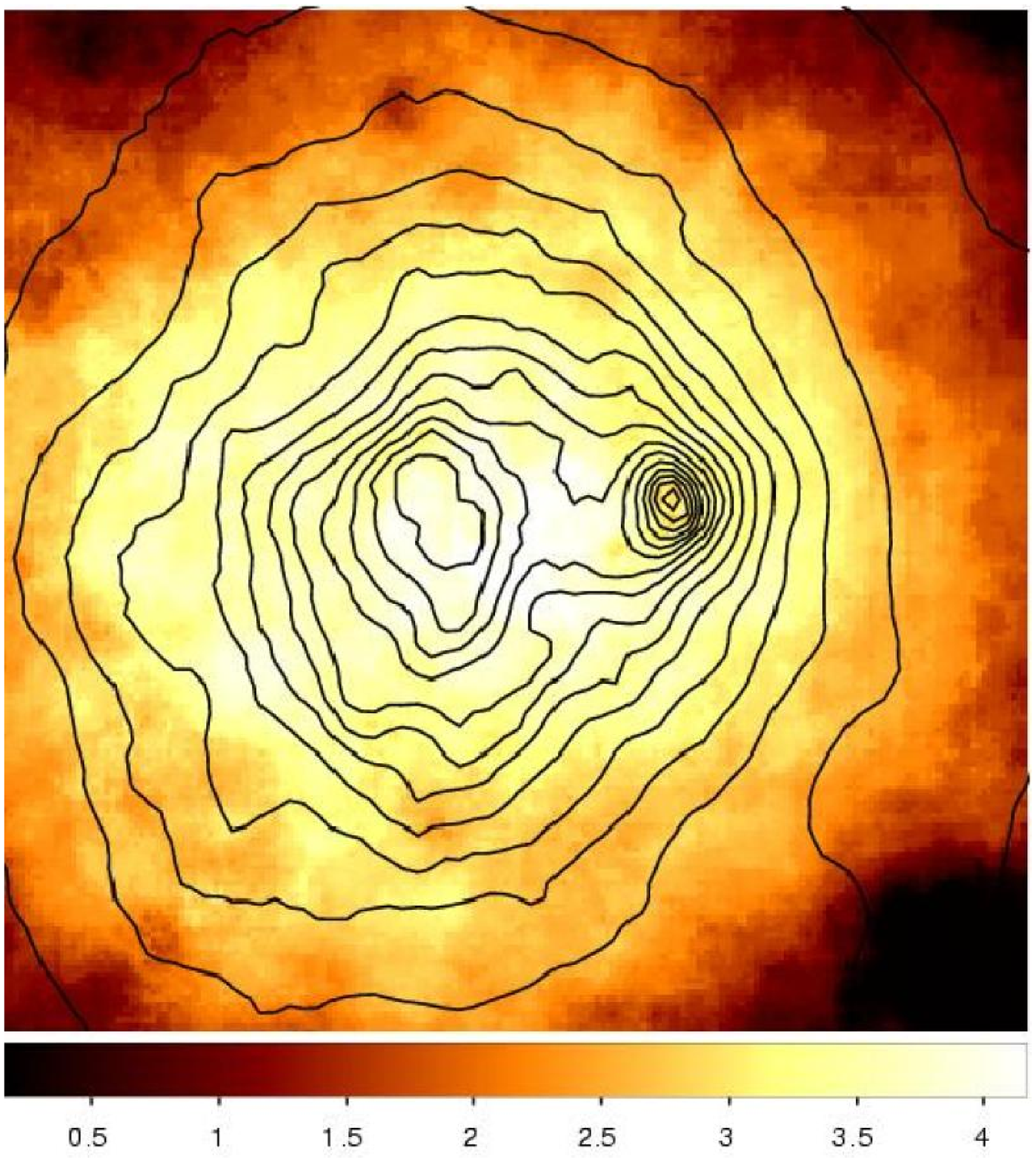} & 
      \includegraphics[width=2.2in,angle=0]{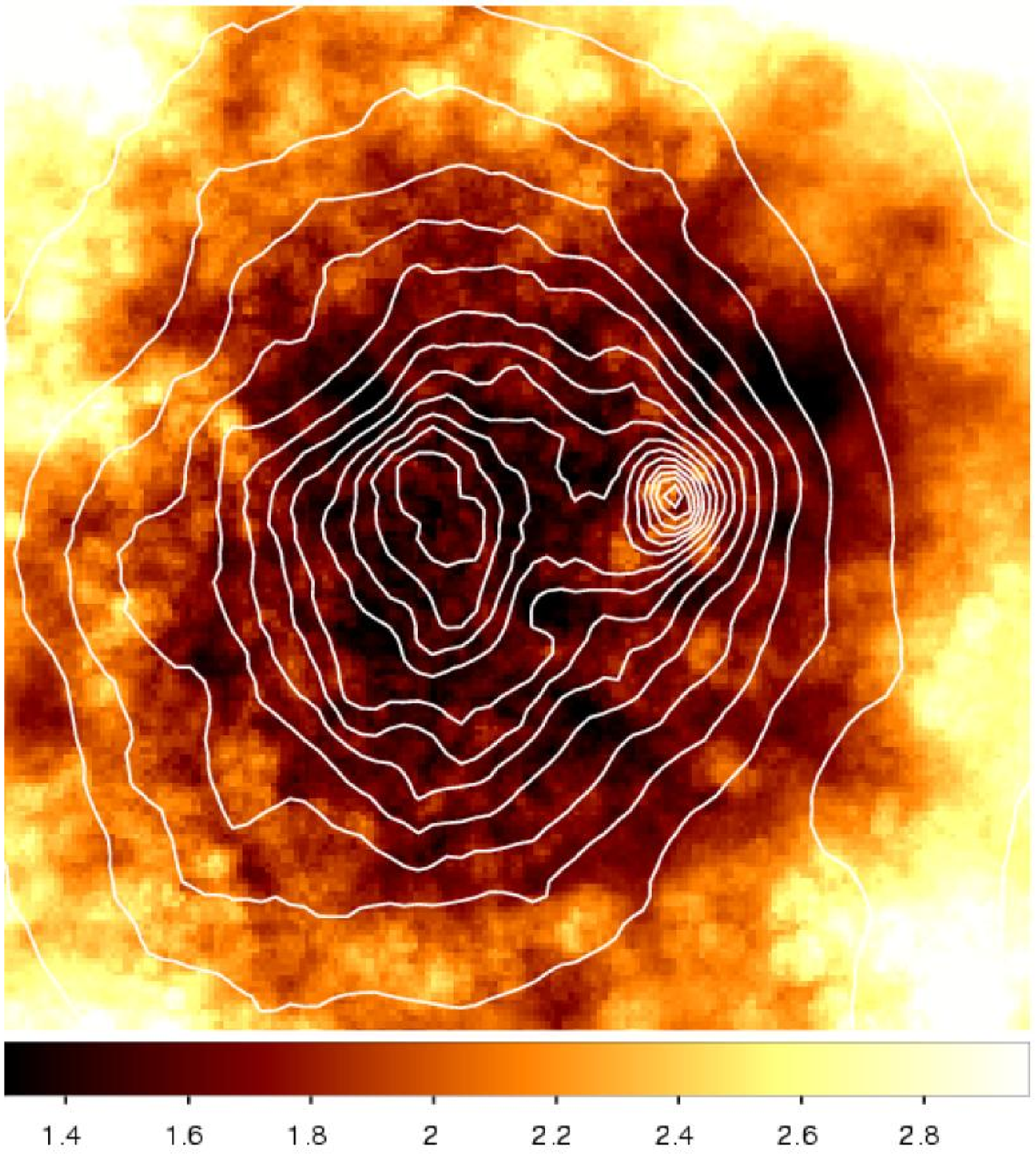} \\ 
    \end{tabular}
  \end{center}
  \caption{Results of the analysis for RX J0658-55, showing the central $5\arcmin \times 5\arcmin$ region.
{\sl Top left,} temperature map (in units of keV) constructed
using the mode of the distribution for all samples at each spatial point;
{\sl top right,} temperature map (in units of keV) constructed using the median of the
distribution for all samples at each spatial point;
{\sl bottom left,} temperature uncertainty map (in units of keV) showing the 1 $\sigma$ variation
of the temperature mode in the model sample;
{\sl bottom right,} temperature uncertainty map (in units of keV) showing the 1 $\sigma$ variation
of the emission-weighted temperature in the model sample.
\label{bulletTerr}}
\end{figure*}

We form and plot the luminosity weighted 
temperature map as well as a map based on the dominant temperature 
mode, in Figure \ref{bulletTerr} ({\sl top right and top left}). 
In conjunction, the associated uncertainty maps are shown ({\sl bottom right and bottom left}).
Both maps clearly show the cold core remnant of 
the ``bullet'' and the hot shocked gas in front of it. 
Note that the mode and weighted temperature maps display different 
properties of the temperature structure and are not comparable.
This of course is due to the approach used to model the 
plasma in our method. Instead of using just a single temperature, as 
is customary, in every spatial region, we use a number of phases that 
over the model sample form a nearly continuous distribution.  
We discuss this 
further in the paragraph discussing isothermal simulations below.  The mode is 
often biased towards lower temperatures for plasmas with temperatures 
of 9 keV and lower, due to the fact that within the bandpass of EPIC, 
at higher $T$, the spectra are more similar to each other and therefore are more 
difficult to distinguish. 

We do not see a clear decrease in temperature in front of the 
shock, as is expected, since the gas should be undisturbed here. It is possible 
that emission from the post-shock gas is smeared by the {\sl XMM-Newton} PSF and 
completely dominates the pre-shock emission, which in turn 
would be very faint in comparison. The cold gas of the bullet 
shows an apparent tail stretching south-east from the bullet center. 
By looking at the temperature map alone, one would conclude that 
this tail might reveal the movement history of the bullet 
in the merger. However, the luminosity map and, much more clearly, the 500 ks 
{\sl Chandra} exposure (M. Markevitch et al. in preparation), show a symmetrical 
Mach cone directed westward, indicating that this is the direction of motion.  
The weak-lensing analysis \citep{bra06} also shows the western dark 
matter halo just west of the bullet. It still cannot be ruled out that 
the bullet core entered to the north-east of the 
main cluster (see the ``entry hole'' 
devoid of emission in this region), proceeding south-west through the 
main cluster and eventually slingshotting around to the northwest, thus creating the tail 
visible in the temperature maps. The dark-matter dominated regions 
and the gas-dominated ones, of course, do not need to have similar 
merger paths. This scenario is also supported by the apparently 
previously shock-heated region of gas located directly south 
of the cluster. This part seems to have almost 
as hot a gas as the shock in front of the bullet. 
We select this region (No. 6) along with the six other regions shown in Figure 
\ref{bulletLumTemp}, to do a more detailed analysis. 

\begin{figure*}[!htb]
  \begin{center}
    \begin{tabular}{ccc}
      \includegraphics[width=1.5in,angle=-90]{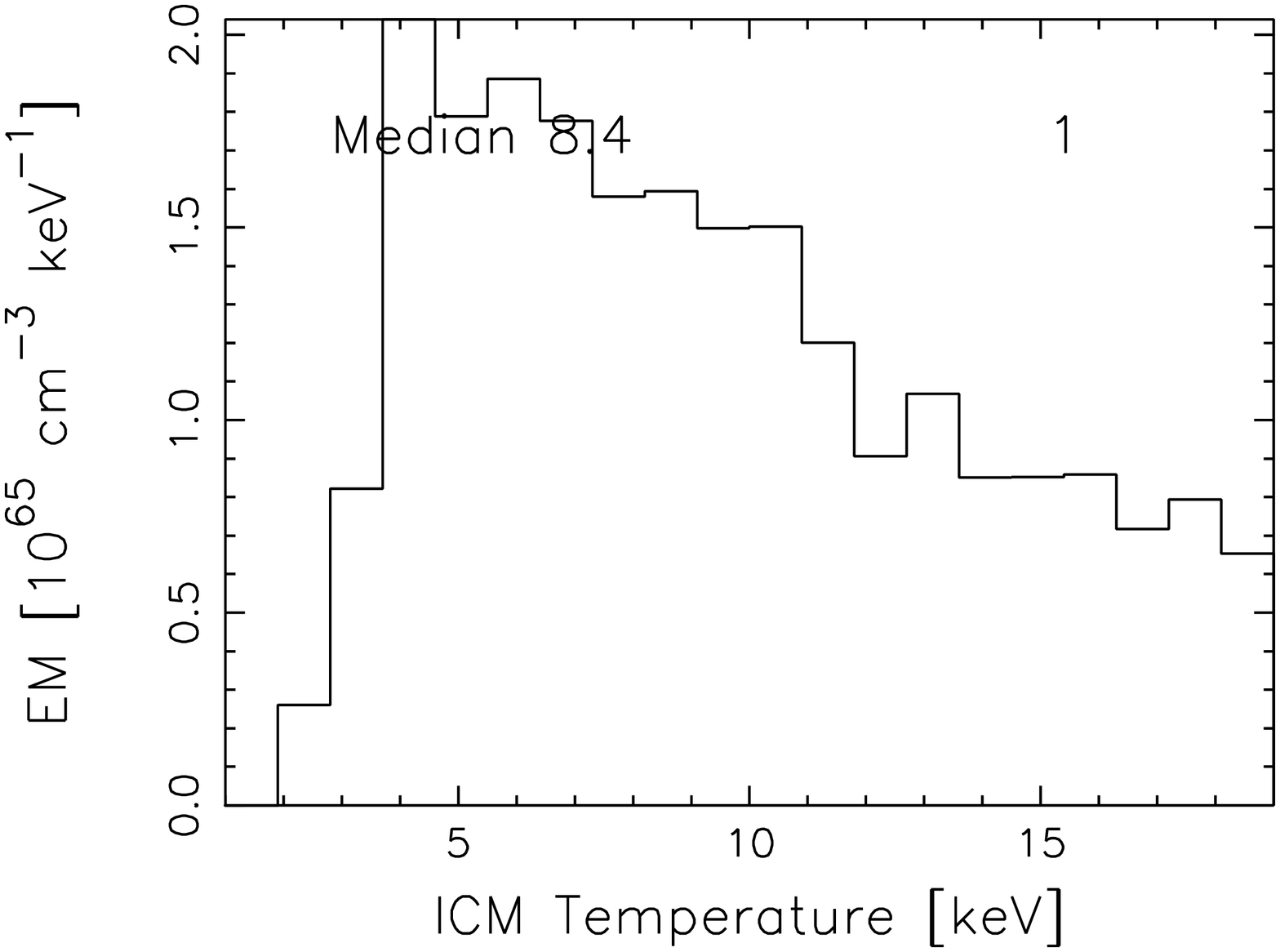} & 
      \includegraphics[width=1.5in,angle=-90]{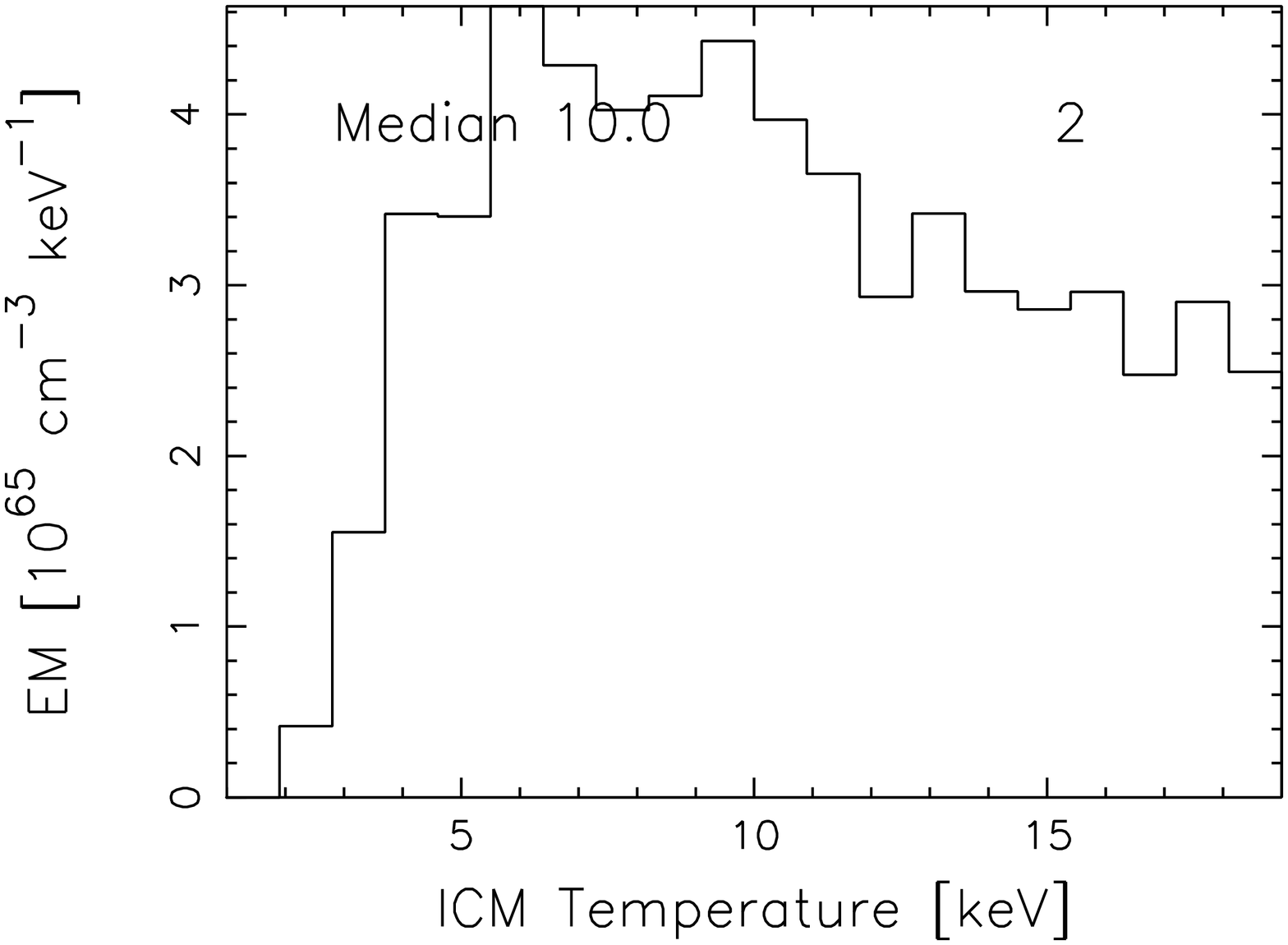} &
      \includegraphics[width=1.5in,angle=-90]{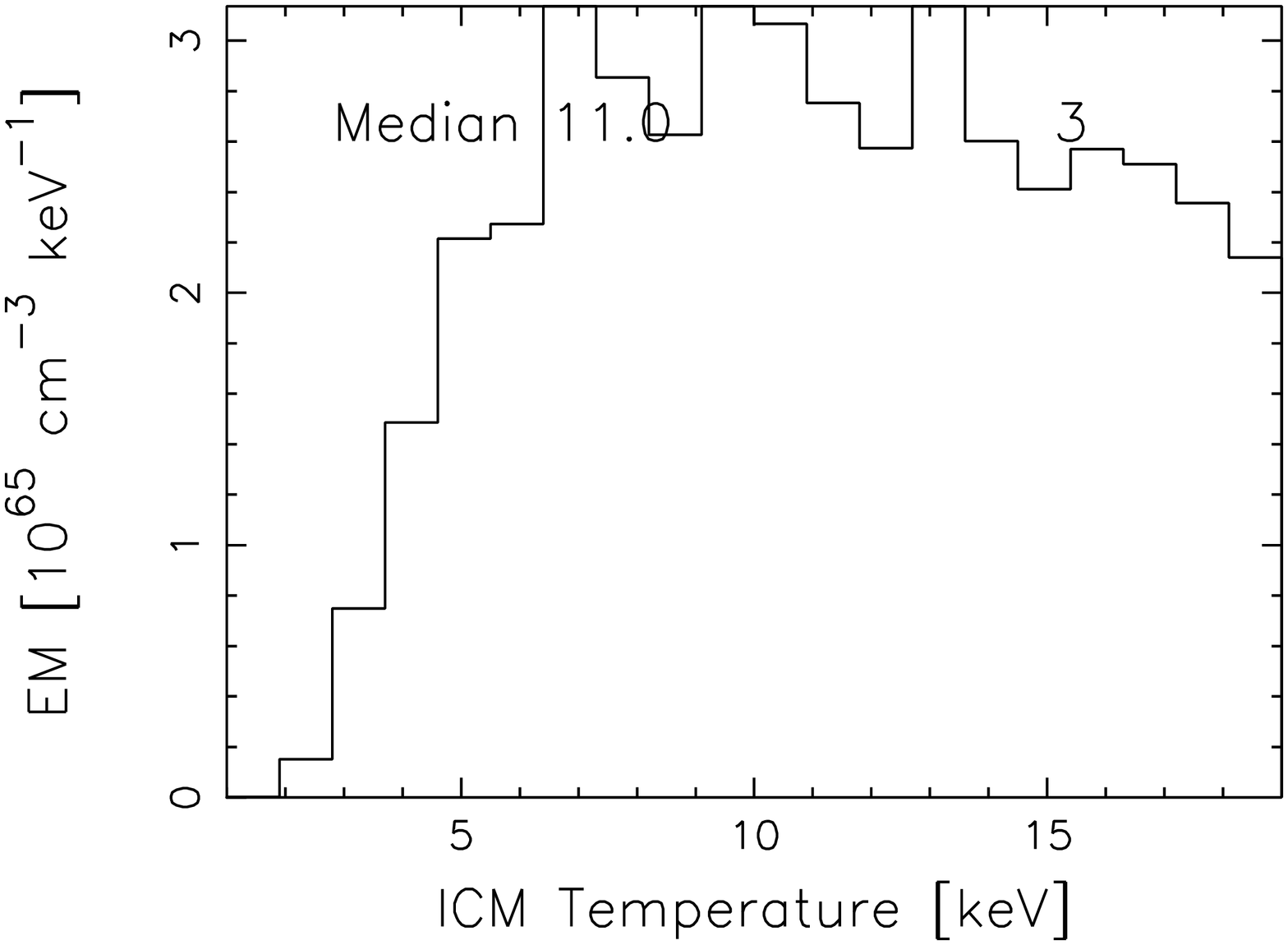} \\
      \includegraphics[width=1.5in,angle=-90]{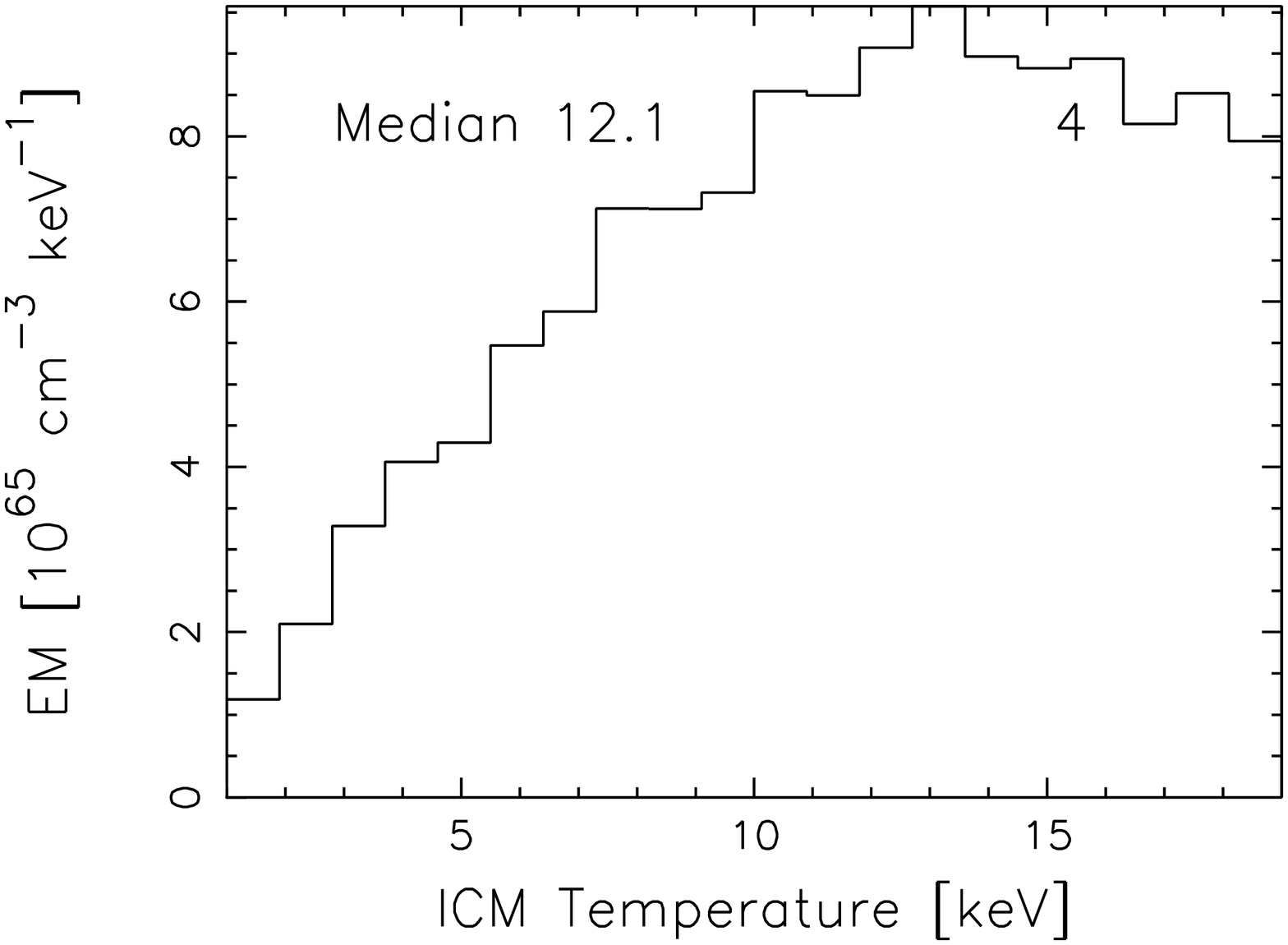} & 
      \includegraphics[width=1.5in,angle=-90]{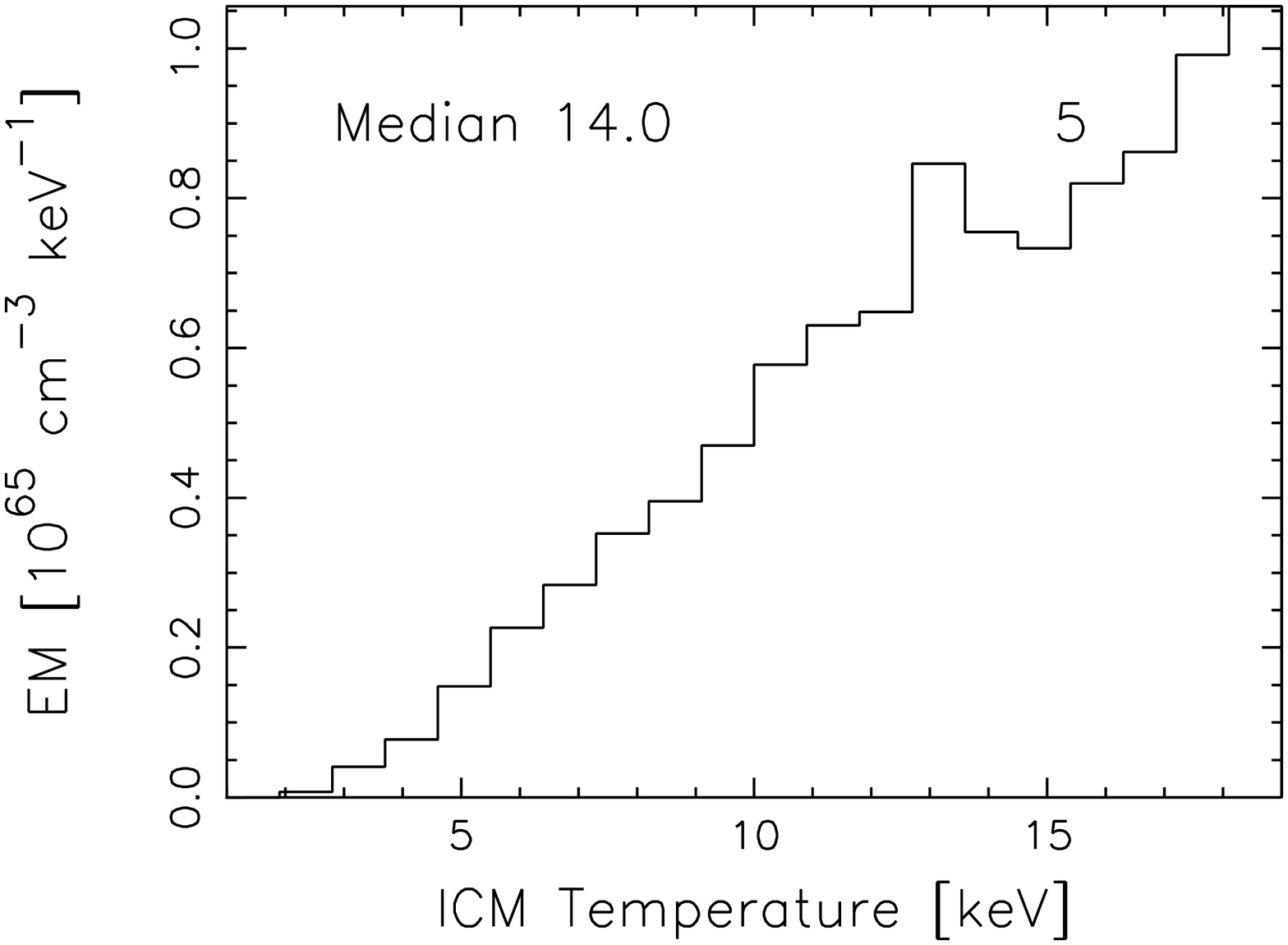} &
      \includegraphics[width=1.5in,angle=-90]{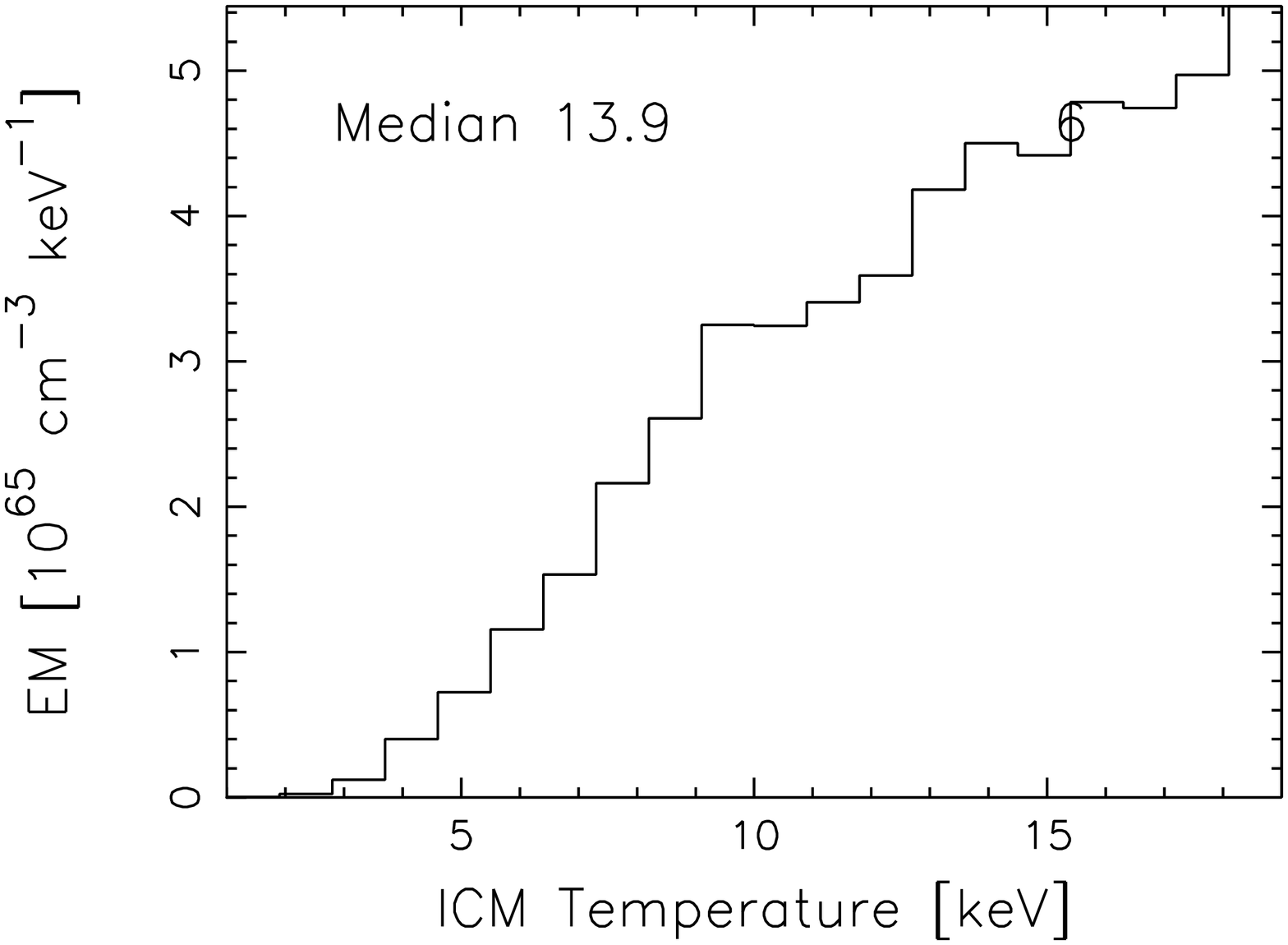} \\
      \includegraphics[width=1.5in,angle=-90]{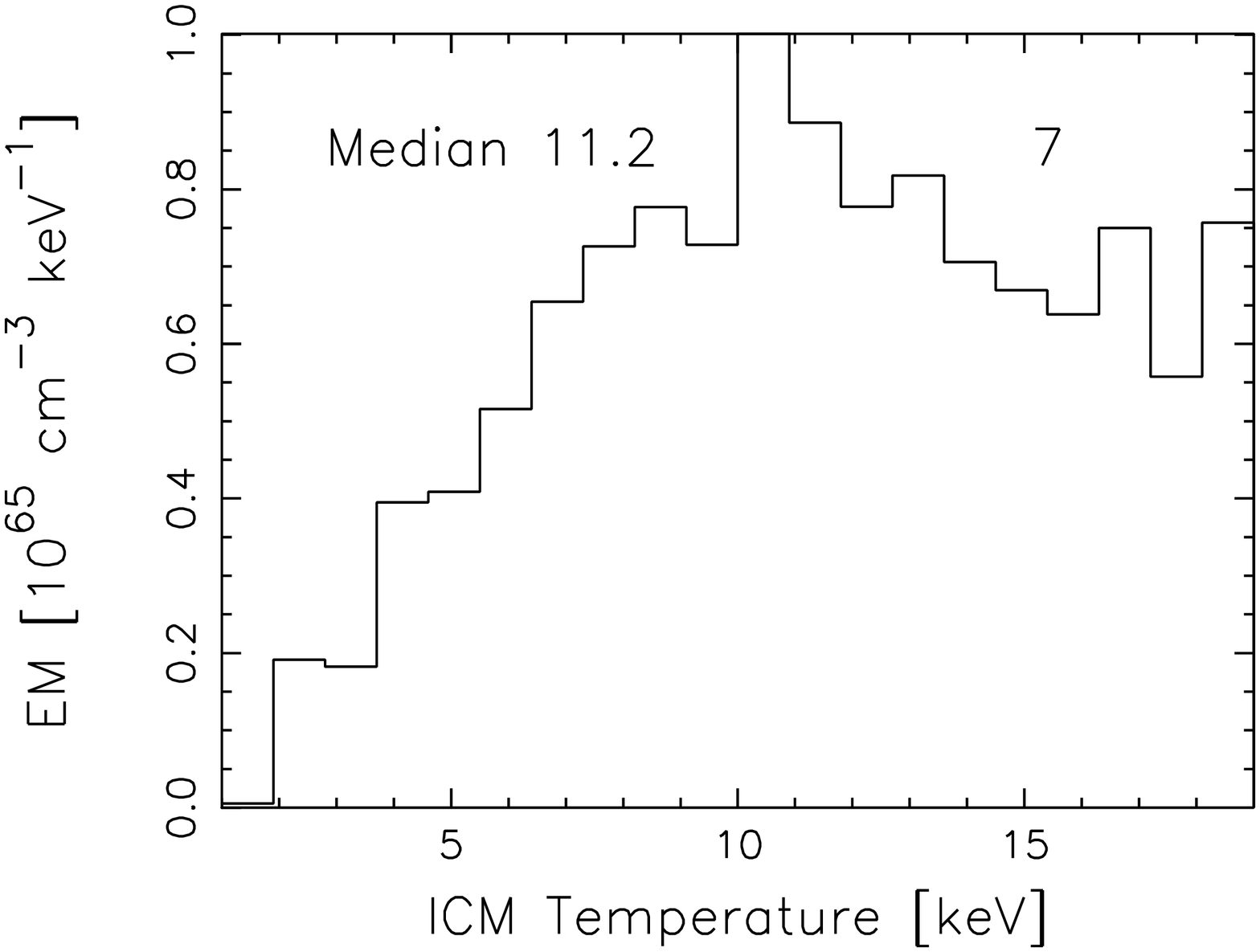} \\
    \end{tabular}
  \end{center}
  \caption{Iteration-averaged distribution of temperatures for RX J0658-55 
in regions 1-7, also showing the median.  \label{bulletTdist} }
\end{figure*}

In Figure \ref{bulletTdist} we show the detailed distributions of 
temperatures, in numerical order, from the selected regions above. 
In region 1 we have extracted 
the central emission from the cluster core. The distribution clearly shows the 
signatures of a $kT = 7~$keV plasma, with possible contamination from higher 
temperatures (seen as peaks around 11 and 14 keV), most likely from projection 
effects. Regions 2-7 show the distributions of various regions in the 
cluster. 

\begin{figure*}[!htb]
  \begin{center}
    \begin{tabular}{ccc}
      \includegraphics[width=1.5in,angle=-90]{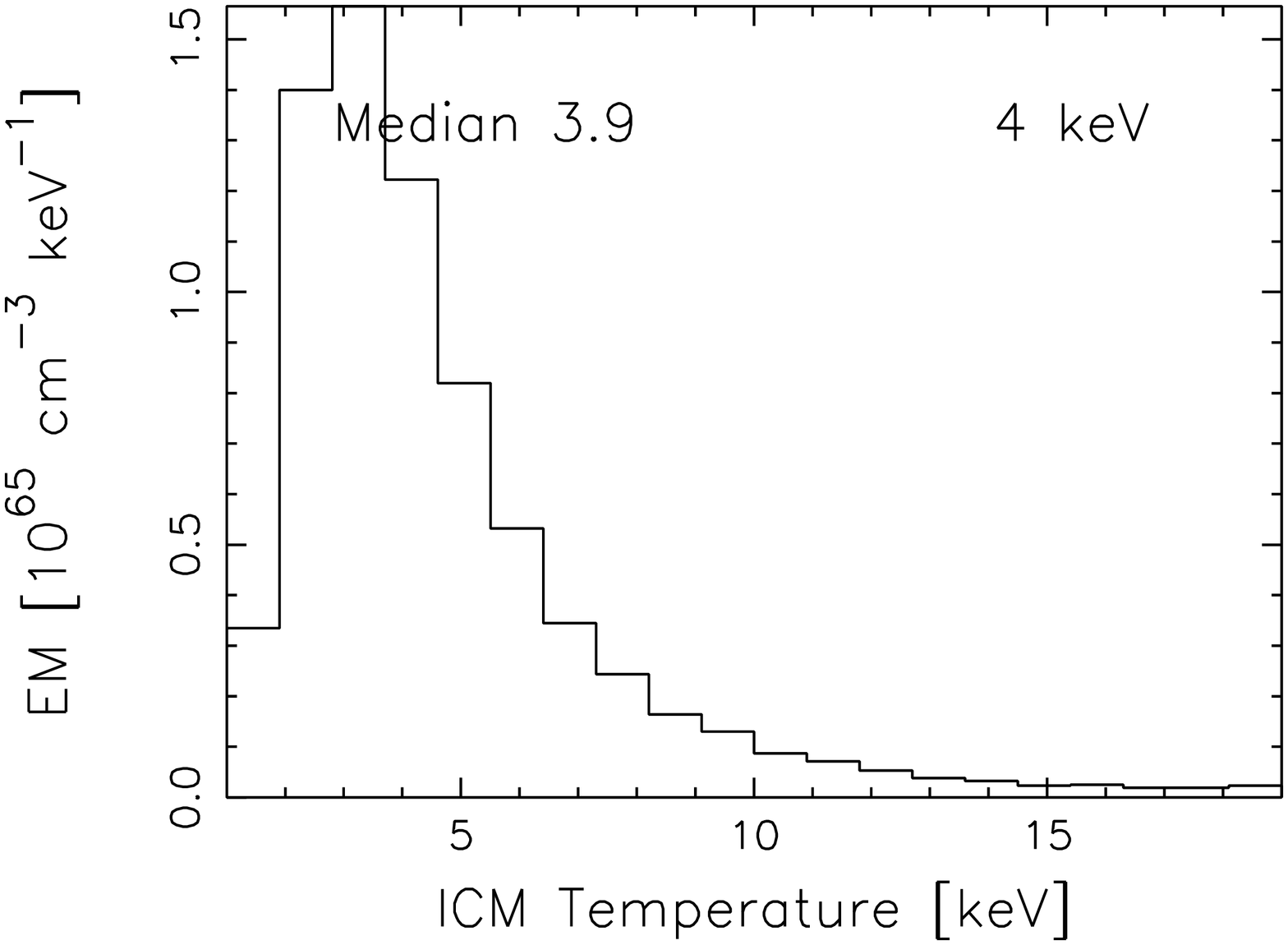} & 
      \includegraphics[width=1.5in,angle=-90]{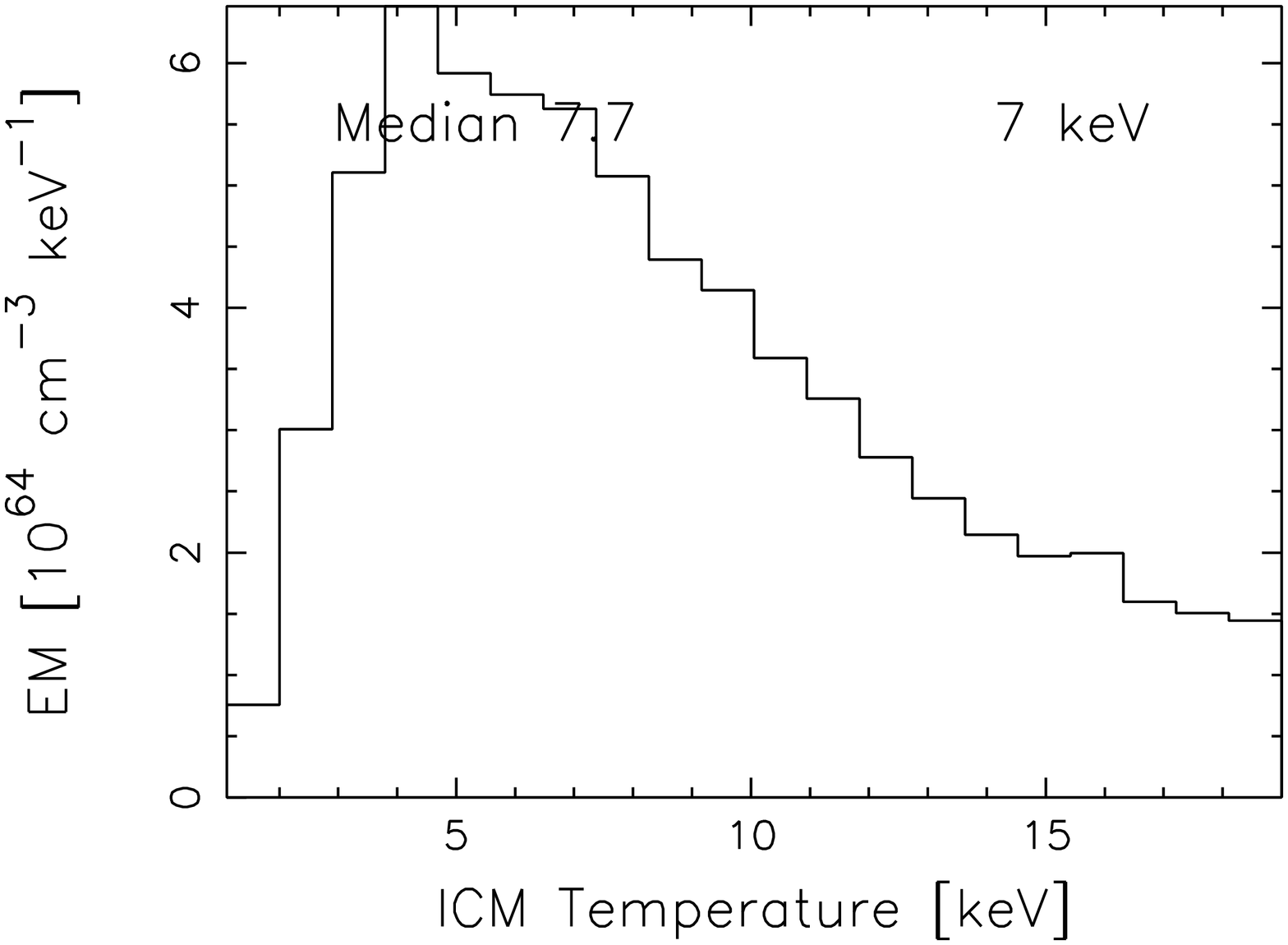} &
      \includegraphics[width=1.5in,angle=-90]{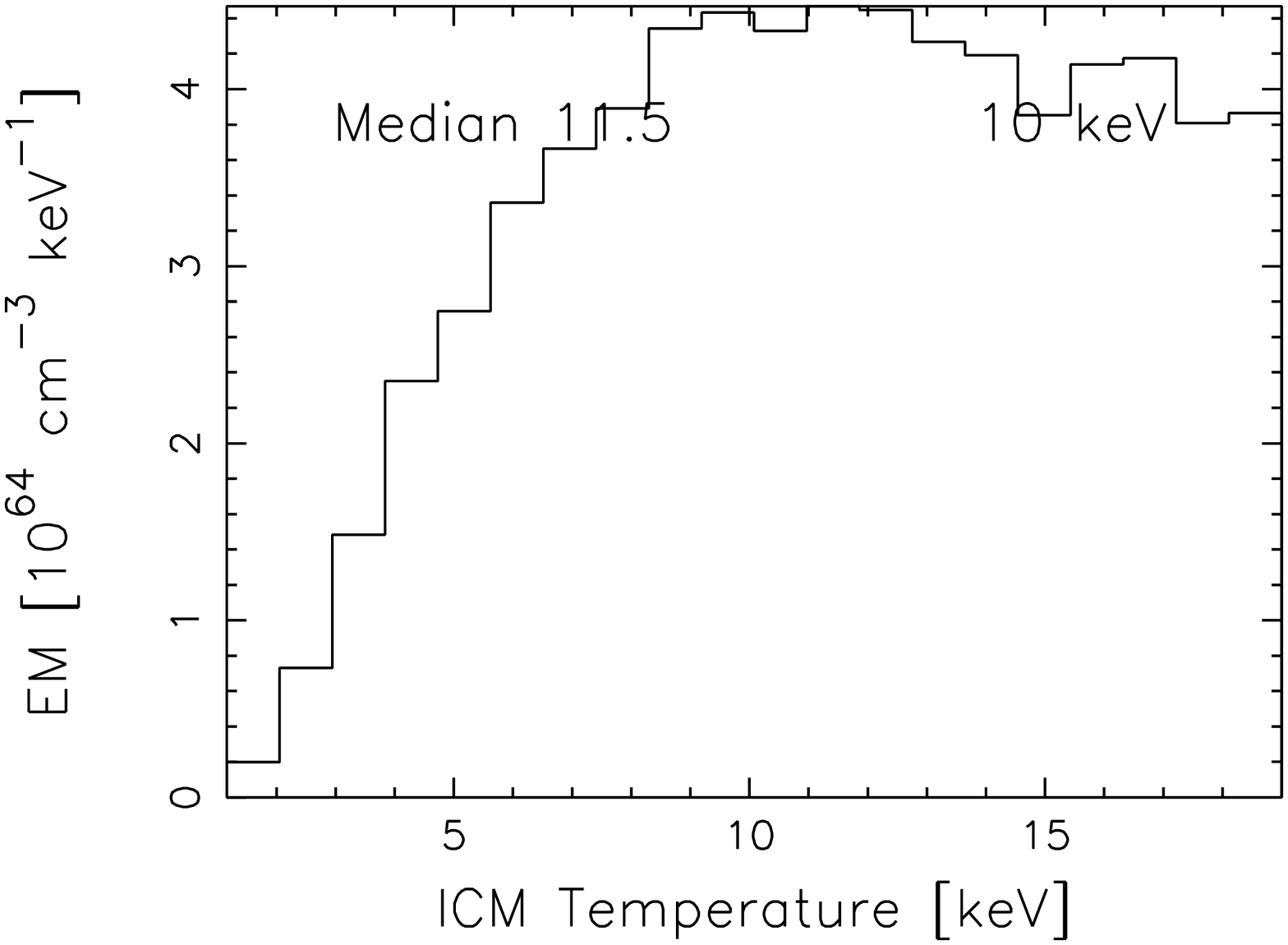} \\
      \includegraphics[width=1.5in,angle=-90]{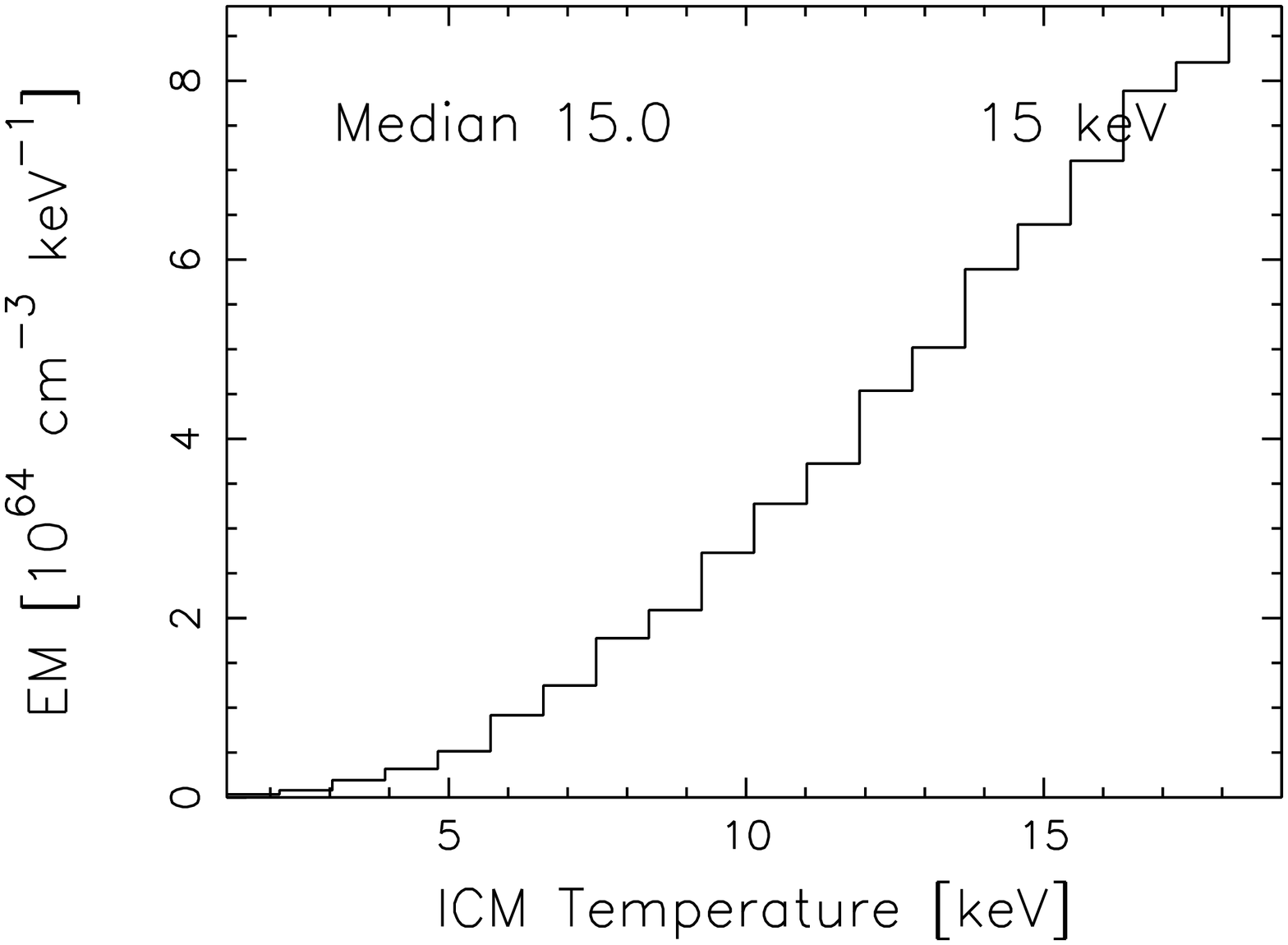} \\
    \end{tabular}
  \end{center}
  \caption{Iteration-averaged distribution of temperatures for the reconstruction of 
isothermal simulations of RX J0658-5557 due to 4, 7, 10, 
and 15 keV plasmas. \label{isothermal} }
\end{figure*}

In order to determine the deviation from isothermality of the plasma in the 
selected regions we 
generated isothermal data sets of a cluster resembling RX J0658-55 with the same number of photons 
and the same spatial structure.  The isothermal models were reconstructed 
using the exact same method as in the reconstruction of the real data. 
The distribution of temperatures from the isothermal reconstructions are shown 
for 4, 7, 10, and 15 keV plasmas in Figure \ref{isothermal}.
None of the distributions in Figure \ref{bulletTdist} can conclusively be distinguished 
from an isothermal plasma. 

\begin{figure*}[!htb]
  \begin{center}
    \begin{tabular}{c}
      \includegraphics[width=5.5in]{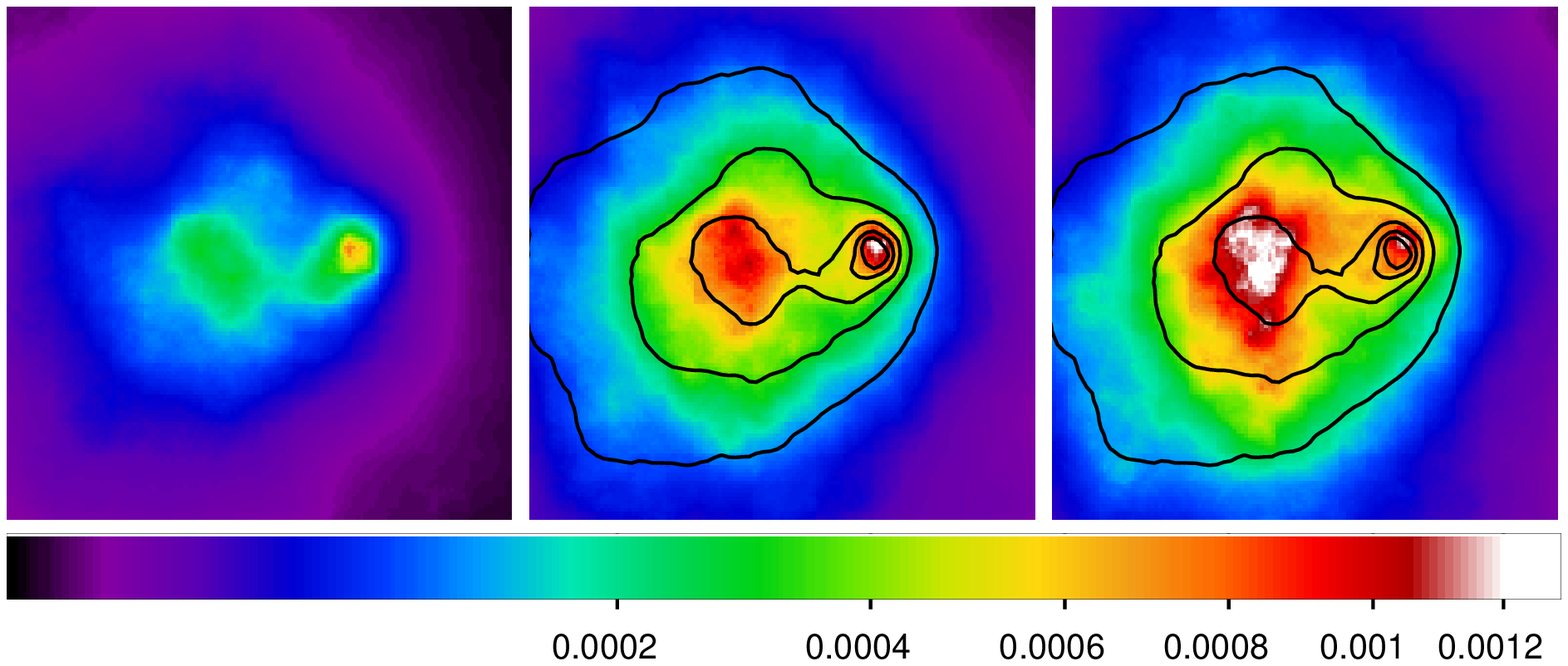} \\ 
    \end{tabular}
  \end{center}
  \caption{Luminosity maps ($L_{bol}/10^{44}~$erg$~$s$^{-1}$ per 
$2\arcsec \times 2\arcsec$) of RX J0658-55 in three different temperature 
bands. From left to right, the ranges are 
1 - 7, 7 - 13 and 13 - 19 keV. \label{3phase}}
\end{figure*}

Finally we have created luminosity maps of RX J0658-55 using emission components 
in separate bands of temperature. Figure \ref{3phase} shows 
the cluster luminosity 
in the $kT = 1 - 7$, $7 - 13$ and $13 - 19$ keV bands. The 
contours are iso-luminosity contours from the $kT = 1 - 7$ keV map.
These maps show an interesting property hinted at by the temperature maps: 
the bullet appears to move further north with higher temperature.  
It is possible that this feature indicates an increased compression to the 
north-west resulting in a higher temperature. This is supported by 
the {\sl Chandra} image which shows a shorter distance between the bullet 
and shock front to the northwest than to the southwest. 

The three-phase map also clearly shows the extension of hotter gas 
to the south that is causing the high-temperature region in the temperature 
maps (region 6).

\subsection{The Centaurus Cluster}
The model spectrum of the Centaurus cluster as inferred from the model sample 
is shown with the data spectrum and 
the ratio of the two in Figure \ref{modelvdata}.  
The model spectrum 
can be seen to fit the data well, and this is a consequence of the fact that 
we have a low value of the overall Poisson $\chi^2$, as can be seen in Figure \ref{statrej}. 
However, there are some residuals at $> 5~$keV energies due to some inadequately 
fitted spectral lines. Since most of the data are in the low-energy range, these 
data tend to drive the fit, leading to some unwanted excesses and deficits at 
higher energies. These, however, do not have a major impact on the statistic. 

\begin{figure*}[!htb]
\begin{center}
  \begin{tabular}{cc}
    \includegraphics[width=2.2in]{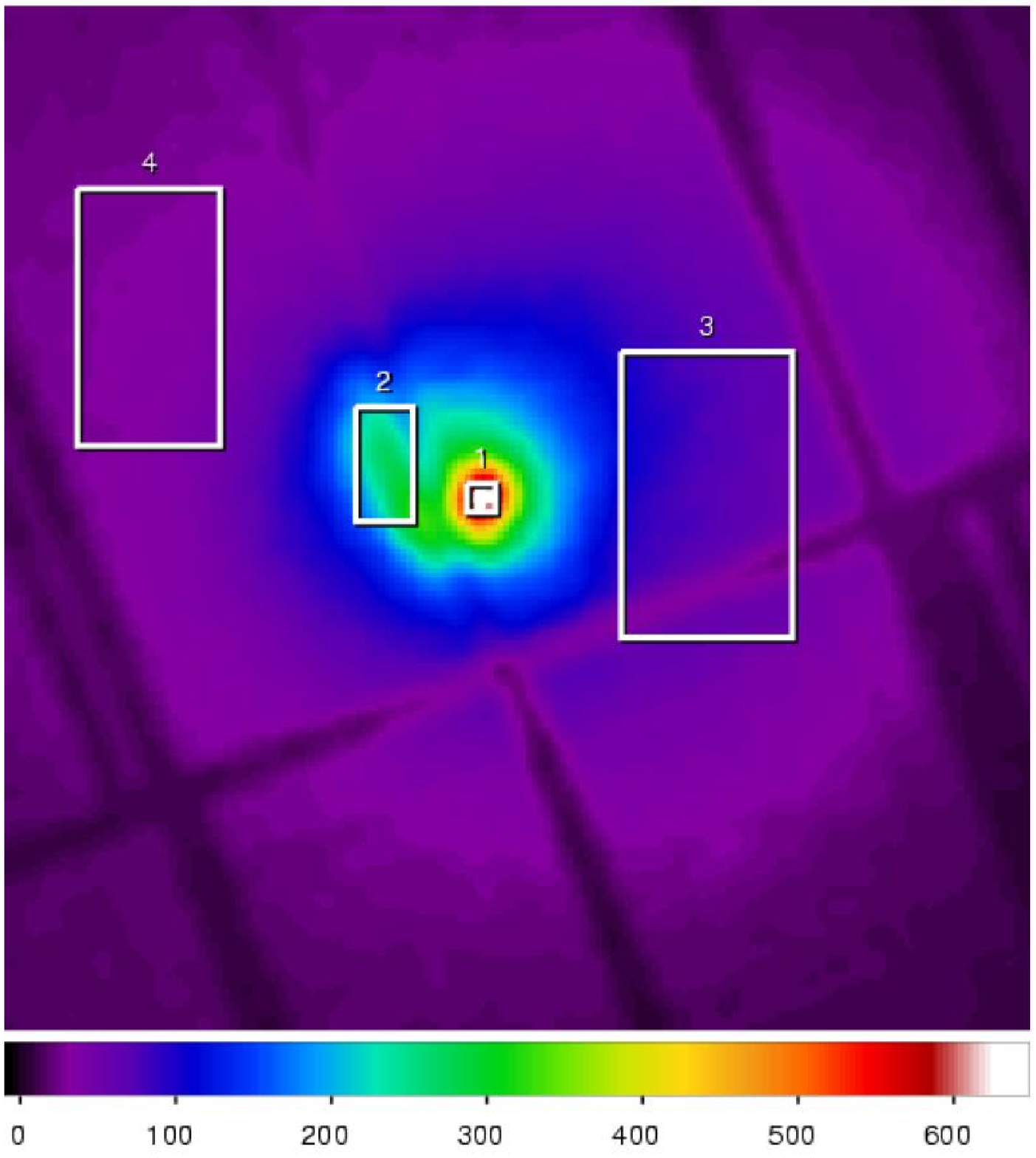} & 
    \includegraphics[width=2.2in]{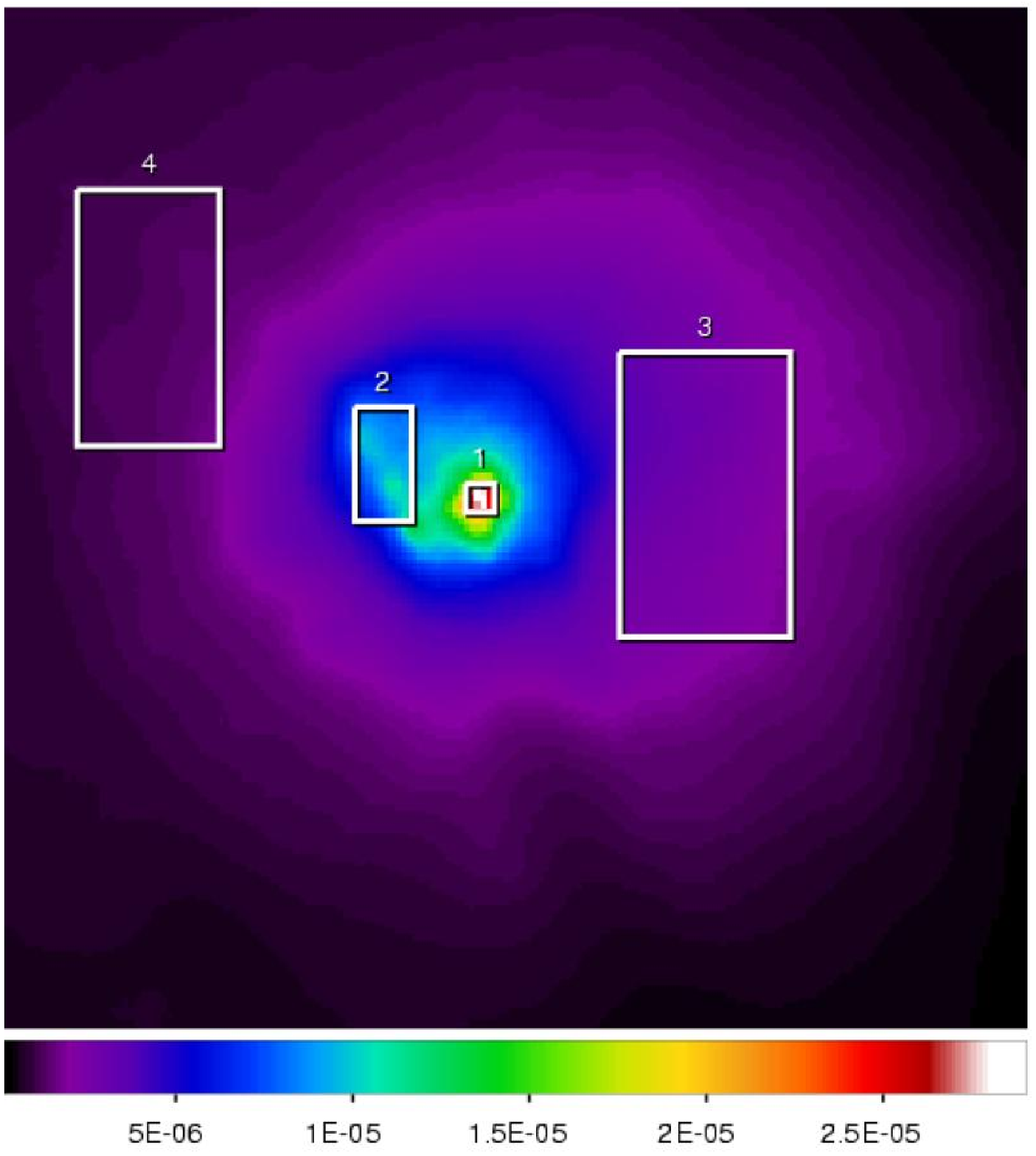} \\
  \end{tabular}
\end{center}
  \caption{Results of the analysis for the Centaurus cluster, showing the central $6\arcmin \times 6\arcmin$ region.
{\sl Left:} Raw data smoothed by a $4\arcsec$ kernel Gaussian 
(counts per $2\arcsec \times 2\arcsec$ pixel).
{\sl Right:} Luminosity reconstruction using the median of all samples at
each spatial point ($L_{bol}/10^{44}~$erg$~$s$^{-1}$ per $2\arcsec \times 2\arcsec$). \label{cenLumTemp}}
\end{figure*}

\begin{figure*}[!htb]
  \begin{center}
    \begin{tabular}{cc}
      \includegraphics[width=2.2in]{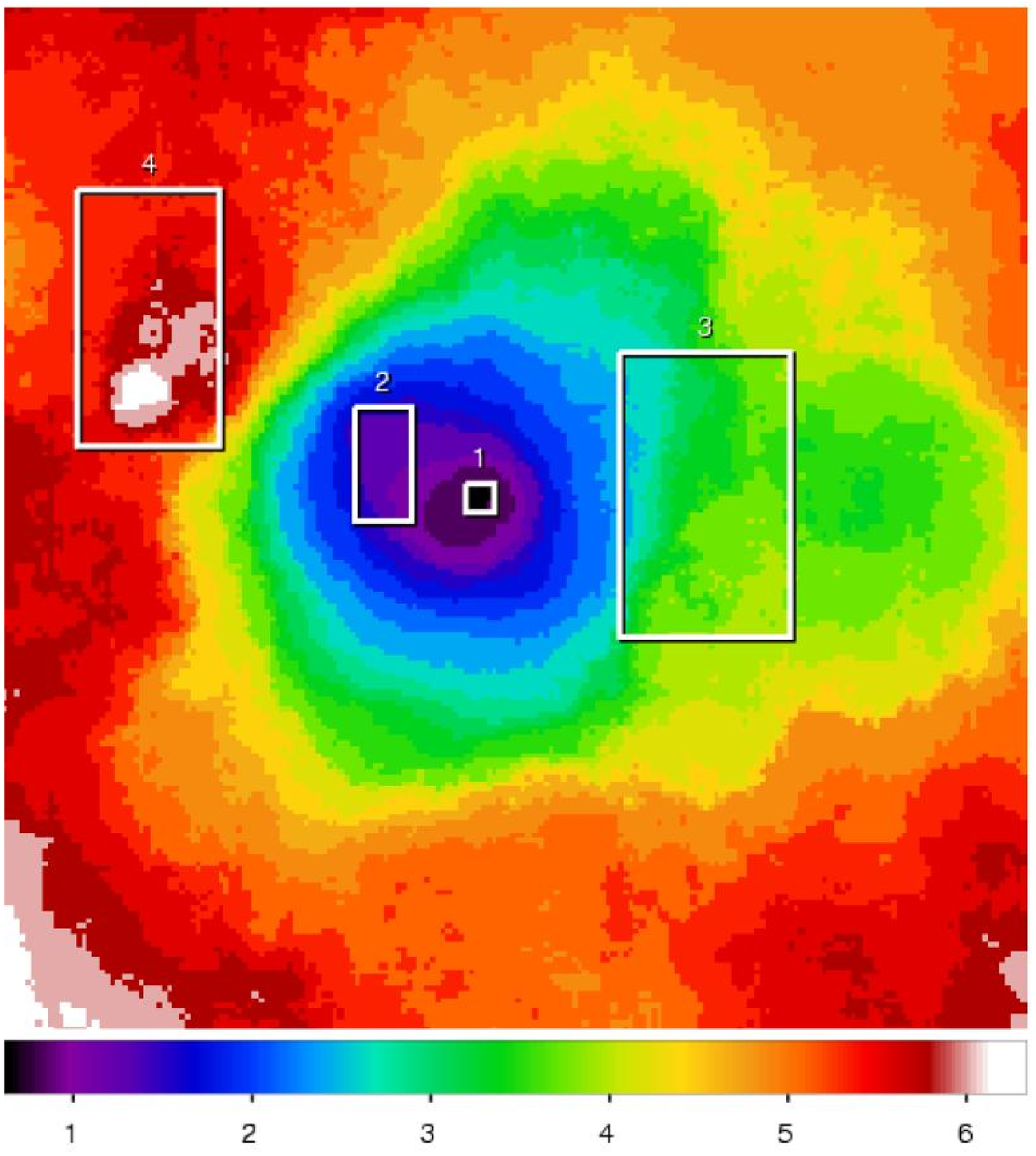} &
      \includegraphics[width=2.2in]{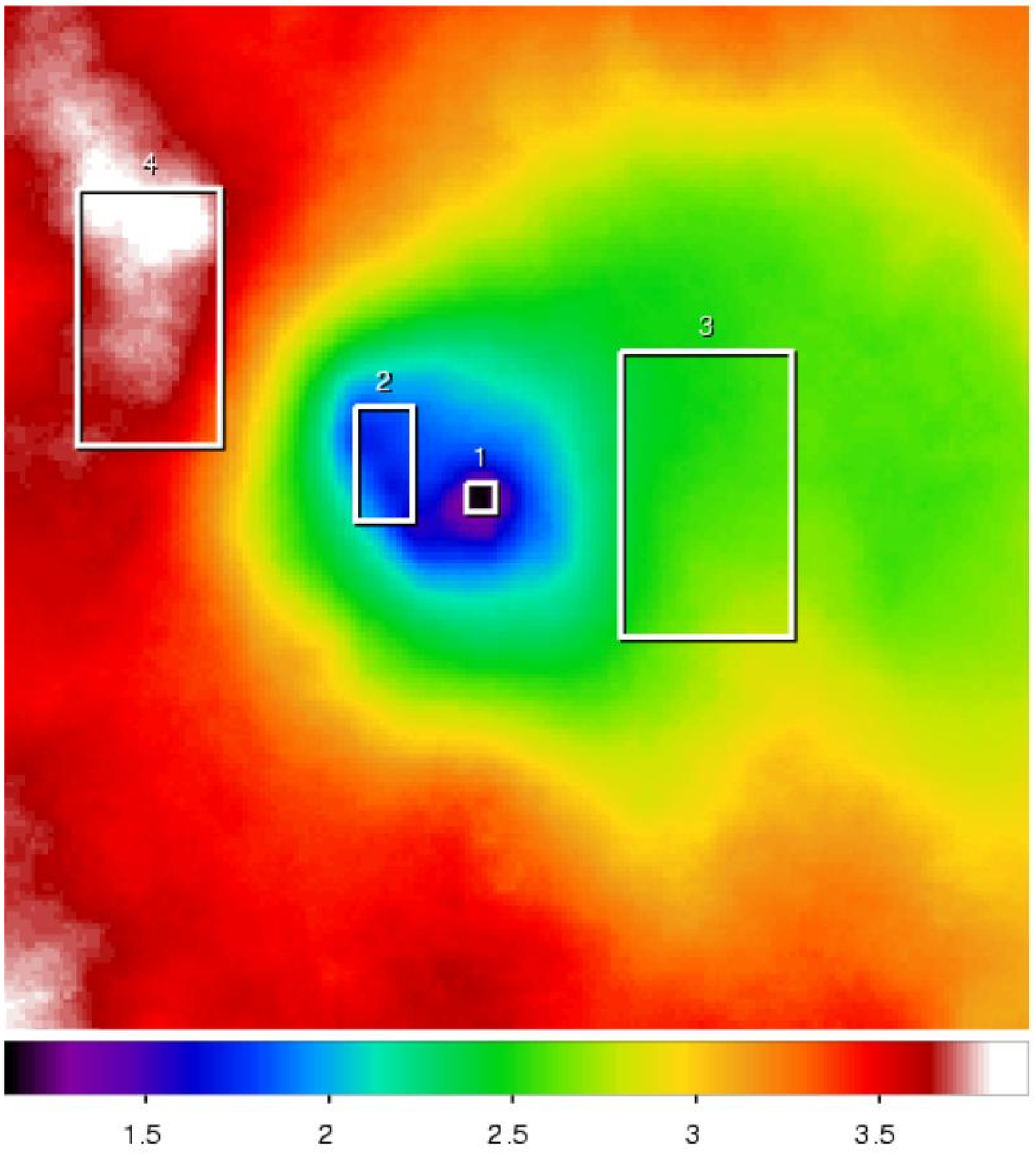} \\
      \includegraphics[width=2.2in,angle=0]{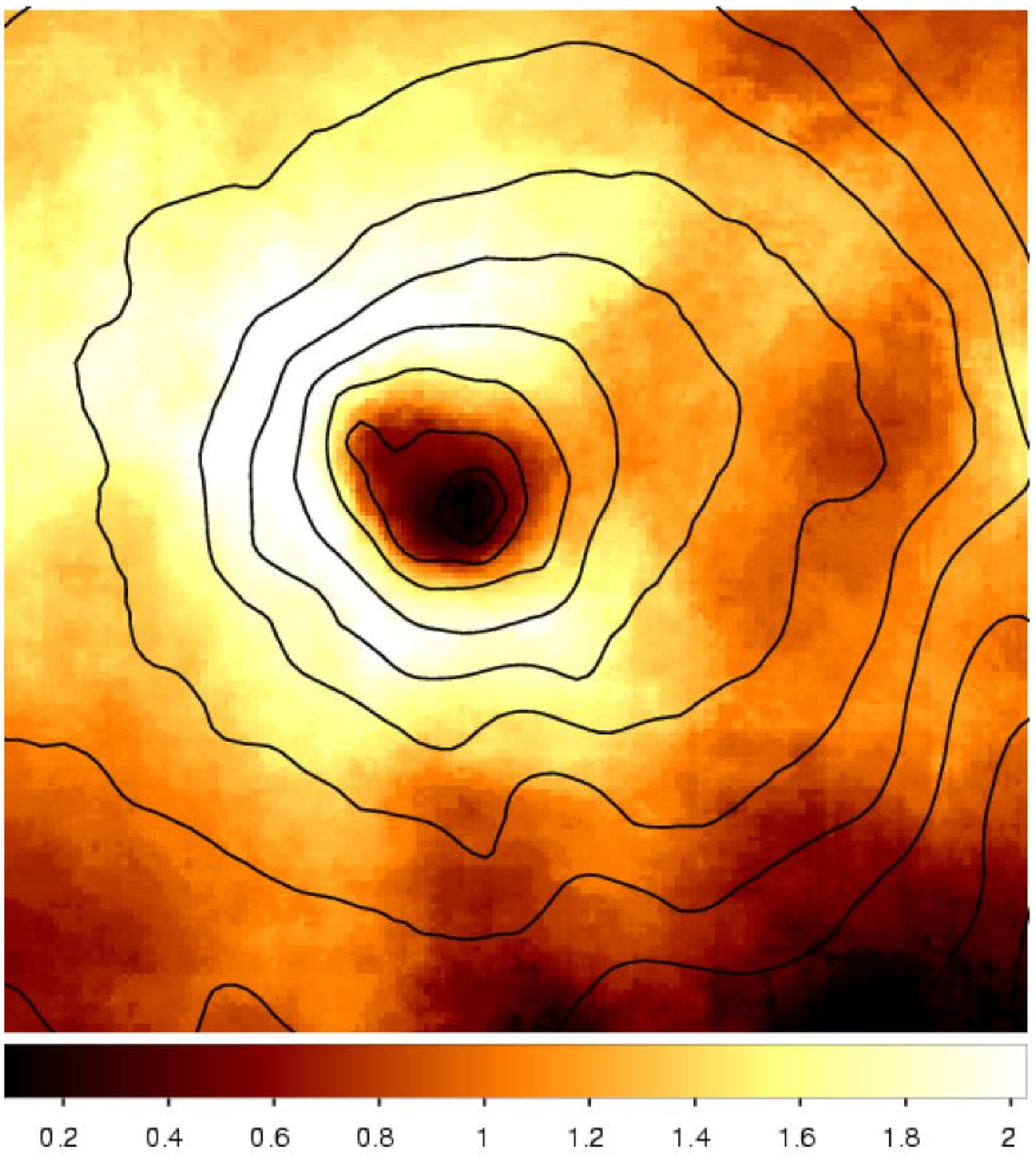} & 
      \includegraphics[width=2.2in,angle=0]{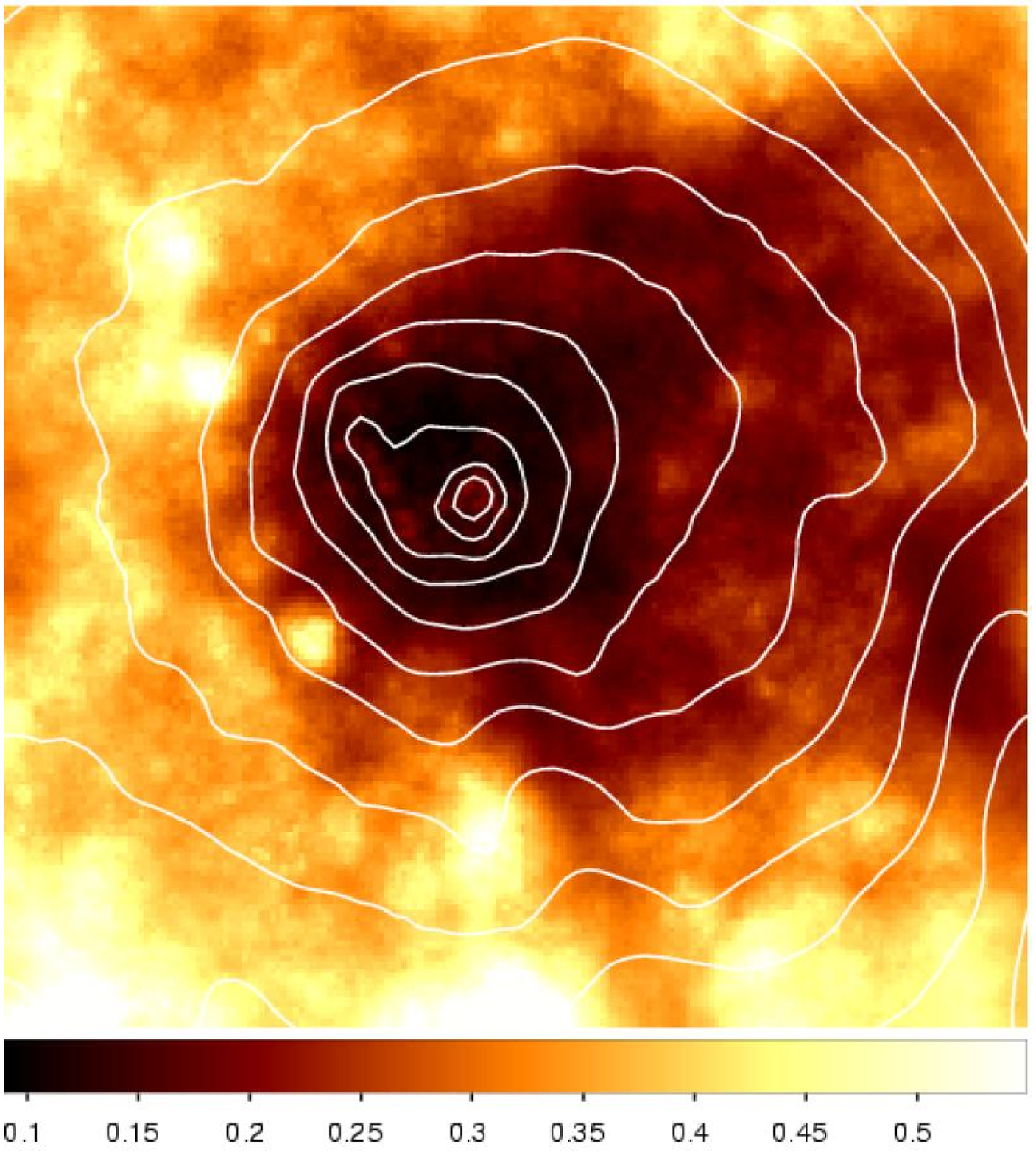} \\ 
    \end{tabular}
  \end{center}
  \caption{Results of the analysis for the Centaurus cluster, showing the central $6\arcmin \times 6\arcmin$ region.
{\sl Top left,} temperature map (in units of keV) constructed
using the mode of the distribution for all samples at each spatial point;
{\sl top right,} temperature map (in units of keV) constructed using the median of the
distribution for all samples at each spatial point;
{\sl bottom left,} temperature uncertainty map (in units of keV) showing the 1 $\sigma$ variation
of the temperature mode in the model sample;
{\sl bottom right,} temperature uncertainty map (in units of keV) showing the 1 $\sigma$ variation
of the emission weighted temperature in the model sample.
\label{cenTerr}}

\end{figure*}

We form both luminosity and temperature maps, analogous to our procedures in previous sections. 
In Figure \ref{cenLumTemp} the raw count map of the {\sl XMM-Newton} data ({\sl left}) is 
shown as well as the luminosity reconstruction ({\sl right}). 
Even though the gaps from dead pixel rows are taken into account 
via the exposure map, some artifacts can be seen. The model aligns 
itself with the chip gap where a filament extends north-east of the 
cluster core. Figure \ref{cenTerr} shows a temperature mode 
map ({\sl top left}), as well as a median temperature map ({\sl top right}), 
with the relevant uncertainty maps shown below.
The dominant temperature mode can be seen to be elevated with respect 
to the median in a hot region north-east in the direction of the filament. 

\begin{figure*}[!htb]
  \begin{center}
    \begin{tabular}{cc}
      \includegraphics[width=1.5in,angle=-90]{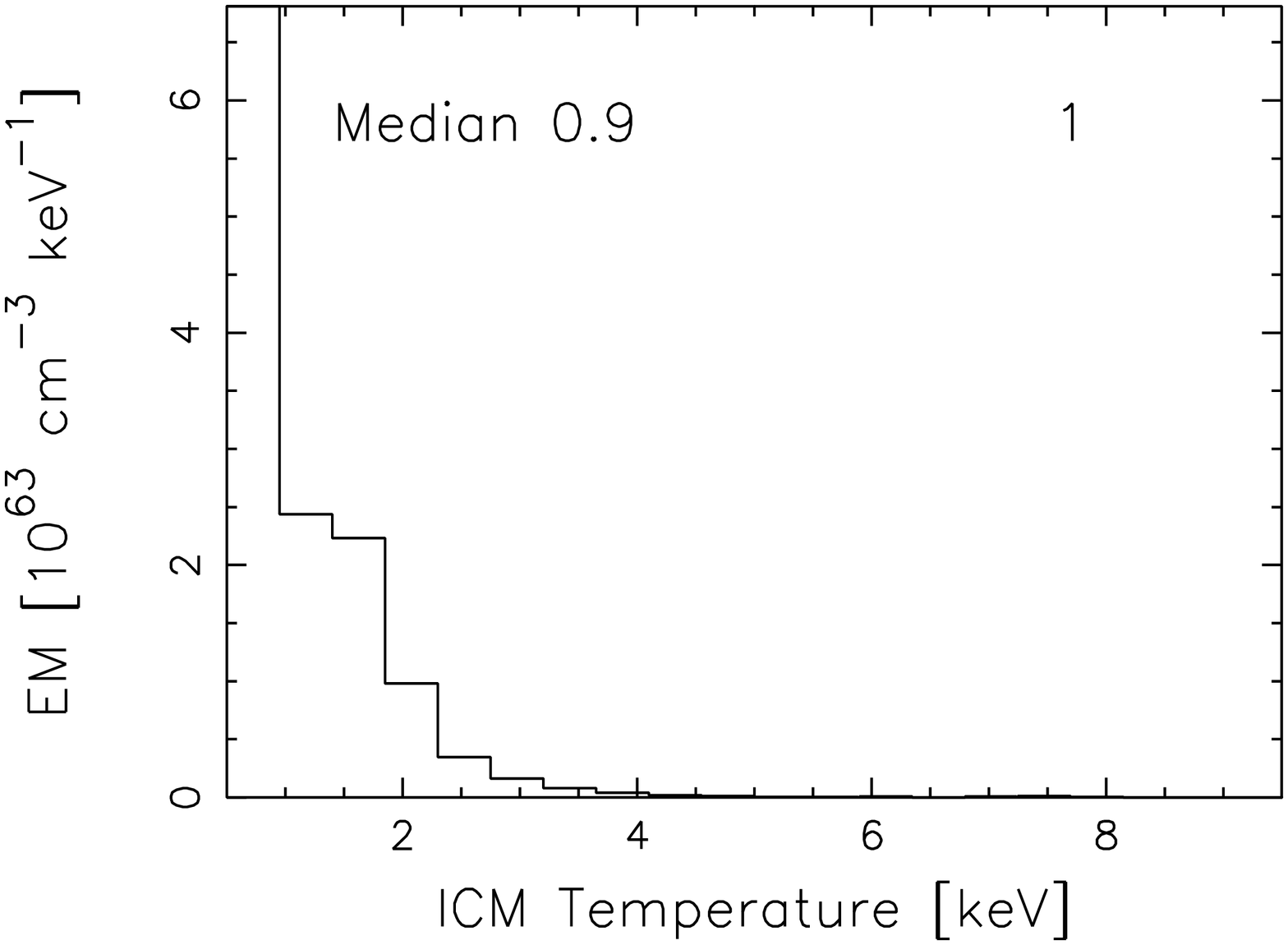} & 
      \includegraphics[width=1.5in,angle=-90]{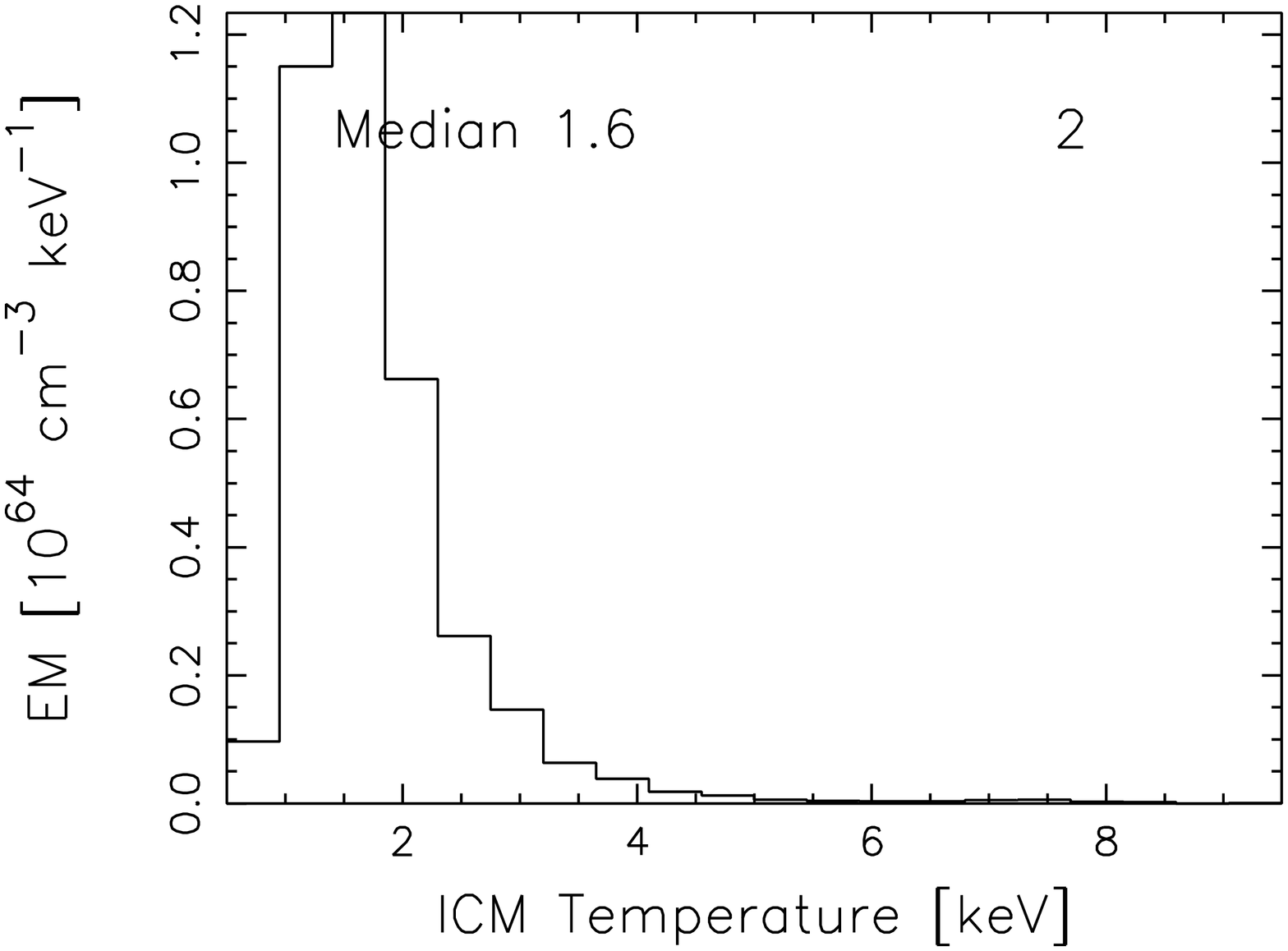} \\
      \includegraphics[width=1.5in,angle=-90]{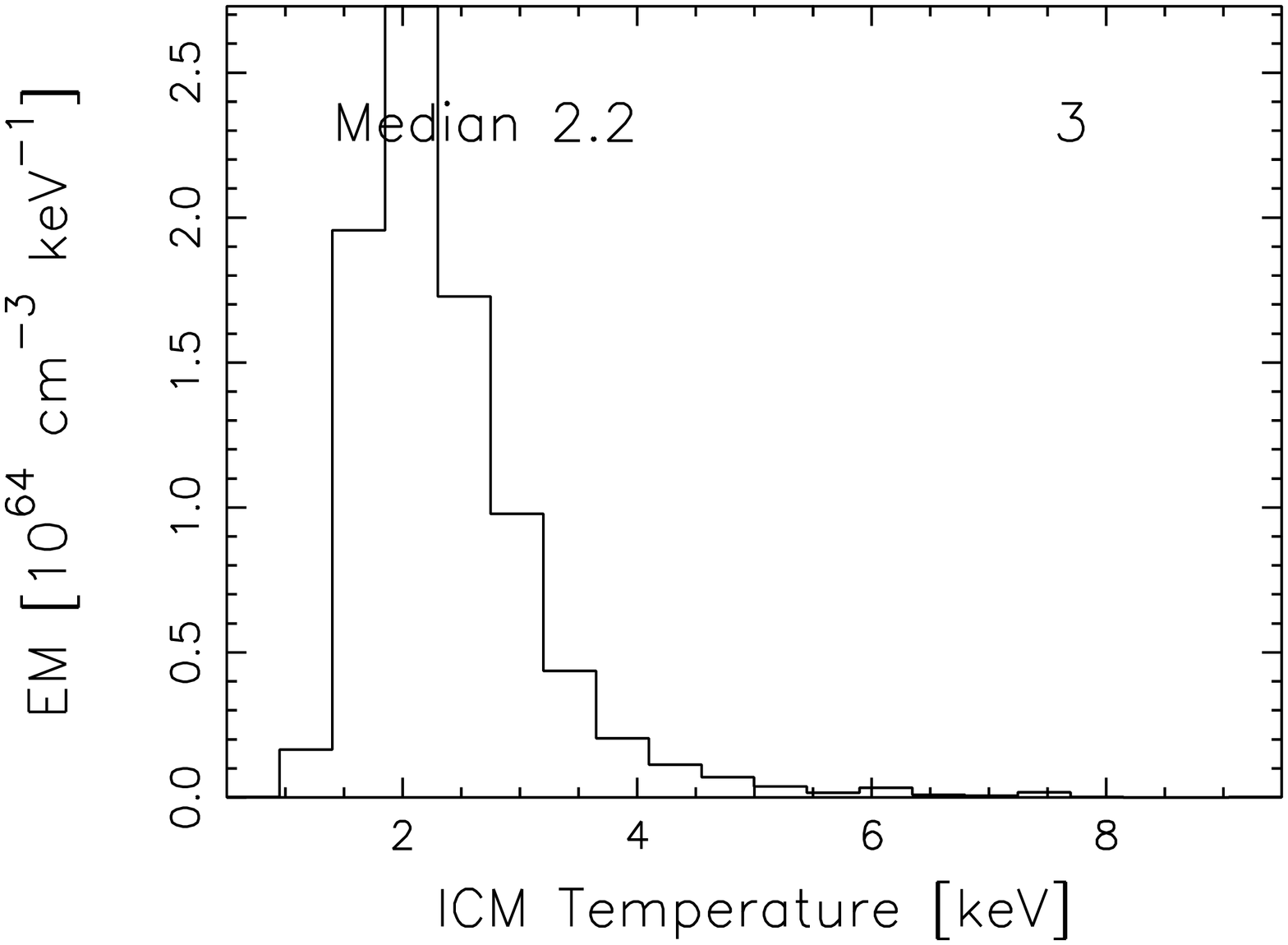} &
      \includegraphics[width=1.5in,angle=-90]{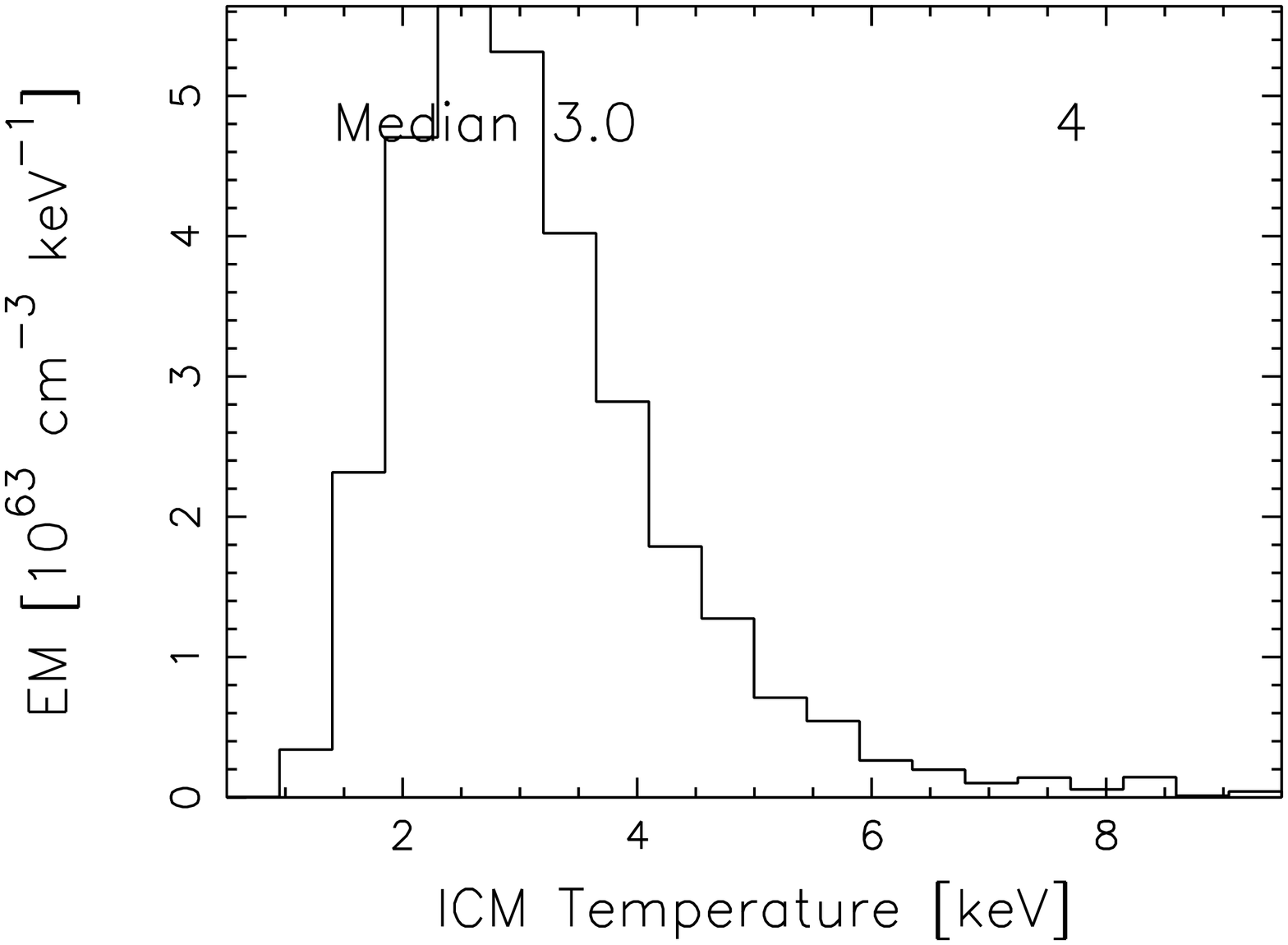} \\ 
    \end{tabular}
  \end{center}
  \caption{Iteration-averaged distribution of temperatures for Centaurus 
in regions 1-4 also showing the median.  \label{cenTdist} }
\end{figure*}

To investigate the temperature in some of the more interesting regions 
in detail, we calculated the temperature distributions in four regions 
and compared them. These plots are shown in Figure \ref{cenTdist}.
The order of the regions corresponds to the numbers in Figure \ref{cenLumTemp}. 
Selected regions correspond to the cluster core (No. 1), the extended 
filament (No. 2), the ambient temperature directly to the west of the core 
(No. 3), and the anomalously hot region to the north-east (No. 4). 
The core shows signs of nonisothermality; the distribution includes 
a narrow peak at 0.5 keV, as well as a bump around 1.5 keV, very 
probably due to projection. In the other plots it is harder to 
distinguish the distributions from isothermal. However in region 4 there 
is a hint of a hotter $\sim 8 $keV phase. 

\begin{figure*}[!htb]
  \begin{center}
    \begin{tabular}{ccc}
      \includegraphics[width=1.8in,angle=0]{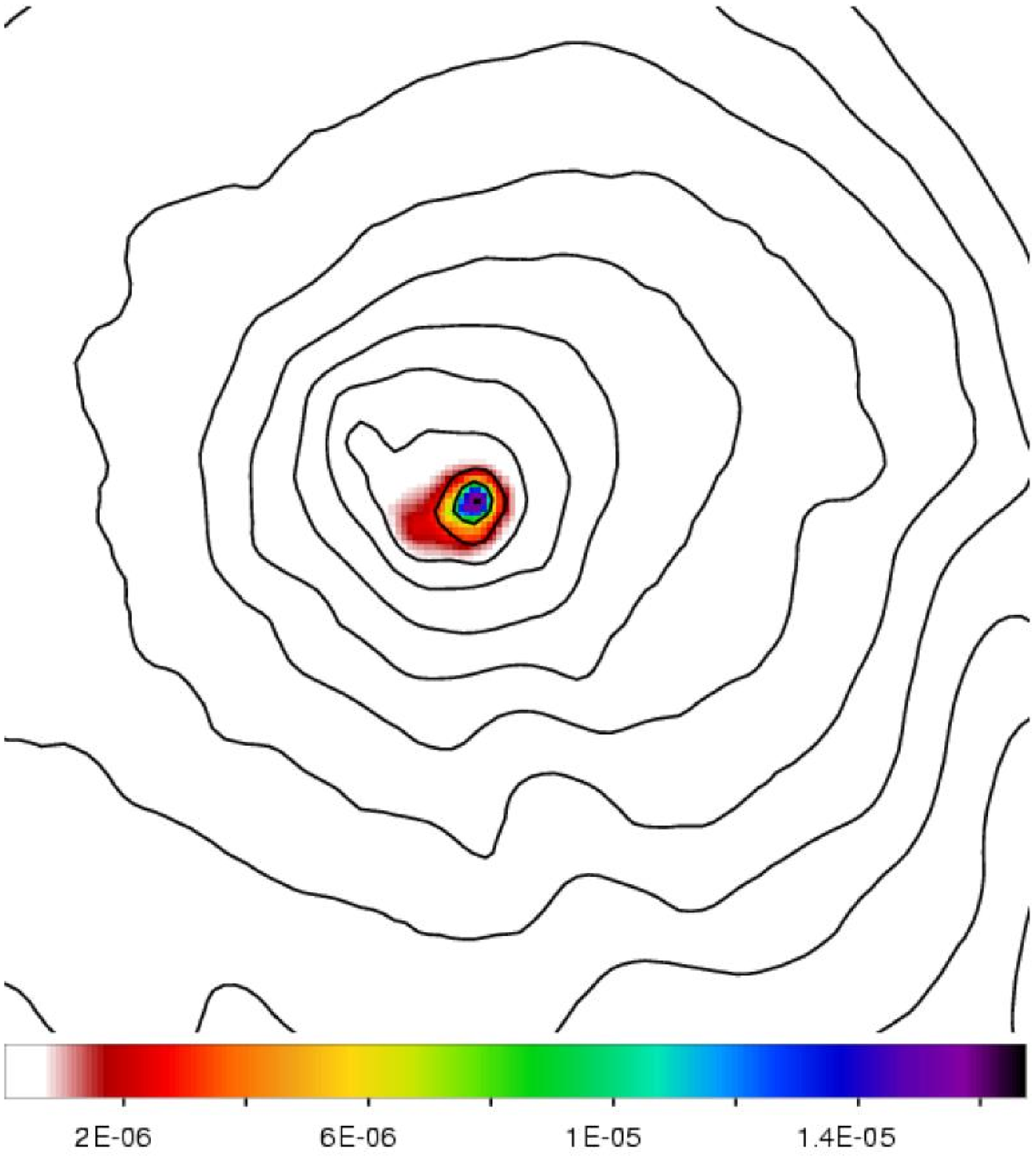} & 
      \includegraphics[width=1.8in,angle=0]{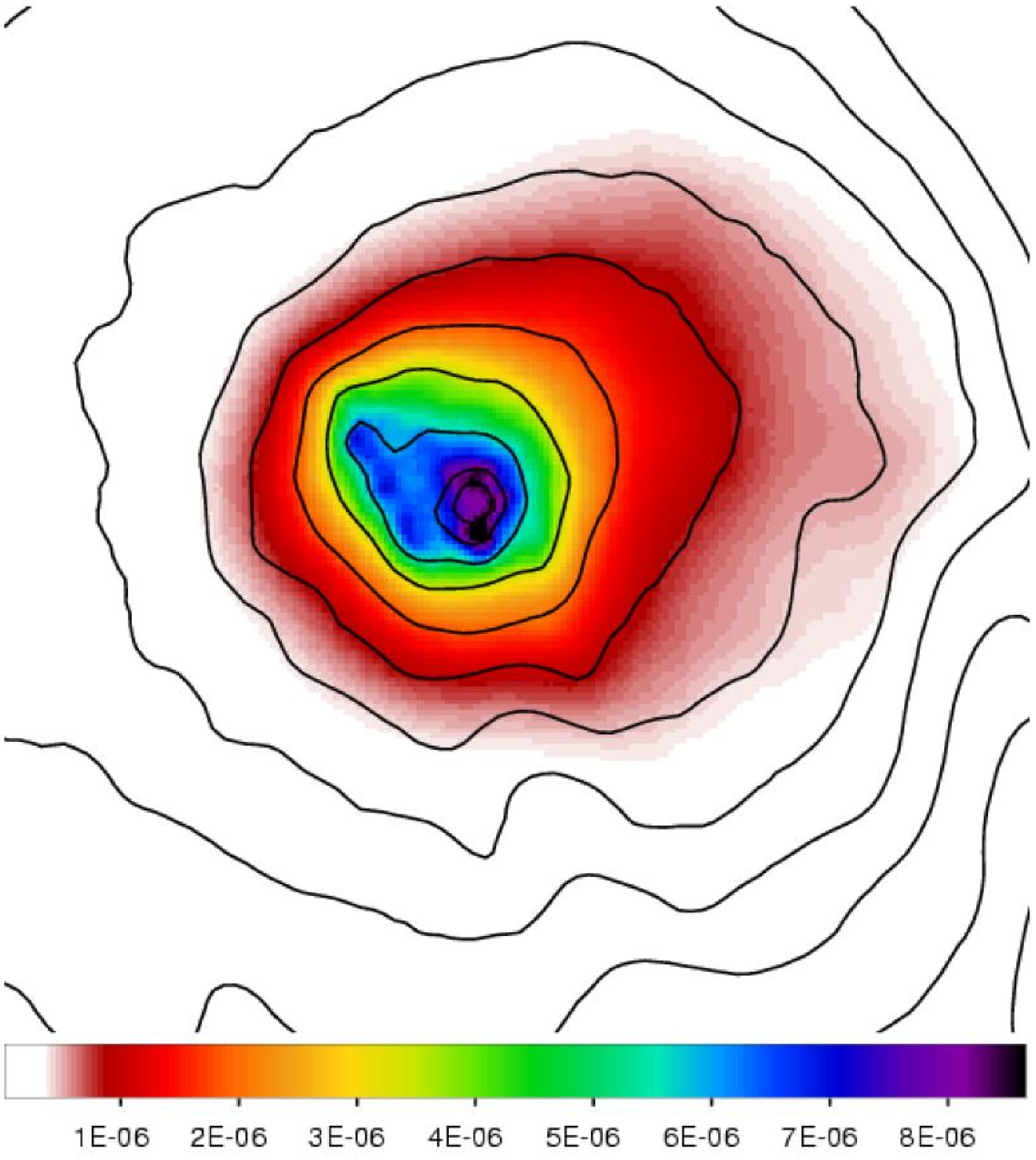} &
      \includegraphics[width=1.8in,angle=0]{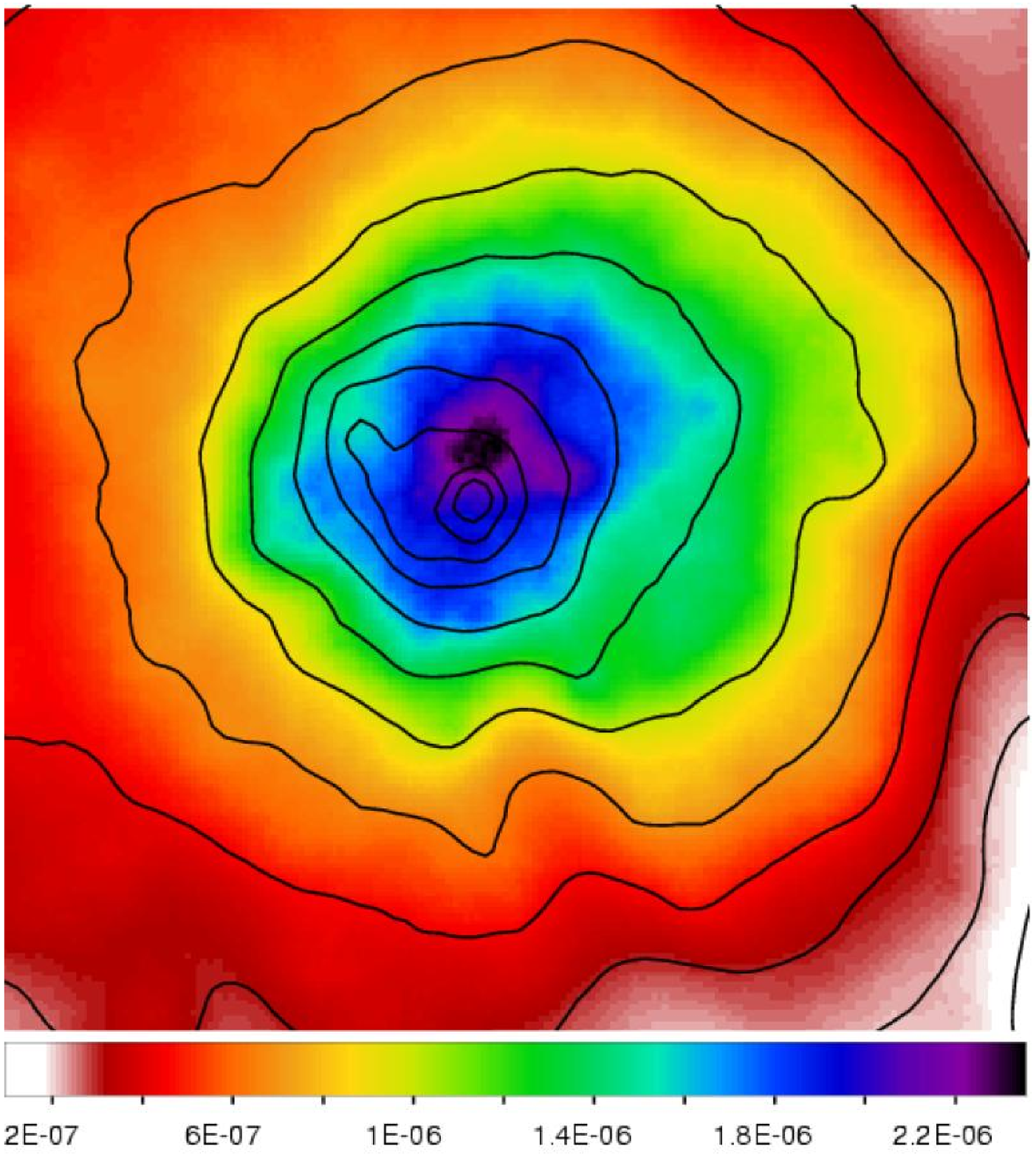} \\
      \includegraphics[width=1.8in,angle=0]{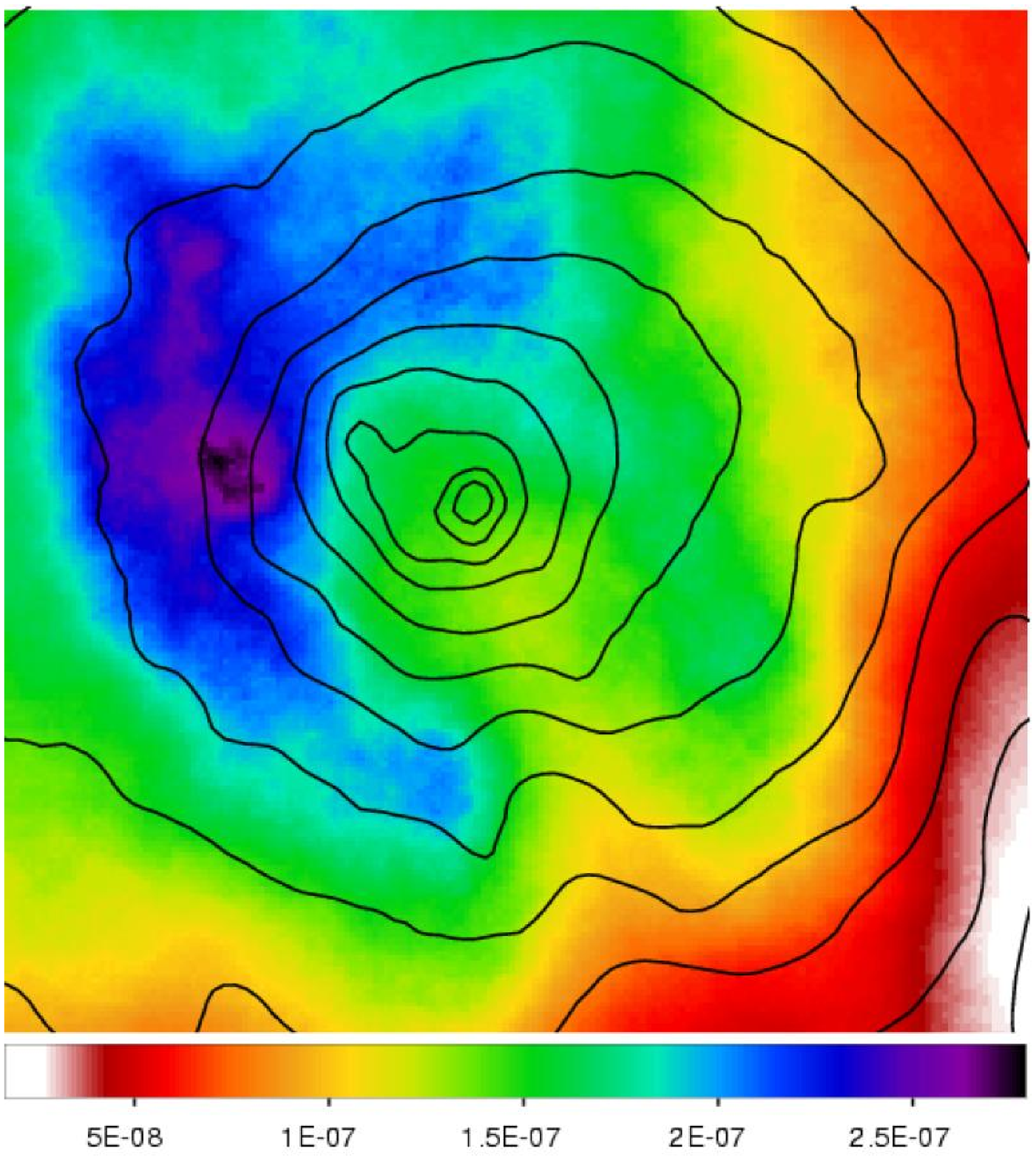} & 
      \includegraphics[width=1.8in,angle=0]{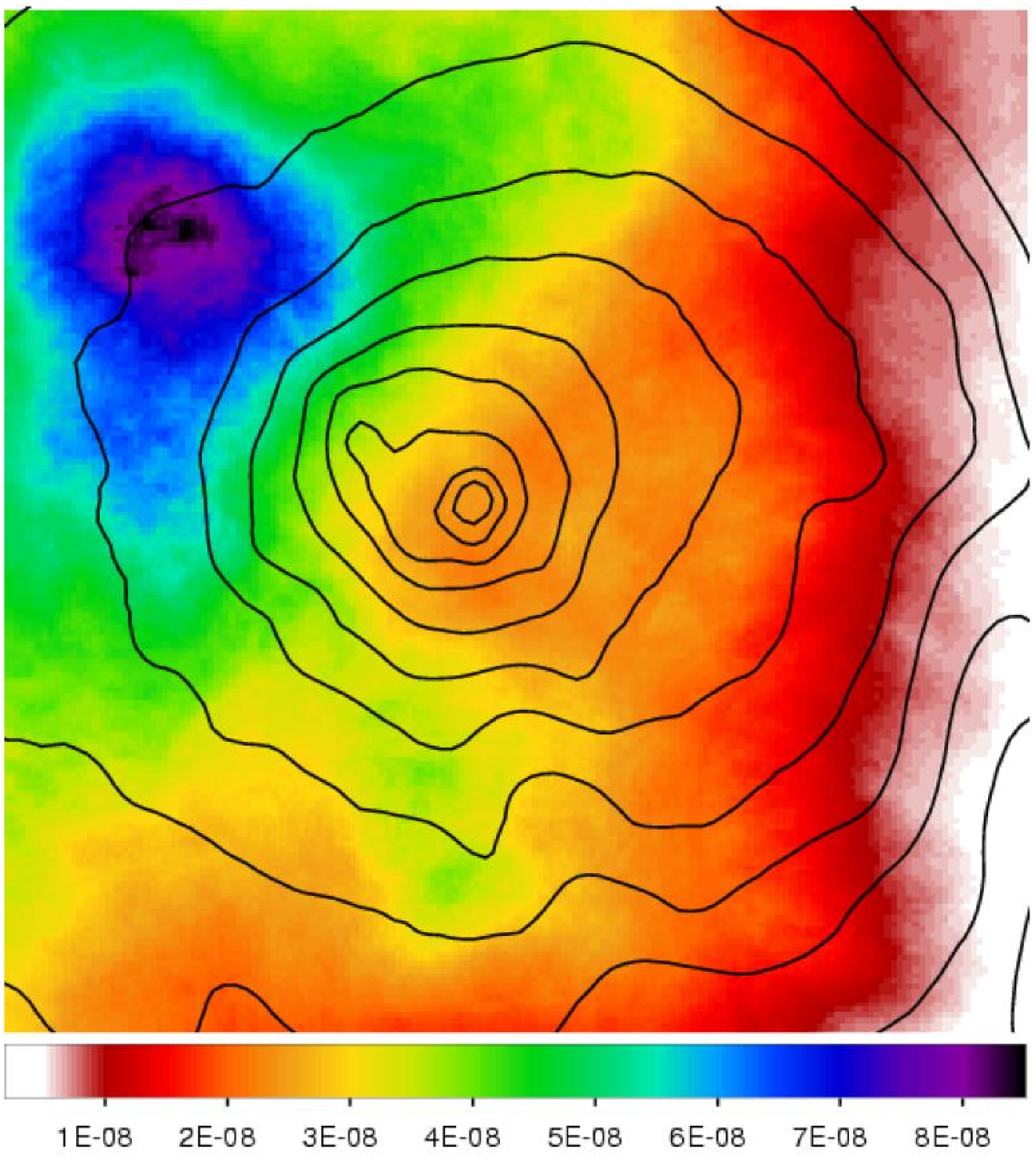} &
      \includegraphics[width=1.8in,angle=0]{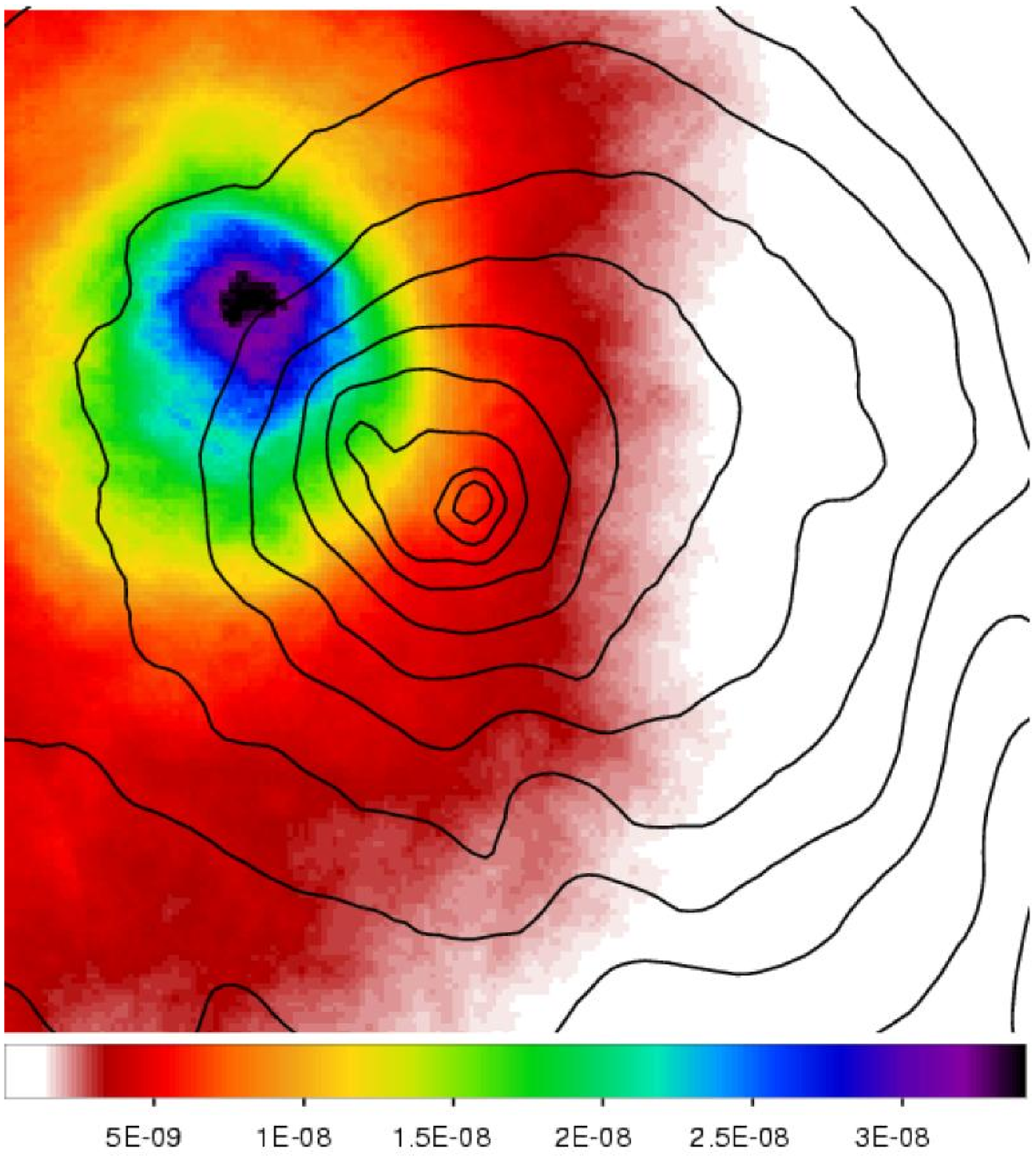} \\
    \end{tabular}
  \end{center}
  \caption{Luminosity maps ($L_{bol}/10^{44}~$erg$~$s$^{-1}$ per
$2\arcsec \times 2\arcsec$) of the Centaurus cluster in the 0 - 1, 1 - 2, 2 - 4, 
4 - 6, 6 - 8 and 8 - 10 keV temperature bands. 
  The superposed contours represent full band luminosity. \label{cenlband}}
\end{figure*}

For the Centaurus cluster we have focused on a new approach in 
analyzing cluster structure. In Figure \ref{cenlband} we have produced 
median maps for the luminosity in bands of different temperatures. 
This is similar to the method of \citet{fabian06}, who fit a six-phase 
plasma model to finely spatially binned data of the Perseus cluster 
in order to make maps of gas mass in six different temperature bands.
The main differences with our modeling are that here we use a smooth 
cluster model and allow for a very large number of variable temperature 
phases. 
From left to right, Figure \ref{cenlband} shows the cluster luminosity in the 0 - 1, 
1 - 2, 2 - 4, 4 - 6, 6 - 8, and 8 - 10 keV temperature bands. 

This subdivision reveals some striking features: the cluster core 
is dominant in the 0-1 keV band, and apparently it has moved in 
from the northeast, leaving a tail of colder gas. The 1 - 2 keV temperature 
band map shows a bi-polar nature of this emission around the core. 
With higher temperatures, the emission becomes more and more offset, and 
the 8 - 10 keV map shows a concentration of superheated gas in 
an isolated region to the north-east, possibly the remnant of a merger. 
This was noted in the analysis of the {\sl Chandra} data for this cluster 
by \citet{crawford05}.
Interestingly, this feature is 
aligned with a filament extending from the cluster core. Possibly 
these two irregular phenomena are related. 

\begin{figure*}[!htb]
\begin{center}
  \begin{tabular}{c}
    \includegraphics[width=2.5in]{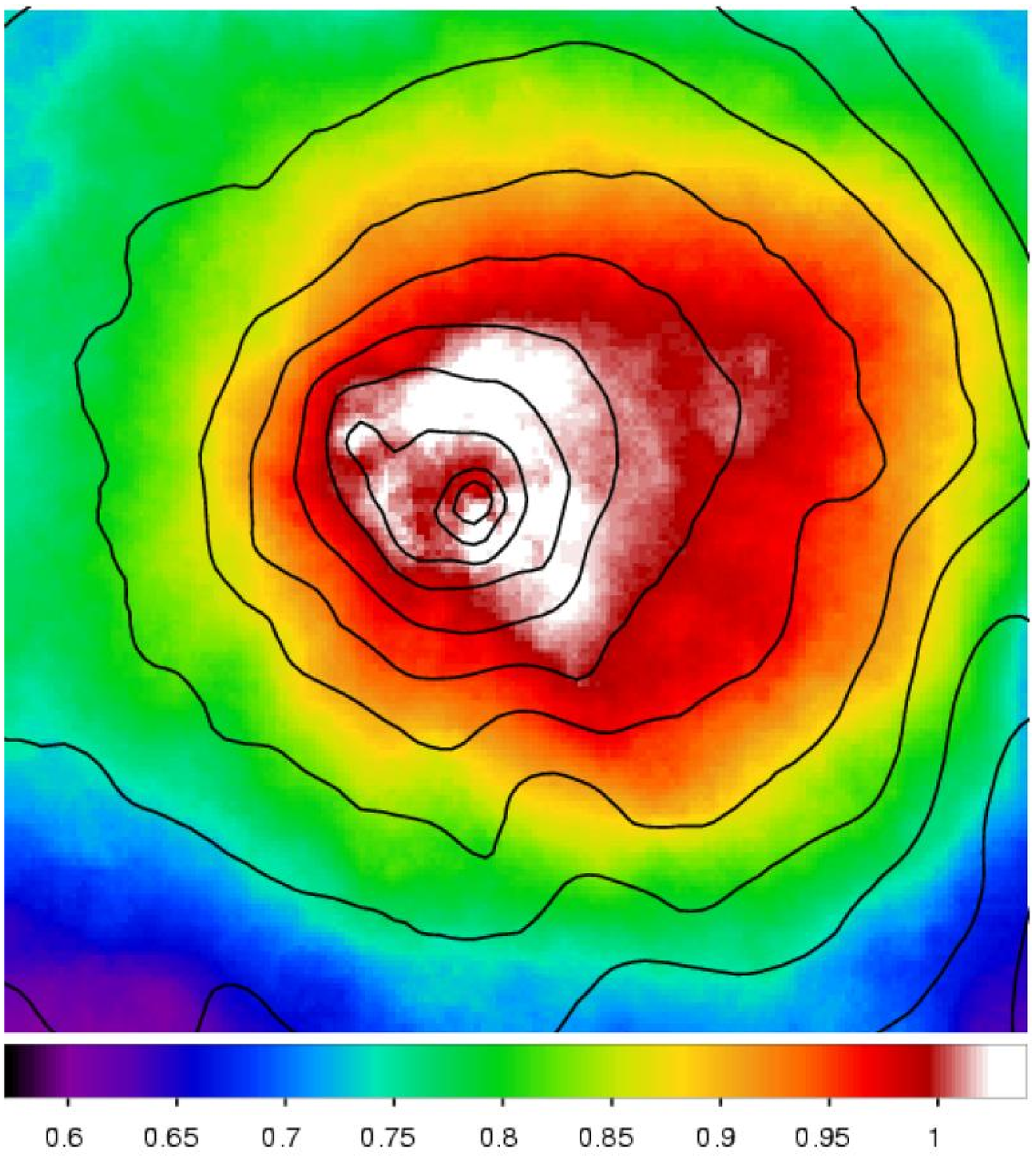} \\   
  \end{tabular}
\end{center}
  \caption{Metallicity map of the central $6\arcmin \times 6\arcmin$ region of Centaurus 
showing the metallicity with respect to solar, and with contours 
representing luminosity. \label{cenabun}}
\end{figure*}

Finally, we explore the metallicity structure of Centaurus by 
creating a median metal abundance map, analogous to our construction 
of median temperature maps. The metallicity map is shown in 
Figure \ref{cenabun} and confirms previous findings by \citet{fabian05} 
and \citet{sanders06} that the metallicity increases by about a factor of 
2 (from 0.5 to 1.0 solar) towards the center. However, in contrast to 
\citet{fabian05}, we do not find metal abundance levels as high as 
twice the Solar value and we also do not see a sharp decrease towards 
0 in the very center. A minor dip can however, be distinguished. 
It is most likely that both the absence of very high metallicity and 
the sharp central feature are due to PSF smearing effects. 
We also confirm a tight correlation between temperature and metallicity 
in the gas, with lower temperature plasma generally having a higher 
metallicity. Calculating the Pearson product-moment correlation coefficient 
for the median values of temperature and metal abundance per $2\arcsec \times 2\arcsec$ 
bin, as shown in Figures \ref{cenTerr} and \ref{cenabun}, weighted by the 
luminosity in that bin gives a coefficient of $-0.87$, hence confirming a strong 
correlation. 
We do not find any significant discrepancy in the spatial distribution of 
the redshift of the gas, in agreement with \citet{ota}.

\section{Discussion}
This paper describes an application of the Markov chain Monte Carlo 
technique developed by us for the analysis of observations with the {\sl XMM-Newton} imaging 
instruments.  We demonstrate the flexibility and power of this technique -- 
employing smoothed particle inference -- 
via studies of three very different clusters, Abell 1689, RX J0658-55 and 
Centaurus, 
especially regarding the ability to determine the spatial distribution of 
temperature of the radiating plasma, which is difficult via more traditional 
techniques.  

We found evidence for cluster
merger activity in all these systems, but
in each case, the signature was quite distinct.  The bullet of RX J0658-55,
the remnant of a merger in Centaurus and the asymmetry of temperature in
A1689 may roughly correspond to the early, middle, and late stages
of cluster merging. In all cases the core of the cluster seemed
relatively unaffected. Further systematic studies in temperature structure
may add to our understanding of the effect of mergers on cluster
properties.

The most important difference of our technique compared to conventional 
modeling is that we use overlapping emission components.  This facilitates 
the use of a large number of phases describing the X-ray emission at each 
spatial coordinate.   This scenario is much more physical than when one 
adapts the assumption that any given point in the projected cluster image 
can be described by a single-temperature phase.  However, it also poses the challenge 
of constraining a distribution of phases when the spectra of high-temperature plasmas in 
particular are very similar. 
Comparison with other analyses requires the projection of a distribution into 
a single value.  We have chosen in this paper to show the mode and the median of that 
distribution in our temperature maps along with the associated 1 $\sigma$ variations 
(see Figures \ref{A1689Terr}, \ref{bulletTerr}, and \ref{cenTerr}).  

The difference in the assumptions in modeling precludes any direct comparison with other 
measurements, and it is more than likely that the results will be different.  
The median of our distributions in particular will depend on the choice of the 
allowed range of temperatures in the model.  This is especially true for 
the choice of the upper boundary in high-temperature sources.  
The selection of the boundaries must be based on prior knowledge of the 
source, such as measurements in the hard X-ray band.  
This introduces a systematic uncertainty that has to be taken 
into account.  

Another advantage of this technique is the way we treat the background. Since 
the background is modelled instead of subtracted, difficult aspects 
of background subtraction, such as weighing the different vignetting of different 
components, are avoided.  The background model has been calibrated against a 
number of observations in which the filter wheel was closed, as well as against 
numerous ``blank sky'' observations (see Section \ref{backsec}).  Even though we allow the normalizations 
of different background components to be optimized by the MCMC, there is still 
a small systematic uncertainty associated with residual calibration uncertainties.

In this work we have opted to simultaneously fit the spatial distributions 
of the temperature, metallicity, and redshift of the X-ray-emitting, plasma 
illustrating the broad power of the method. 
This is not necessarily the optimal approach to answer a specific question 
about the cluster, such as resolving a redshift difference across A1689. 
This may explain the larger statistical uncertainty obtained in this 
work compared to our conventional analysis \citep{andersson}.
For such a problem the method could be designed specifically for this purpose.  
Since temperature and metallicity variations are small across this cluster 
these could be global parameters in the modeling significantly simplifying 
the overall problem.  For such tasks that are highly dependent on detailed 
spectral features, the number of cluster particles should also be kept 
to a minimum to avoid the risk of smearing the spectral feature 
in question. 
This is, however, a compromise, 
since a lower number of particles would limit the spatial 
resolution of the spectral feature that one wants to resolve. 

In summary, the statistical uncertainties in this work, which are inherent in 
problems with large numbers of degrees of freedom, may seem larger 
than those that are achieved in traditional analyses, but the uncertainties 
can always be reduced by simplifying the problem.  This work attempts 
to constrain all cluster characteristics into a single model.  
There will be other applications in which a specific model can be tested 
to answer a specific problem.

\section{Future work}

We expect that this method will prove to be powerful in the determination
of cluster gas density and entropy. Since the model is multi parametric and analytical, 
based on the Markov chain posterior, the smoothed particles can be manipulated 
in order to construct a three-dimensional cluster model.
In principle, the $z$-coordinate can be chosen for each particle on the basis of some assumption
of spherical symmetry. This can be done separately in ranges of particle plasma
temperature, assuming that particles of similar temperatures exhibit the same structure
as in the two-dimensional case. This will accurately produce the temperature gradient in the
three-dimensional case while preserving the two-dimensional observed structure.
This is likely the most accurate method to determine the three-dimensional 
spectrally resolved structure of galaxy clusters in the X-ray band.

Since the likelihood
calculation does not have to be limited to a single instrument, but can
potentially include multiple X-ray instruments or even data from other
kinds of measurements (such as gravitational lensing, Sunyaev-Zeldovich
data, or optical velocity dispersions), the method is quite
general.  This method is also well suited to analyze other complex,
spatially resolved objects where
the observed X-ray emission might consist of superposed separable
components with varying spectral parameters, such as in supernova remnants.


\section{Acknowledgments}

We acknowledge many helpful discussions on the suitability of smooth
particles for modeling cluster data with Phil Marshall. 
Financial support for KA is
provided  by the G\"{o}ran Gustavsson Foundation for Research in Natural
Sciences  and Medicine. This work was supported in part by the U.S.
Department of Energy under contract number DE-AC02-76SF00515.



\end{document}